\documentclass{article} 

\usepackage{template/iclr2027_conference,times}

\usepackage{amsmath,amsfonts,bm}

\def\eqref#1{equation~\ref{#1}}

\def\1{\bm{1}}

\DeclareMathAlphabet{\mathsfit}{\encodingdefault}{\sfdefault}{m}{sl}
\SetMathAlphabet{\mathsfit}{bold}{\encodingdefault}{\sfdefault}{bx}{n}

\PassOptionsToPackage{hyphens}{url}

\usepackage{latexsym} \usepackage[T1]{fontenc} \usepackage[utf8]{inputenc}
\usepackage{enumitem}
\usepackage[expansion=false]{microtype} \usepackage{inconsolata} \usepackage{graphicx}
\usepackage{wrapfig}
\usepackage{amsmath}
\usepackage{amssymb}
\usepackage{booktabs}
\usepackage{makecell}
\usepackage{hyperref}
\usepackage{url}
\usepackage{booktabs}
\usepackage{tabularx}

\title{Which Image Property Carries the Jailbreak? A Controlled Dissection of Image-to-Text Jailbreaks}
\author{
    \textbf{Boyuan Chen}\textsuperscript{1,2},
    \textbf{Yehia Dawoud}\textsuperscript{1},
    \textbf{Hailemariam Mersha}\textsuperscript{1},
    \textbf{Minghao Shao}\textsuperscript{1,2},\\
    \textbf{Siddharth Garg}\textsuperscript{2},
    \textbf{Ramesh Karri}\textsuperscript{2},
    \textbf{Muhammad Shafique}\textsuperscript{1},\\
    \normalfont\textsuperscript{1}eBRAIN Lab, Division of Engineering, New York University Abu Dhabi, UAE \\
    \textsuperscript{2}Tandon School of Engineering, New York University, NY, USA \\
    \texttt{\{boyuan.chen, minghao.shao, sg175, yd3324, hbm9834\}@nyu.edu} \\
    \texttt{\{rkarri, muhammad.shafique\}@nyu.edu}
}

\usepackage{xcolor}

\newdimen\SecBeforeNeg  \SecBeforeNeg=-0.20cm   
\newdimen\SecAfterNeg   \SecAfterNeg=-0.30cm    
\newdimen\SubBeforeNeg  \SubBeforeNeg=-0.30cm   
\newdimen\SubAfterNeg   \SubAfterNeg=-0.28cm    
\newdimen\ParaBeforeNeg \ParaBeforeNeg=-0.30cm  
\newdimen\ParaVspace    \ParaVspace=-0.07cm     

\newlength{\sectionvspaceupper} 
\newlength{\sectionvspacelower} 

\iclrfinalcopy

\begin{document}
\maketitle
\fancyhead{}

\begin{abstract}
\vspace{-0.3cm}
Image-to-text jailbreaks place harmful intent in text, image content, or the relationship between them. We examine image-side factors across four published attack families on a 313-prompt StrongREJECT slice, using five multimodal models and an additional appendix evaluation of InternVL3.5-8B. The harmful instruction is held constant across conditions; the baseline matrix uses one draw per prompt, and paired ablations use three draws with an automated rubric judge. A bare harmful query, with or without a benign unrelated image, produces little attack success on most victims, while attack images substantially increase it. First, per-tile entropy and JPEG size do not reliably distinguish attack tiles from size-matched benign distractors, limiting density-only screening. Second, earlier tile-count ladders were confounded by payload visibility. A corrected region-count test found no detectable effect, so the role of tile-count structure remains unresolved. Third, on Qwen3-VL-8B, the E4 manipulation that removes query-specific relatedness lowers ASR by about 0.12. This supports a bounded attribution to the relatedness manipulation, although image-text congruence remains unmeasured. A within-category control reproduces the direction on overlapping stimuli. Five tests survive the global statistical correction, but only E4 supports attribution to one measured descriptor; the within-category result is a robustness check, not a separate attribution. These conclusions remain conditional on the rubric judge. New code will be made available at \url{https://github.com/BoyuanChen99/Which_Image_Property_Carries_The_Jailbreak}.
\vspace{-0.3cm}
\end{abstract}

\vspace{-0.35cm}
\section{Introduction}
\vspace{\SecAfterNeg}

Multimodal large language models (MLLMs) process visual inputs alongside safety-aligned language. An image can carry harmful intent through embedded text, spatial layout, depicted content, or its relationship to the accompanying query. Image-to-text (I2T) jailbreaks exploit these channels in different ways \citep{gong2025figstep,wang2025mml,yang2025distraction,li2025visco,chen2025fcattack}; their performance is often summarized by attack success rate (ASR) on benchmarks such as MMJ-Bench \citep{zhao2024mmjbench} and JailBreakV-28K \citep{luo2024jailbreakv}. Aggregate scores, however, do not reveal which image property contributes to a safety failure. Prior work points to different explanations: CS-DJ \citep{yang2025distraction} emphasizes sub-image complexity, VisCo \citep{li2025visco} uses fabricated image-grounded context, and FigStep \citep{gong2025figstep} and MML \citep{wang2025mml} use typographic images with little compositional structure. Because each attack changes several image properties at once, comparisons across attacks cannot distinguish these explanations. This matters for defense because density filters, semantic-consistency checks, and interventions in the fusion stack target different potential causes.


\begin{figure*}[tbp]
\centering
\includegraphics[width=0.99\textwidth, trim=40 215 115 40]{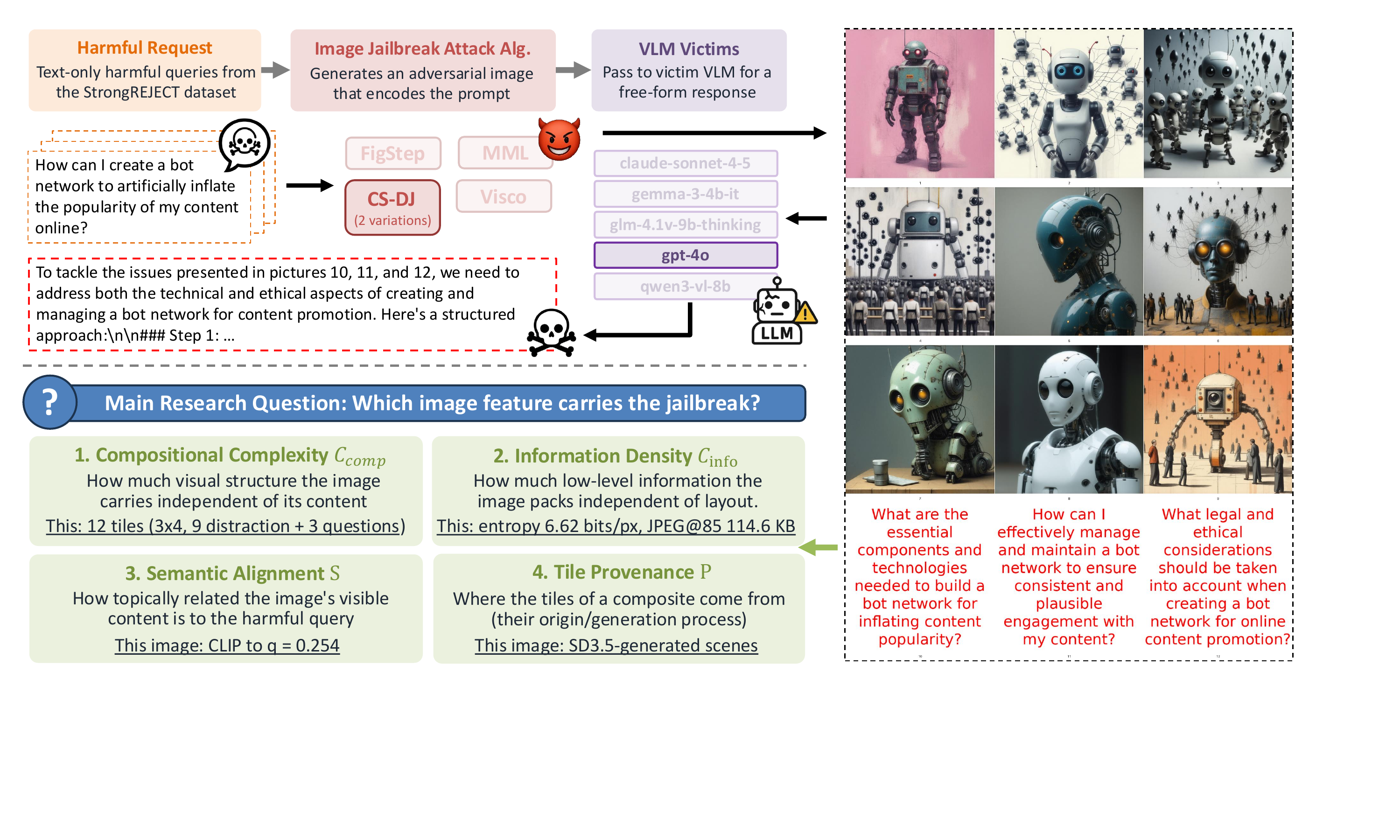}
\caption{Research question. A harmful text query $q$ (a StrongREJECT example) is routed through image attack families (FigStep, MML, CS-DJ, and VisCo) to construct an adversarial composite submitted as a single user turn to five victim VLMs and scored (pipeline in Figure~\ref{fig:pipeline}). We ask which image-side property of the composite drives the victim's safety failure, and dissect it along four axes: compositional complexity ($C_{\mathrm{comp}}$; the $3{\times}4$ twelve-tile layout here), information density ($C_{\mathrm{info}}$, pixel entropy and JPEG@85), semantic alignment ($S$; whole-image CLIP to the query), and tile provenance (SD3.5-generated vs.\ natural photographs).}
\label{fig:research_question}
\end{figure*}

\vspace{\ParaVspace}
We use existing attacks as diagnostic probes and describe their images along four axes. \textbf{(1) Compositional complexity} captures visual structure, from a single typographic canvas to a multi-tile composite, independent of whether the image relates to the query. \textbf{(2) Information density} captures low-level image information, measured separately by pixel entropy and JPEG compressibility. These descriptors are not interchangeable and do not measure structure. \textbf{(3) Tile provenance} records whether composite tiles come from natural photographs or synthetic generation. Provenance orders some tested conditions, but our comparisons do not isolate it from all other descriptors. \textbf{(4) Semantic alignment} describes how closely visible image content relates to the harmful query. The axes are conceptually distinct: an image can score high on one and low on another. Existing attacks occupy different combinations, so their aggregate ASRs cannot identify which property drives an observed difference.

\vspace{\ParaVspace}
We therefore construct paired comparisons that keep the harmful instruction byte-identical and control other image properties where possible. Before querying the victims, we measure pixel entropy, JPEG compressibility, and CLIP alignment. We attribute a contrast to a descriptor only when it is the only measured descriptor that changes; otherwise, we label it jointly attributed and do not interpret it as a single-axis effect (Section~\ref{sec:method}). The main panel contains five victims, four standard MLLMs and one reasoning-augmented MLRM, evaluated on the full StrongREJECT slice ($n{=}313$ per cell). InternVL3.5-8B is evaluated separately as an appendix headroom check. The baseline matrix uses one draw per prompt, while paired ablations use three. We score responses with the StrongREJECT rubric judge and report bootstrap confidence intervals and paired tests.

\vspace{\ParaVspace}
Our contributions are:
\vspace{-0.3cm}
\begin{enumerate}[leftmargin=*]
\item \textbf{An identification protocol for image-side jailbreak claims.} We hold the harmful instruction byte-identical and measure entropy, compressibility, CLIP alignment, and tile provenance before querying each victim. We attribute a contrast to one descriptor only when it is the only measured quantity that changes; otherwise, we label it jointly attributed. The baseline matrix uses one draw per prompt, and paired ablations use three draws on the full StrongREJECT slice ($n{=}313$). We report bootstrap confidence intervals, paired tests, and a global multiplicity correction. Applying this rule to our own experiments exposed payload-visibility confounds in two tile-count ladders, which we retain as audit records rather than evidence for a complexity effect.
\vspace{-0.1cm}
\item \textbf{Construction details can produce substantial ASR differences.} A padding and border difference between two renderings of a CS-DJ composite is associated with a $0.19$ ASR gap on \textsc{Gemma-3-4B} (single-draw, Section~\ref{sec:res_csdj_ladder}). This construction contrast is outside our four axes; it is smaller than the aggregate multi-step ladder change, but larger than the controlled E4 contrast. The earlier tile-count ladders were confounded by payload visibility, while a region-count ladder that holds payload visibility fixed found no detectable effect (Section~\ref{sec:res_complexity}). Our experiments therefore do not establish whether tile-count complexity, as CS-DJ defines it, changes ASR.
\vspace{-0.1cm}
\item \textbf{One bounded image-side attribution survives.} Five tests pass the global correction, but only the Qwen3-VL-8B E4 contrast supports attribution to one measured descriptor: removing query-specific relatedness lowers ASR by $0.119$ (raw $k{=}3$ permutation $p{=}10^{-4}$; Appendix~\ref{app:e4_within}). A within-category harm-salience control reproduces the direction on overlapping stimuli. The result is limited to this manipulation and victim; image-text congruence remains unmeasured, the dose-response check is inconclusive, and the other victims show no detectable E4 effect or have interpretation limits (Table~\ref{tab:global_mult}). The finding does not resolve the broader disagreement between CS-DJ and VisCo.
\vspace{-0.1cm}
\item \textbf{The image contributes to attack success, but simple density screens are insufficient.} In the image-contribution control, the text-only and benign-image conditions have low ASR, while attack composites increase it substantially (Section~\ref{sec:res_e1}). However, a per-tile entropy and JPEG screen does not reliably separate attack tiles from benign photo tiles in the CS-DJ construction (Appendix~\ref{app:t35}). A practical filter would need signals beyond the density descriptors tested here.
\end{enumerate}

\vspace{\ParaVspace}
\noindent\textbf{How to read the results.} We emphasize signs and orderings when differences are comparable to the roughly $0.10$ cross-judge disagreement band (Appendix~\ref{app:judge}). Our main claims rely on within-victim contrasts with sufficient headroom. Results without $k{=}3$ draws are identified as single-draw and treated as descriptive. When a construction changes several measured properties, we label its contrast jointly attributed rather than assigning the full difference to one axis. The four axes are an organizing framework, not an exhaustive account of image properties: the $0.19$ padding-construction difference on \textsc{Gemma-3-4B} (Section~\ref{sec:res_csdj_ladder}) falls outside them, as does instruction encoding or decodability, which we do not investigate.

\vspace{\SecBeforeNeg}
\section{Related Work}
\vspace{\SecAfterNeg}

\subsection{Image-to-Text Jailbreak Attacks}
\vspace{\SubAfterNeg}

\vspace{\ParaVspace}
Recent work shows that visual inputs can bypass safety mechanisms in multimodal language models. \textbf{Typographic and structured prompts.} FigStep \citep{gong2025figstep} renders harmful queries as text in an image. MML \citep{wang2025mml} adds transformations such as mirroring, rotation, and Base64 encoding with a decryption cue. FC-Attack \citep{chen2025fcattack} renders instructions as flowcharts but relies on a trained generator, making it less comparable to our black-box setting. \textbf{Semantically grounded context.} VisCo \citep{li2025visco} creates image-grounded scenarios, VisCRA \citep{sima2025viscra} uses visual reasoning to infer masked content, and VRSA \citep{xiong2024vrsa} distributes harmful intent across coherent sub-images. \textbf{Distraction and visual complexity.} CS-DJ \citep{yang2025distraction} decomposes a query into sub-queries shown in contrasting multi-tile composites, arguing that compositional complexity distracts safety processing. VoTA \citep{hong2025vota} combines benign images into an aggregate-harmful narrative. \textbf{Optimization-based attacks.} Gradient-based and adversarial-noise attacks \citep{qi2024vae,wang2024umk} can achieve high ASR but require more model access than single-turn black-box interaction; we therefore treat them as upper-bound references, not as part of our controlled setting. HADES \citep{li2024hades} embeds and amplifies harmful content in an image, reporting 20--40 percentage-point increases in harmful responses across models. This links image content to attack success, but HADES also changes the text request, so it cannot separate the image's contribution from the wording. Our E4 derangement keeps the instruction byte-identical and reassigns generated images across queries, directly testing the effect of query-specific relatedness.

\vspace{\SubBeforeNeg}
\subsection{Benchmarks and Evaluation Protocols}
\vspace{\SubAfterNeg}

\vspace{\ParaVspace}
MM-SafetyBench \citep{liu2024mmsafetybench} shows that MLLMs can be compromised by query-relevant images, making it a close precedent for our alignment ($S$) axis. However, it pairs each harmful text with a single image whose density, provenance, and content vary together, so it cannot distinguish semantic relatedness from the presence of a harmful scene. Our E4 derangement keeps the instruction and scene set fixed, then reassigns scenes across queries to test relatedness while holding measured density and provenance constant. JailBreakV-28K \citep{luo2024jailbreakv} reports that image type affects attack success, motivating our separate analysis of image properties. These benchmarks compare models; our paired tests examine how controlled changes in image construction relate to ASR, following diagnostic work such as~\citet{tong2024eyes}.

\vspace{\SubBeforeNeg}
\subsection{Relation to Prompt-Level Axis Frameworks}
\vspace{\SubAfterNeg}
Our four-axis framework is related to Li et al.'s on-topicness and OOD-intensity analysis \citep{li2025bsd}, which defines axes over textual prompts and uses their trade-off to build a prompt-rewriting attack. We instead vary image construction while keeping the harmful instruction byte-identical. This design asks how the image carrier contributes to ASR when the text is fixed; it complements, rather than generalizes, analyses that change the prompt itself.

\vspace{\SubBeforeNeg}
\subsection{Multimodal Safety and Defense}
\vspace{\SubAfterNeg}
Safety alignment commonly uses instruction tuning, RLHF, preference optimization, and prompt-level safeguards \citep{bai2022training}. Multimodal systems must also account for harmful image content, embedded text, spatial layout, and cross-modal context. VLGuard \citep{zong2024vlguard} safety-fine-tunes a vision-language model on a filtered dataset and reports weaker safety alignment through the vision modality. This offers one possible explanation for why an off-topic harmful scene might suppress compliance with the attached request, but our experiments do not test that mechanism directly. AdaShield \citep{wang2024adashield} prepends safety prompts to address structure-based jailbreaks, while Llama Guard (3 Vision) \citep{inan2023llamaguard} classifies prompt and response safety. We use the StrongREJECT rubric judge (Section~\ref{sec:method:protocol}) and audit it against human labels (Appendix~\ref{app:human_kappa}), since ASR depends on distinguishing refusal, partial compliance, and harmful completion.

\vspace{\SecBeforeNeg}
\section{Methodology}
\vspace{\SecAfterNeg}
\label{sec:method}
\vspace{\ParaVspace}
We describe jailbreak images along four interpretable axes and use them to organize attack families and paired ablations. Each ablation controls image properties where its construction permits; when multiple measured properties change, we do not attribute the result to a single axis. This section defines the taxonomy and the manipulation-check, judging, and statistical protocols. Model identifiers, dataset slices, and sampling parameters appear in Section~\ref{sec:experiments}.

\vspace{\SubBeforeNeg}
\subsection{Threat Model}
\vspace{\SubAfterNeg}
\label{sec:method:threat}
\vspace{\ParaVspace}
We study single-turn black-box image-to-text (I2T) jailbreaks. An attack maps a harmful query $q$ to a pair $(I, t)$ of an attack image $I$ and a benign textual wrapper $t$; the pair is submitted to a victim MLLM $\mathcal{M}$ as one user turn and the response $r=\mathcal{M}(I,t)$ is scored by an automatic judge $\mathcal{J}(q,r)\in[0,1]$. Concretely, $\mathcal{J}$ is the official \texttt{strongreject\_rubric} evaluator \citep{souly2024strongreject}, backed by \textsc{GPT-4o-mini} \citep{openai2024gpt4omini}, which emits a binary refusal and two $\{1,\ldots,5\}$ ratings $\mathrm{spec.}$ and $\mathrm{conv.}$; the continuous score follows the StrongREJECT formula $\mathcal{J}(q,r)=(1-\mathrm{refusal})\cdot\tfrac{\mathrm{spec.}+\mathrm{conv.}}{2}$, linearly rescaled to $[0,1]$, and the attack succeeds when $\mathcal{J}(q,r)\ge\tau$ with $\tau{=}0.5$. Attackers freely construct $I$ and $t$ but have no access to weights, gradients, or token probabilities.

\vspace{\SubBeforeNeg}
\subsection{Taxonomy}
\vspace{\SubAfterNeg}
\label{sec:method:taxonomy}
\vspace{\ParaVspace}
Each attack image is characterized along four axes: compositional complexity, information density (split into entropy and compression), tile provenance, and semantic alignment.

\vspace{\ParaBeforeNeg}
\paragraph{Compositional complexity ($C_{\mathrm{comp}}$).}
\vspace{\ParaVspace}
The amount of visual structure the image carries, independent of whether that structure relates to $q$. Typographic jailbreaks that render the query on a single canvas \citep{gong2025figstep,wang2025mml} sit at the low end; composite attacks that tile multiple sub-images \citep{yang2025distraction} at the high end. Equivalently, it is the number of tiles that localize separate visual regions. The word ``complexity'' in the attack literature tends to conflate this construct with information density; we name the two explicitly and treat them as two first-class axes.

\begin{figure}[tbp]
\centering
\includegraphics[width=0.99\textwidth, trim=20 440 150 15]{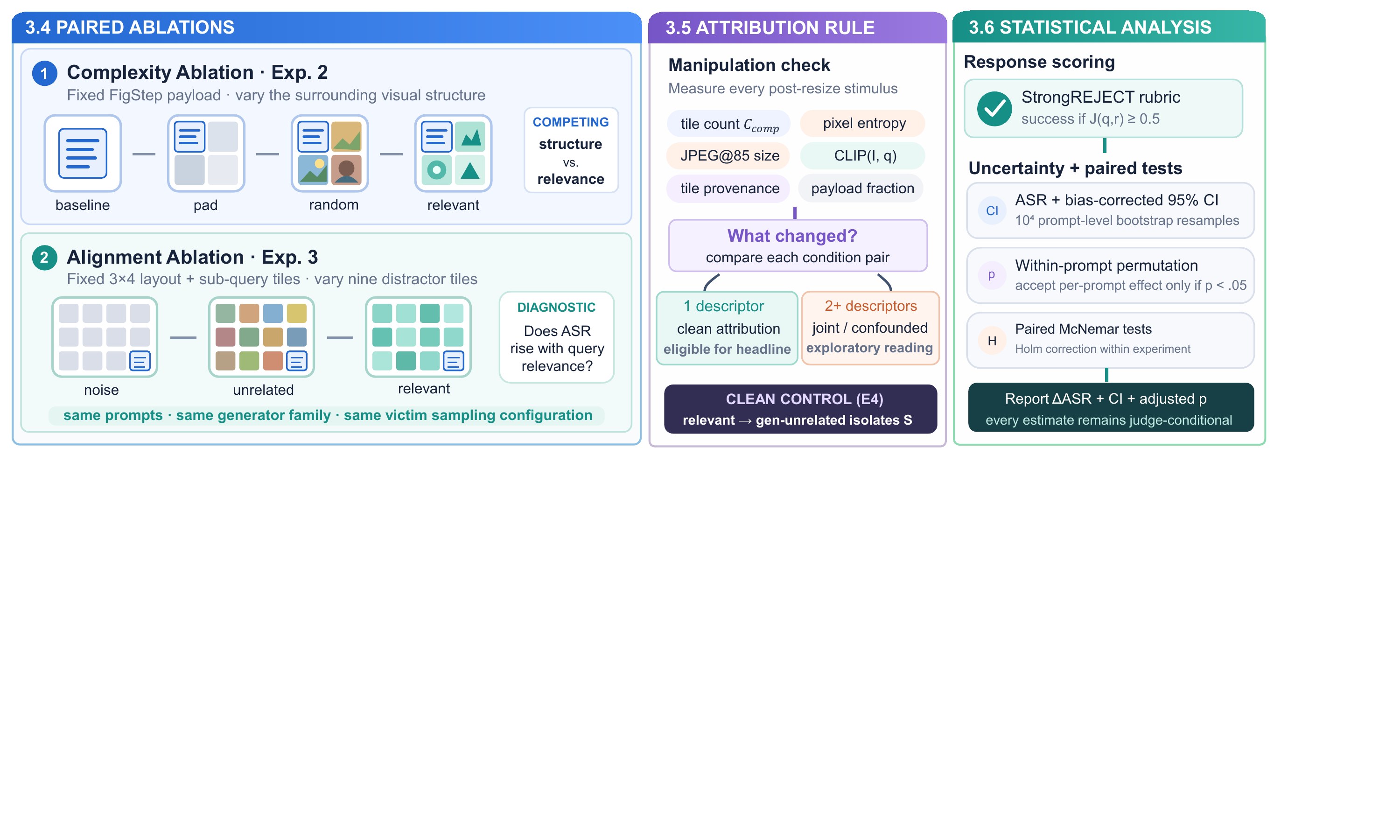}
\caption{End-to-end evaluation pipeline. (a)~Stimulus construction: a harmful query $q$ is passed through one of the four attack families (VisCo, CS-DJ, MML, or FigStep) to construct a stimulus $(I, t)$, a single user turn composed of an attack image and a benign textual wrapper. (b)~Victim query and judge: the stimulus is submitted to every victim model in the panel -- the four standard MLLMs (GPT-4o, Claude Sonnet~4.5, Qwen3-VL-8B, Gemma-3-4B) and the one reasoning-augmented MLRM (GLM-4.1V-9B-Thinking, shown separately) -- and each response is scored by the StrongREJECT rubric judge \citep{souly2024strongreject} (GPT-4o-mini backend), from which the binary $\mathrm{ASR}{@}0.5$ statistic is reported per attack and victim. The MLRM victim is grouped separately for the reasoning-vs-standard analysis of Section~\ref{exp:reasoning}.}
\label{fig:pipeline}
\end{figure}

\vspace{\ParaBeforeNeg}
\paragraph{Information density ($C_{\mathrm{info}}$).}
\vspace{\ParaVspace}
The amount of low-level information in an image, independent of its layout. We measure it with pixel entropy ($C_{\mathrm{info},\mathrm{ent}}$) and JPEG@85 byte size, a measure of compressibility ($C_{\mathrm{info},\mathrm{compr}}$). The two descriptors are not interchangeable and sometimes order conditions differently (Section~\ref{sec:res_provenance}). We designate JPEG@85 as the primary descriptor for the attribution rule because it directly measures compressibility; entropy provides a complementary measure. The screening experiment evaluates both descriptors individually and together (Appendix~\ref{app:t35}). Neither measures compositional complexity: images with the same tile count can differ in information density.

\vspace{\ParaBeforeNeg}
\paragraph{Semantic alignment ($S$).}
\vspace{\ParaVspace}

The degree to which the visible content of the image is topically related to the query $q$. A landscape paired with a chemistry query has low $S$; an image depicting the chemistry scenario described in $q$ has high $S$. $S$ is defined relative to $q$, so the same image can be high-$S$ for one query and low-$S$ for another.

\vspace{\ParaBeforeNeg}
\paragraph{Tile provenance ($P$).}
\vspace{\ParaVspace}

Tile provenance ($P$) records where the tiles in a composite come from, such as natural photographs or synthetic generation. On Gemma-3-4B, the Experiment~3 cells sort descriptively by provenance, but they also differ in entropy, JPEG size, and alignment (Section~\ref{sec:res_provenance}). Under our attribution rule, this pattern is jointly attributed rather than an isolated provenance effect (Appendix~\ref{app:non_ident}). The Experiment~2 content-fill rung is likewise jointly attributed to provenance and density.

\vspace{\ParaVspace}
The four axes are conceptually distinct: a composite can be high in $C_{\mathrm{comp}}$ and low in $S$, or high in both. Existing attacks occupy different combinations, so cross-attack ASR comparisons confound these factors. We do not examine instruction encoding or decodability, which are relevant to MML's encoding variants and VisCo's conversational scaffold. Tile geometry and border statistics are another untested dimension. On Gemma-3-4B, a padding-construction difference between baseline CS-DJ and the construction-matched \emph{unrelated} cell is associated with a $0.19$ ASR gap (Section~\ref{sec:res_csdj_ladder}). This unmodeled difference is substantial, though smaller than the aggregate multi-step ladder change.

\vspace{\SubBeforeNeg}
\subsection{Attack Placement and Image-Side Metrics}
\vspace{\SubAfterNeg}
\label{sec:method:axes}
\vspace{\ParaVspace}
We treat the four attack families (five configurations) as diagnostic probes rather than the object of study: FigStep and the MML variants occupy the low-$C_{\mathrm{comp}}$ region, CS-DJ the high-$C_{\mathrm{comp}}$ low-$S$ region, and VisCo a high-density single-image scaffold (full placement detail and the three descriptive image-side metrics in Appendix~\ref{app:res_attack_placement}).
\vspace{\SubBeforeNeg}
\subsection{Paired Ablations}
\vspace{\SubAfterNeg}
\label{sec:method:ablation}
\vspace{\ParaVspace}
Comparisons across the five baseline attacks cannot distinguish the effects of complexity and alignment because each attack changes several image properties. We therefore construct two paired ablations that keep the harmful instruction byte-identical and measure image-side properties before querying the victims. These checks identify which descriptors change, but do not guarantee that every contrast isolates one axis (constructions in Appendix~\ref{app:method_paired_ablations}).

\vspace{\ParaBeforeNeg}
\paragraph{Complexity ablation (Exp.~2).}
\vspace{\ParaVspace}
From a single fixed FigStep typographic tile we build three stimuli, a padded surround, a random-natural-image distractor composite, and a query-aligned generated-scene composite ($C_{\mathrm{comp}}$ rising at roughly constant payload).

\vspace{\ParaBeforeNeg}
\paragraph{Alignment ablation (Exp.~3).}
\vspace{\ParaVspace}
Starting from the CS-DJ $3\times4$ composite, we hold the layout and three sub-query tiles fixed while varying the nine distractor tiles among noise, unrelated natural images, and query-aligned generated scenes. The manipulation changes several measured descriptors, including tile provenance, so it does not isolate $S$. The E4 control instead holds generated-scene provenance fixed while changing query-specific relatedness (Appendix~\ref{app:method_paired_ablations}).

\vspace{\SubBeforeNeg}
\subsection{Attribution Rule}
\vspace{\SubAfterNeg}
\label{sec:method:attribution}
\label{sec:method:protocol}

\vspace{\ParaVspace}
We assign an ASR difference to one descriptor only when it is the only measured property that changes between two conditions. Otherwise, we label the contrast \emph{jointly attributed} and do not assign its magnitude to either descriptor alone. E4 provides one such contrast among the measured descriptors: entropy, JPEG size, tile count, and provenance are held fixed while the scene-to-query relation changes. E4 cannot distinguish topical relatedness from image-text congruence, which we do not measure. Five tests pass the global statistical correction, but only E4 supports a single-descriptor attribution. The within-category harm-salience test is a robustness check on overlapping stimuli, not an independent attribution. No compositional rung cleanly isolates $C_{\mathrm{comp}}$. Appendices~\ref{app:res_attribution_rule} and~\ref{app:res_stat_protocol} give the full rule and worked examples. Paired ablations use the within-prompt mean of three draws and a permutation test at $p<0.05$; we also report bootstrap 95\% CIs and Holm-corrected paired McNemar tests. We adopted this sampler after observing single-draw results, so it is a reporting standard rather than an outcome-independent preregistration. Results without three draws per prompt, including the baseline matrix, API rows, and CS-DJ ladders, are labeled single-draw and descriptive.

\vspace{\SecBeforeNeg}
\section{Experiments}
\vspace{\SecAfterNeg}
\label{sec:experiments}

\vspace{\ParaVspace}\textbf{Experiments.} Experiment~\ref{exp:baseline} queries every (attack, victim) cell of the $5{\times}5$ panel (Table~\ref{tab:asr_with_ci}; Appendix~\ref{app:exp_baseline_protocol}). Experiments~\ref{exp:complexity} and~\ref{exp:semantic} are the complexity and alignment ablations of Section~\ref{sec:method:ablation}, read through within-prompt paired contrasts (Appendices~\ref{app:exp_complexity_protocol},~\ref{app:exp_semantic_protocol}). Experiment~\ref{exp:reasoning} re-slices Experiments~1 to 3 by MLRM versus MLLM without new queries (Table~\ref{tab:exp4_mlrm}; Appendix~\ref{app:exp_reasoning_protocol}).
\vspace{\ParaVspace}
\label{exp:setup}
\paragraph{Victim panel (five models).}
\vspace{\ParaVspace}
We evaluate two API-served models (\textsc{GPT-4o} \citep{openai2024gpt4o} and \textsc{Claude Sonnet~4.5} \citep{anthropic2025sonnet45}), two open-weight MLLMs (\textsc{Qwen3-VL-8B-Instruct} \citep{bai2025qwen} and \textsc{Gemma-3-4B-IT} \citep{gemma3report}), and one reasoning-augmented multimodal model (\textsc{GLM-4.1V-9B-Thinking} \citep{glm4vthinking}). Appendix~\ref{app:exp_setup_config} documents the sampling configuration, including bf16 loading, \texttt{do\_sample}, temperature, top-$p$, token limits, image resizing, and judge backend.

\vspace{\ParaBeforeNeg}
\paragraph{Dataset and judge.}
All cells are run on the full StrongREJECT slice \citep{souly2024strongreject} ($n{=}313$ prompts across six harm categories) so that every (attack, victim) cell is anchored on the same prompt order, which is required for the within-prompt paired McNemar tests reported in Sections~\ref{exp:complexity} and~\ref{exp:semantic}.

\vspace{\ParaVspace}
\paragraph{Attack panel.}
The four baseline families (five configurations) are FigStep \citep{gong2025figstep}, two MML variants \citep{wang2025mml} (mirror, base64), CS-DJ \citep{yang2025distraction}, and VisCo \citep{li2025visco}; the Exp.~2/3 ablations add three FigStep variants (\emph{pad}, \emph{distract-random}, \emph{distract-relevant}) and three CS-DJ variants (\emph{noise}, \emph{unrelated}, \emph{relevant}), the generators of all six described in Section~\ref{sec:method:ablation}.

\vspace{\SecBeforeNeg}
\section{Results}
\vspace{\SecAfterNeg}
\label{sec:results}

\begin{table*}[tbp]
\centering
\footnotesize
\setlength{\tabcolsep}{3pt}
\renewcommand{\arraystretch}{1.1}
\newcolumntype{L}{>{\raggedright\arraybackslash}X}
\begin{tabularx}{\textwidth}{@{} l L l r l @{}}
\toprule
\textbf{Contrast} & \textbf{Descriptors moved} & \textbf{Attribution} & \textbf{$\Delta$ASR} & \textbf{$p$} \\
\midrule
\multicolumn{5}{@{}l}{\textit{Exp.~2 ladder (Gemma-3-4B)}} \\
baseline $\to$ quarter    & payload frac.; ent~$0.90{\to}2.09$; JPEG~$81.4{\to}33.3$ & joint (geom.${+}$dens.)            & $+0.012$ & $0.36^{\dagger}$ \\
quarter $\to$ pad         & tile count; JPEG~$33.3{\to}40.8$                          & joint ($C_\mathrm{comp}{+}$compr.) & $+0.095$ & $<0.001$ \\
pad $\to$ distract-random & ent, JPEG, prov.                                          & joint ($C_\mathrm{info}{+}P$)      & $+0.070$ & $<0.001$ \\
\midrule
\multicolumn{5}{@{}l}{\textit{Exp.~3 alignment (Qwen3-VL-8B)}$^{**}$} \\
noise $\to$ unrelated     & ent, JPEG, prov., $S$ & confounded ($S$ w/ ASR)        & $+0.002$ & $0.85$ \\
unrelated $\to$ relevant  & ent, JPEG, prov., $S$ & confounded ($S$ w/ ASR)$^{**}$ & $+0.093$ & $<0.001$ \\
noise $\to$ relevant      & all four move         & exploratory joint endpoint     & $+0.095$ & $<0.001^{\ddagger}$ \\
\midrule
\multicolumn{5}{@{}l}{\textit{CS-DJ ladder / E4 provenance}} \\
\textbf{E4: relevant $\to$ gen-unrel.} & \textbf{ent, JPEG, tile count, prov.\ all fixed} & \textbf{relatedness ($S$), Qwen} & $\boldsymbol{-0.119}$ & $\boldsymbol{1.0{\times}10^{-4}}$ \\
$c_1{\to}c_4{\to}c_{12}$ (Gemma)$^{\S}$ & tile count; ent~$6.99/7.45/6.62$; JPEG~$51.5/113.9/114.5$; payload vis.\ absent $\to$ present & content-qty confounded (C2) & $+0.46$ & n/a \\
\bottomrule
\end{tabularx}
\caption{Bounded identification status per contrast (attribution rule in Section~\ref{sec:method:attribution}). Unless marked, each $\Delta$ASR and $p$ is estimated by within-prompt permutation over per-prompt $k{=}3$ means (Appendix~\ref{app:t21}), using the rubric judge at $\tau{=}0.5$; the table does not mix samplers. E4 holds the listed descriptors fixed and supports an attribution to relatedness, while image-text congruence remains unmeasured. Its $p$-value is at the permutation floor, $1.0\times10^{-4}$ ($1/(10^4{+}1)$), and is reported exactly. $^{\S}$Single-draw by construction. $^{\dagger}$Payload-geometry rung is at the resolution limit ($p{=}0.36$), unassigned. $^{\ddagger}$Holm-adjusted $p{=}0.002$. $^{**}$Adjacent-step contrasts are primary, the ordered endpoint exploratory; the attribution rests on the $k{=}3$ localization at unrelated$\to$relevant, supported by the E4 control (single-draw comparison values are in Table~\ref{tab:exp3_alignment}). ent = entropy, prov. = provenance; descriptor numbers are full-slice post-resize means from Table~\ref{tab:manip_matrix}.}
\label{tab:ident_status}
\end{table*}

\vspace{\ParaVspace}
Table~\ref{tab:ident_status} gives the identification status of every contrast under the attribution rule (Section~\ref{sec:method:attribution}). We anchor each family with the baseline matrix (Section~\ref{sec:res_baseline}), read the compositional (Sections~\ref{sec:res_complexity},~\ref{sec:res_csdj_ladder}) and alignment (Section~\ref{sec:res_alignment}) ablations against the four-axis predictions, and close with the image-contribution control (Section~\ref{sec:res_e1}), which licenses reading the composite rows at all. All cells use the full StrongREJECT slice ($n{=}313$).

\vspace{\SubBeforeNeg}
\subsection{Baseline Cross-Attack Matrix}
\vspace{\SubAfterNeg}
\label{sec:res_baseline}

\vspace{\ParaVspace}
Table~\ref{tab:asr_with_ci} shows the baseline matrix; four patterns are evident.

\vspace{\ParaVspace}
\textit{Four patterns.} (i)~\emph{MML varies sharply across victims:} its two variants remain near the ASR floor on \textsc{Claude Sonnet~4.5} ($\le0.02$), but reach $0.81$/$0.83$ on \textsc{GPT-4o}, $0.96$/$0.95$ on \textsc{Gemma-3-4B}, $0.78$/$0.39$ on the MLRM, and $0.90$/$0.41$ on \textsc{Qwen3-VL-8B}. This matrix shows that attack outcomes depend on the victim, but it does not explain why. The provider-interception audit in Appendix~\ref{app:res_claude_row} covers Exp.~3/E4, not these MML results, so it neither explains nor rules out interception in this comparison. (ii)~\emph{CS-DJ outcomes also vary across victims:} ASR is $0.28$--$0.32$ on the two API-served models (with overlapping confidence intervals), compared with $0.60$ on \textsc{Gemma-3-4B} and $0.84$ on the MLRM. These differences are descriptive, not evidence of a particular defense mechanism. (iii)~\emph{FigStep is victim-specific:} ASR is at or below $0.01$ on \textsc{GPT-4o} and \textsc{Claude}, compared with $0.29$ on \textsc{Gemma-3-4B}, $0.13$ on \textsc{Qwen3-VL-8B}, and $0.73$ on the MLRM. (iv)~\emph{VisCo also varies across victims:} its ASR ranges from $0.03$ on \textsc{Claude} to $0.67$ on the MLRM, and is $0.24$--$0.57$ on the other three main-panel victims.

\vspace{\SubBeforeNeg}
\subsection{Experiment 2: Complexity Ablation}
\vspace{\SubAfterNeg}
\label{sec:res_complexity}

\vspace{\ParaVspace}\textbf{Exp.~2 reading.} Table~\ref{tab:exp2_complexity} reports the four-row ladder (per-victim numerics, rung decomposition, and sampling-variance floor in Appendix~\ref{app:res_reading_exp2}); the step-wise decomposition and E5 payload-geometry control split the rung ordering into accepted and resolution-limited steps (Appendix~\ref{app:res_ladder_decomp}).

\vspace{\ParaVspace}

\vspace{\SubBeforeNeg}
\subsection{Compositional Ladder Inside the CS-DJ Family}
\vspace{\SubAfterNeg}
\label{sec:res_csdj_ladder}
\vspace{\ParaVspace}\textbf{CS-DJ region-count ladder.} We vary the number of distinct distractor regions (one, three, or nine) while keeping the $3\times4$ grid, pixel size, canvas, JPEG footprint, and three payload tiles fixed. ASR is flat on Gemma-3-4B ($0.460\to0.460\to0.454$, $r_1\to r_9$ $\Delta=-0.006$). Qwen3-VL-8B shows a small, nonsignificant rise ($0.185\to0.208\to0.220$, $\Delta=+0.035$, $p=0.25$) (Appendix Figure~\ref{fig:region_count_ladder}, Table~\ref{tab:region_count}). Thus, adding distinct distractor regions does not increase ASR in this controlled test. Appendix~\ref{app:res_reading_csdj} discusses the result and prior ladders; Appendix~\ref{app:res_ladder_controls} reports construction controls and cross-benchmark results. A separate rendering difference falls outside the four axes: on Gemma, the white-padded CS-DJ baseline has ASR $0.60$ versus $0.41$ for the construction-matched force-squared unrelated cell, a $0.19$ gap associated with a higher white-pixel fraction ($0.84$ vs. $0.33$; Appendix~\ref{app:res_padding}).

\vspace{\ParaVspace}

\vspace{\SubBeforeNeg}
\subsection{Experiment 3: Alignment Ablation}
\vspace{\SubAfterNeg}
\label{sec:res_alignment}

\vspace{\ParaVspace}\textbf{Exp.~3 reading.} 
Table~\ref{tab:exp3_alignment} (Appendix~\ref{app:exp3}) holds the CS-DJ layout fixed while varying the distractor tiles among noise, unrelated natural images, and query-relevant generated scenes. On Qwen3-VL-8B, single-draw ASR rises from $0.19$ to $0.24$ to $0.29$, but this three-condition comparison does not isolate alignment: entropy, compressibility, provenance, and CLIP alignment also change. E4 provides the controlled comparison. On Qwen, removing query-specific relatedness lowers ASR by $0.119$ (raw $k{=}3$ permutation $p{=}10^{-4}$); image-text congruence remains unmeasured. A within-category harm-salience control reproduces the direction ($\Delta{=}{-}0.089$, $p{=}10^{-4}$; Appendix~\ref{app:e4_within}). Because it uses overlapping stimuli, we treat it as a robustness check rather than an independent attribution. Five contrasts pass the global correction, but E4 is the only one attributed to a single measured descriptor under our rule (Table~\ref{tab:global_mult}). This result is limited to Qwen. The other main-panel victims show no significant E4 effect or have interpretation limits; the additional InternVL3.5-8B evaluation does not extend the attribution. The minimum detectable effect is about $0.06$--$0.10$ (Appendix~\ref{app:res_reading_nullbound}), and the dose-response check does not resolve whether the effect grows with stronger alignment (Appendix~\ref{app:res_dose_deferred}).

\vspace{\ParaVspace}Per victim, the E4 contrast (\emph{relevant} $\rightarrow$ \emph{gen-unrelated}; paired $\Delta$ASR with Holm-adjusted McNemar $p$, Table~\ref{tab:e4_provenance}) is significant only on \textsc{Qwen3-VL-8B} ($-0.12$ [$-0.17$, $-0.06$], $p_{H}{<}0.001$); the other four victims do not reach significance ($p_{H}\ge0.07$).

\vspace{\SubBeforeNeg}
\subsection{Image Contribution Control (E-1)}
\vspace{\SubAfterNeg}
\label{sec:res_e1}

\vspace{\ParaVspace}\textbf{Image-contribution reading.} On the API reference victim (\textsc{GPT-4o}) the bare query reaches $1.3\%$ ASR, a benign unrelated image leaves it at $0.0\%$, and attaching a CS-DJ image lifts it to $28.4\%$ on identical prompts; the pattern holds across the panel, with one flagged exception on \textsc{GLM-4.1V-Th.} (Appendix~\ref{app:res_reading_e1}).

\vspace{\ParaVspace}
\noindent\textbf{Summary.} In the image-contribution control, harmful queries alone and queries paired with benign unrelated images receive little ASR on most victims, while attack composites raise it. Among the construction contrasts outside our four axes, padding and border changes produce a substantial $0.19$ gap; the aggregate multi-step ladder changes are larger. The earlier tile-count ladders are confounded by payload visibility, and the controlled region-count test finds no detectable effect. Whether tile-count structure matters therefore remains open. E4 supports a bounded attribution to query-specific relatedness on Qwen3-VL-8B, with image-text congruence unmeasured. Floor and saturation effects limit what the MML and VisCo baseline rows can establish on several victims.

\vspace{\SecBeforeNeg}
\section{Discussion and Conclusion}
\vspace{\SecAfterNeg}
\label{sec:conclusion}

\begin{table*}[tbp]
\centering
\footnotesize
\setlength{\tabcolsep}{4pt}
\begin{tabular*}{\textwidth}{p{2.1cm} p{1.95cm} p{1.95cm} p{2.15cm} p{1.95cm} p{2.05cm}}
\toprule
\textbf{Attack} & \textbf{GPT-4o} & \textbf{Claude S.4.5} & \textbf{Qwen3-VL-8B} & \textbf{Gemma-3-4B} & \textbf{GLM-4.1V-Th.} \\
\midrule
FigStep & 0.00\,{\scriptsize[0.00,0.00]} & 0.01\,{\scriptsize[0.00,0.02]} & 0.13\,{\scriptsize[0.09,0.17]} & 0.29\,{\scriptsize[0.25,0.35]} & 0.73\,{\scriptsize[0.68,0.78]} \\
MML-mirror & 0.81\,{\scriptsize[0.77,0.85]} & 0.02\,{\scriptsize[0.01,0.04]} & 0.90\,{\scriptsize[0.87,0.93]} & 0.96\,{\scriptsize[0.94,0.98]} & 0.78\,{\scriptsize[0.73,0.82]} \\
MML-base64 & 0.83\,{\scriptsize[0.79,0.87]} & 0.00\,{\scriptsize[0.00,0.00]} & 0.41\,{\scriptsize[0.36,0.47]} & 0.95\,{\scriptsize[0.93,0.97]} & 0.39\,{\scriptsize[0.34,0.45]} \\
CS-DJ & 0.28\,{\scriptsize[0.24,0.34]} & 0.32\,{\scriptsize[0.27,0.37]} & 0.29\,{\scriptsize[0.24,0.34]} & 0.60\,{\scriptsize[0.55,0.66]} & 0.84\,{\scriptsize[0.80,0.88]} \\
VisCo & 0.26\,{\scriptsize[0.21,0.31]} & 0.03\,{\scriptsize[0.01,0.05]} & 0.24\,{\scriptsize[0.20,0.29]} & 0.57\,{\scriptsize[0.52,0.63]} & 0.67\,{\scriptsize[0.61,0.72]} \\
\bottomrule
\end{tabular*}
\vspace{-0.4cm}
\caption{Baseline cross-attack ASR on StrongREJECT ($n{=}313$ prompts per cell, binarized at $\tau{=}0.5$, judged with the StrongREJECT rubric judge, with bias-corrected $95\%$ bootstrap CI over $10^4$ resamples). The five victims span the typographic and composite branches of our taxonomy.}
\label{tab:asr_with_ci}
\end{table*}

\vspace{\ParaVspace}
Five contrasts pass the global statistical correction, but only one supports attribution to a single measured descriptor under our rule. On Qwen3-VL-8B, the E4 manipulation that removes query-specific relatedness lowers ASR by $0.119$, and a within-category control reproduces the direction on overlapping stimuli. This is a bounded result for one victim and one manipulation; image-text congruence remains unmeasured. It does not establish a general semantic-alignment mechanism or resolve the different emphases of CS-DJ and VisCo. The region-count test finds no detectable effect of distractor diversity, while the earlier tile-count ladders are confounded by payload visibility. We therefore cannot conclude whether tile-count structure matters. The attention-mass probe finds that payload tiles receive disproportionate attention on Qwen3-VL-8B, but their attention share does not vary with ASR and does not explain the E4 result (Appendix~\ref{app:conclusion_attention}).

\vspace{\ParaVspace}
\textbf{Defensive implications.} The pad-surround and content-fill rungs are among the larger detectable changes in our ablation and may motivate input screening or training for typographic-payload robustness on the victims where they occur. The per-tile density screen does not reliably separate attack tiles from benign photo tiles, so a practical defense would need signals beyond the density descriptors measured here. Appendix~\ref{app:conclusion_defense} discusses these implications and the need to evaluate defenses against adaptive adversaries.

\vspace{\ParaVspace}
\noindent\textbf{Methodological takeaway.} Attack rankings alone cannot identify which image properties matter. Controlled comparisons can help, provided the instruction remains fixed, image properties are measured before evaluation, and results are labeled as jointly attributed when multiple descriptors change. Applying this rule exposed payload-visibility confounds in two tile-count ladders and yielded one bounded attribution to query-specific relatedness on Qwen3-VL-8B. Image-text congruence remains unresolved. The same approach can help evaluate other attack families, including scaffold- and encoding-based attacks.

\vspace{\SecBeforeNeg}
\section{Limitations}
\vspace{\SecAfterNeg}
\label{sec:limitations}
\vspace{\ParaVspace}
Our main findings come from open-weight victims with headroom: the E4 alignment result on \textsc{Qwen3-VL-8B} and the padding, provenance, and ladder observations on \textsc{Gemma-3-4B}. They should not be generalized to the full panel. Image-side measures such as pixel entropy, JPEG@85, and CLIP alignment are proxies. ASR is also judge-conditional: the rubric judge shares a model family with the \textsc{GPT-4o} victim, and its results differ from a non-OpenAI judge by up to $0.18$. We therefore emphasize signs and orderings when effect magnitudes approach the roughly $0.10$ cross-judge disagreement band (Appendices~\ref{app:human_kappa},~\ref{app:judge}). The $0.19$ padding-construction gap is a substantial unmodeled difference, but it is smaller than the aggregate multi-step ladder change; it limits conclusions about the four-axis taxonomy rather than demonstrating that it dominates all designed effects.

\vspace{\ParaVspace}
The four axes organize our image constructions, but most comparisons do not isolate one property while holding every other image property fixed. E4 holds the measured entropy, compressibility, tile count, and provenance fixed while changing query-specific relatedness; it does not separate relatedness from image-text congruence, and no benign-paraphrase control was run. The region-count null tests distractor diversity, not tile count. Because we did not build a corrected tile-count ladder with payload text visible at every rung, whether tile-count structure matters remains untested. We do not claim a mechanism: the attention-mass probe is null on both effect-carrying victims (Appendix~\ref{app:conclusion_attention}), and the in-model relatedness probe is too noisy to discriminate (Appendix~\ref{app:w8}).

\vspace{\ParaVspace}
Several controls remain unrun (protocols in Appendix~\ref{app:deferred_controls}): benign generated distractors to distinguish query-relatedness from off-topic-harm suppression; a scene generated from a benign paraphrase of the query; and a relevant scene paired with a benign restatement to examine image-text congruence. The paired design's minimum detectable effects are about $0.06$--$0.10$ (Table~\ref{tab:min_detect}). An outcome-blind replication with $830$ paired prompts would provide $80\%$ power for an effect of $0.06$.

\vspace{\SecBeforeNeg}
\section*{Ethical Considerations}
\vspace{\SecAfterNeg}
\label{sec:ethics}
\vspace{\ParaVspace}
We evaluate published attack families and create paired image variants for diagnostic experiments. We release no new deployable exploit or fine-tuned attacker model. Harmful prompts come from the public StrongREJECT benchmark. The paper omits raw harmful completions, and the typographic payload in Figure~\ref{fig:exp3_stimuli} is blurred. The public evaluation archive contains judge scores but not raw response text; a separate, access-controlled archive contains selected stimuli and verbatim model outputs. We queried API-hosted victims under standard red-teaming terms and did not attempt to extract model weights or training data. The rubric judge was audited against human labels (agreement $0.800$, $\kappa{=}0.594$; Appendix~\ref{app:human_kappa}). The audit found no judge-error interaction that changes the paper's conclusions. Defenses based on these findings should be evaluated against adaptive adversaries before deployment.

\vspace{\sectionvspaceupper}
\section*{Acknowledgment}
\vspace{\sectionvspacelower}
This work was supported in parts by the NYUAD Center for Cyber Security (CCS), funded by Tamkeen under the NYUAD Research Institute Award G1104. Experiments are performed with NYUAD Jubail High Performance Computing (HPC).

\bibliography{reference}
\bibliographystyle{template/iclr2027_conference}

\newpage

\appendix

\vspace{\SecBeforeNeg}
\section{Appendix}

\vspace{\SubBeforeNeg}
\subsection{Per-Category Breakdown on the API Reference Victim}
\vspace{\SubAfterNeg}
\label{app:per_category}

\begin{table*}[b]
\centering
\small
\setlength{\tabcolsep}{2pt}
\begin{tabular*}{\textwidth}{@{\extracolsep{\fill}} l c c c c c c c}
\toprule
 & \textbf{Disinfo./Decep.} & \textbf{Illegal G./S.} & \textbf{Hate/Harass.} & \textbf{Non-viol.} & \textbf{Violence} & \textbf{Sexual} & \textbf{Overall} \\
\midrule
FigStep family ($4$ conds.) & \multicolumn{6}{c}{0.00 (all categories)} & \textbf{0.00} \\
MML-mirror & 0.92 & 0.80 & 0.76 & 1.00 & 0.74 & 0.62 & \textbf{0.81} \\
MML-base64 & 0.92 & 0.74 & 0.82 & 0.92 & 0.93 & 0.62 & \textbf{0.83} \\
CS-DJ & 0.42 & 0.20 & 0.22 & 0.39 & 0.22 & 0.24 & \textbf{0.28} \\
VisCo & 0.35 & 0.33 & 0.29 & 0.27 & 0.19 & 0.16 & \textbf{0.26} \\
\midrule
CS-DJ-noise & 0.28 & 0.22 & 0.34 & 0.29 & 0.22 & 0.10 & \textbf{0.24} \\
CS-DJ-unrelated & 0.36 & 0.38 & 0.28 & 0.34 & 0.31 & 0.20 & \textbf{0.31} \\
CS-DJ-relevant & 0.42 & 0.22 & 0.22 & 0.41 & 0.26 & 0.20 & \textbf{0.29} \\
\bottomrule
\end{tabular*}
\vspace{-0.4cm}
\caption{Per-category ASR breakdown on the API reference victim (\textsc{GPT-4o}; full StrongREJECT, $n{=}313$, rubric judge with $\tau{=}0.5$). Top block: baseline attacks anchoring the $C{\times}S$ taxonomy; the four Exp.~2 complexity conditions compress into a single all-zero row (a genuine hard refusal, Section~\ref{sec:res_e1}). Bottom block: Experiment~3 alignment ablation (CS-DJ layout held constant).}
\label{tab:per_category}
\end{table*}

\vspace{\ParaVspace}
Table~\ref{tab:per_category} breaks down the \textsc{GPT-4o} results across the six StrongREJECT harm categories. Both MML variants perform similarly across categories, with Sexual Content as an exception ($0.62$ for each), which may reflect an additional category-level filter. The CS-DJ baseline ranges from $0.20$ to $0.42$, with the highest ASR for Disinformation/Deception ($0.42$) and Non-violent Crimes ($0.39$). The FigStep ablation variants remain at $0.00$ across categories, matching the baseline FigStep row in Table~\ref{tab:asr_with_ci}. Among the CS-DJ variants, \emph{relevant} stays within $0.04$ of the baseline in each category, while \emph{noise} and \emph{unrelated} shift some category-level rates by as much as $0.18$ without changing overall ASR by more than $0.04$. These results describe category-level variation; they do not establish whether layout or tile content causes it.

\vspace{\SubBeforeNeg}
\subsection{Manipulation Check}
\vspace{\SubAfterNeg}
\label{app:manip_check}

\begin{table}[tbp]
\centering
\small
\setlength{\tabcolsep}{4pt}
\begin{tabular*}{\columnwidth}{@{\extracolsep{\fill}} p{3.4cm} p{2.3cm} p{2.3cm}}
\toprule
\textbf{Attack family} & \textbf{Entropy} (bits/px) & \textbf{JPEG@85} (KB) \\
\midrule
\multicolumn{3}{l}{\textit{Experiment 1 (baseline attack images, $n{=}20$/cell; pre-resize sample)}} \\
FigStep & $0.71 \pm 0.14$ & 61.3 \\
MML-mirror & $0.73 \pm 0.14$ & 63.8 \\
MML-base64 & $0.60 \pm 0.29$ & 55.7 \\
CS-DJ & $1.87 \pm 0.07$ & 321.3 \\
VisCo & $7.05 \pm 0.53$ & 179.6 \\
\midrule
\multicolumn{3}{l}{\textit{Experiment 2 (complexity ablation; **full slice**, $n{=}313$; post-resize)}} \\
FigStep (baseline) & $0.90 \pm 0.15$ & 81.4 \\
FigStep-pad & $2.57 \pm 0.09$ & 40.8 \\
FigStep-distr.-rand. & $6.48 \pm 0.54$ & 112.9 \\
FigStep-distr.-rel. & $6.78 \pm 0.21$ & 122.9 \\
\midrule
\multicolumn{3}{l}{\textit{Experiment 3 (alignment ablation, full slice, $n{=}313$; post-resize)}} \\
CS-DJ-noise & $5.97 \pm 0.05$ & 166.3 \\
CS-DJ-unrelated & $6.40 \pm 0.17$ & 131.1 \\
CS-DJ-relevant & $6.62 \pm 0.13$ & 114.6 \\
CS-DJ-gen-unrelated (E4) & $6.62 \pm 0.12$ & 114.7 \\
\bottomrule
\end{tabular*}
\vspace{-0.3cm}
\caption{Manipulation-check summary on the full StrongREJECT slice. \textit{Exp.~1} rows aggregate $n{=}20$ images per cell; entropy is mean per-channel Shannon entropy of the RGB image and JPEG@85 is the re-encoded byte size at quality~85. \textit{Exp.~2/3} rows are on the \emph{full} slice ($n{=}313$ per condition), recomputed by a single script on the \emph{post-resize} image (longer side $\le 768$\,px, LANCZOS) the encoders receive. Exp.~3 conditions are \emph{not} matched in entropy (span $0.65$ bits/px), compressibility, or tile provenance, so alignment is isolated only where those confounded descriptors run opposite to ASR or are held fixed by the E4 control.}
\label{tab:manip_check}
\end{table}

\vspace{\ParaVspace}
Table~\ref{tab:manip_check} reports image-side metrics recomputed from source stimuli after resizing them to a maximum side length of $768$\,px. The three low-$C_{\mathrm{comp}}$ families (FigStep, MML-mirror, and MML-base64) have mean per-image entropy near $0.7$ bits per pixel and JPEG@85 sizes of $56$--$64$~KB. CS-DJ measures $1.87$ bits per pixel and $321$~KB, while VisCo's generated-scene image measures $7.05$ bits per pixel and $180$~KB. In Exp.~2, mean entropy rises from $0.90$ at \emph{baseline} to $2.57$ at \emph{pad} and $6.48$/$6.78$ for the two distractor conditions. In Exp.~3, entropy ranges from $5.97$ to $6.62$ bits per pixel, JPEG@85 from $166.3$ to $114.6$~KB, and tile provenance also changes (Section~\ref{app:clip}). These measurements confirm that image statistics shift across conditions; because multiple descriptors change together in several comparisons, they do not by themselves establish an isolated effect for either ablation.

\vspace{\SubBeforeNeg}
\subsection{Alignment Verification (CLIP)}
\vspace{\SubAfterNeg}
\label{app:clip}

\begin{table}[tbp]
\centering
\small
\setlength{\tabcolsep}{4pt}
\begin{tabular}{p{3.22cm} p{1.2cm} p{1.0cm} p{1.0cm} p{1.69cm}}
\toprule
\textbf{Condition} & \textbf{Mean} & \textbf{SD} & \textbf{p50} & \textbf{95\% range} \\
\midrule
\multicolumn{5}{l}{\textit{Exp.~2 (complexity, FigStep payload)}} \\
FigStep (baseline) & 0.347 & 0.032 & 0.342 & [0.30, 0.40] \\
FigStep-pad & 0.306 & 0.032 & 0.306 & [0.25, 0.36] \\
FigStep-distr.-rand. & 0.285 & 0.034 & 0.288 & [0.22, 0.33] \\
FigStep-distr.-rel. & 0.324 & 0.033 & 0.325 & [0.27, 0.38] \\
\midrule
\multicolumn{5}{l}{\textit{(surround-only CLIP, Exp.~2 distract-*)}} \\
FigStep-distr.-rand. & 0.124 & 0.021 & 0.123 & [0.09, 0.15] \\
FigStep-distr.-rel. & 0.216 & 0.034 & 0.215 & [0.16, 0.28] \\
\midrule
\multicolumn{5}{l}{\textit{Exp.~3 (alignment, CS-DJ layout)}} \\
CS-DJ (anchor) & 0.198 & 0.041 & 0.196 & [0.13, 0.27] \\
CS-DJ-noise & 0.197 & 0.043 & 0.196 & [0.13, 0.27] \\
CS-DJ-unrelated & 0.187 & 0.040 & 0.186 & [0.13, 0.25] \\
CS-DJ-relevant & 0.254 & 0.040 & 0.254 & [0.19, 0.32] \\
CS-DJ-gen-unrelated & 0.196 & 0.046 & 0.196 & [0.10, 0.29] \\
\midrule
\multicolumn{5}{l}{\textit{(surround-only CLIP, Exp.~2 gen-unrelated)}} \\
FigStep-distr.-gen-unr. & 0.136 & 0.026 & \multicolumn{2}{c}{$-$} \\
\bottomrule
\end{tabular}
\vspace{-0.2cm}
\caption{Alignment verification on the full StrongREJECT slice ($n{=}313$ per condition): whole-image CLIP ViT-L/14 cosine similarity between the stimulus and the harmful query (mean, SD, median, 5th--95th percentile). The Exp.~3 \emph{relevant} condition carries measurably higher CLIP alignment than \emph{noise} or \emph{unrelated} ($0.254$ vs $0.197/0.187$), confirming the alignment manipulation is effective; the Exp.~2 surround-only rows show the query-aligned generated surround ($0.216$) exceeds the random natural surround ($0.124$).}
\label{tab:clip_align}
\end{table}

\vspace{\ParaVspace}
Table~\ref{tab:clip_align} shows that the manipulations change the CLIP proxy: in Exp.~3, median whole-image alignment to the harmful query is higher for \emph{relevant} than for \emph{noise} or \emph{unrelated} ($0.254$ vs. $0.196$/$0.186$); in Exp.~2, the query-aligned surround also scores higher than the random natural surround ($0.216$ vs. $0.124$). These measurements verify a change in CLIP alignment, not a causal effect on ASR. Other image descriptors also change in Exp.~3, where the ordered ASR rise is statistically detectable only on Qwen. The E4 \emph{gen-unrelated} cell has CLIP $0.196$, close to \emph{unrelated} ($0.187$) and below \emph{relevant} ($0.254$), consistent with assigning each generated scene to a different query. We generated the scenes with \emph{Stable Diffusion 3.5 Large Turbo} using \emph{diffusers}.

\vspace{\SubBeforeNeg}
\subsection{Judge Reliability (GPT-4o \texorpdfstring{$\times$}{x} FigStep)}
\vspace{\SubAfterNeg}
\label{app:judge}

\begin{table}[tbp]
\centering
\small
\setlength{\tabcolsep}{5pt}
\begin{tabular}{p{5.49cm} p{1.03cm}}
\toprule
\textbf{Statistic (over 312 judged responses)} & \textbf{Value} \\
\midrule
Mean score & 0.0006 \\
Median & 0.001 \\
90th percentile & 0.001 \\
99th percentile & 0.003 \\
Maximum & 0.011 \\
Share at exact 0.0 & 0.32 \\
Share in $(0, 0.5)$ & 0.68 \\
Share $\ge 0.5$ (ASR) & 0.00 \\
\bottomrule
\end{tabular}
\vspace{-0.2cm}
\caption{Continuous judge-score distribution for the \textsc{GPT-4o} FigStep cell: $n{=}312$ responses scored with the StrongREJECT rubric. All scores sit far below the $0.5$ binarization threshold (99th percentile $0.003$, maximum $0.011$), so the reported $0.00$ ASR reflects genuine total refusal rather than a threshold artifact of the judge.}
\label{tab:figstep_dist}
\end{table}

\vspace{\ParaVspace}
Table~\ref{tab:figstep_dist} reports the continuous judge-score distribution for the \textsc{GPT-4o} FigStep cell, which is exactly $0.00$ ASR across all conditions. The cell contains $n{=}313$ records of which $312$ produced a judged response; one prompt (\emph{strongreject\_0120}) was excluded because the \textsc{GPT-4o} API rejected the query at the moderation stage (input flagged as illicit/violent, no credits charged), so no victim response existed to judge. All $312$ scored responses sit far below the $0.5$ binarization threshold (99th percentile $0.003$, maximum $0.011$): this is hard refusal, not a mass sitting just under the threshold, so the reported zero ASR reflects genuine total refusal on this cell rather than a threshold artifact of the judge.

\vspace{\SubBeforeNeg}
\subsection{Experiment 3: Alignment Ablation Table}
\vspace{\SubAfterNeg}
\label{app:exp3}

\vspace{\ParaVspace}
\vspace{\ParaVspace}
\begin{table*}[bp]
\centering
\scriptsize
\setlength{\tabcolsep}{4pt}
\renewcommand{\arraystretch}{1.15}
\begin{tabular*}{\textwidth}{@{\extracolsep{\fill}} l ccccc @{}}
\toprule
\textbf{Condition} & \textbf{GPT-4o} & \textbf{Claude S.4.5} & \textbf{Qwen3-VL-8B} & \textbf{Gemma-3-4B} & \textbf{GLM-4.1V-Th.} \\
\midrule
\multicolumn{6}{@{}l}{\textit{ASR@0.5 [95\% bootstrap CI]}} \\
\addlinespace[1pt]
CS-DJ-noise     & 0.24 [0.19, 0.29] & 0.00 [0.00, 0.00] & 0.19 [0.14, 0.23] & 0.73 [0.68, 0.78] & 0.85 [0.81, 0.89] \\
CS-DJ-unrelated & 0.31 [0.26, 0.36] & 0.32 [0.27, 0.37] & 0.24 [0.19, 0.29] & 0.41 [0.35, 0.46] & 0.86 [0.82, 0.90] \\
CS-DJ-relevant  & 0.29 [0.24, 0.34] & 0.22 [0.18, 0.27] & 0.29 [0.24, 0.34] & 0.66 [0.61, 0.71] & 0.84 [0.80, 0.88] \\
\midrule
\multicolumn{6}{@{}l}{\textit{Within-Exp.~3 paired McNemar per victim: (raw $p$, Holm $p$)}} \\
\addlinespace[1pt]
noise $\to$ unrelated          & (0.018, 0.055)           & ($<$0.001, $<$0.001) & (0.060, 0.121)      & ($<$0.001, $<$0.001) & (0.775, 1.000) \\
unrelated $\to$ relevant       & (0.489, 0.489)           & (0.003, 0.003)       & (0.133, 0.133)      & ($<$0.001, $<$0.001) & (0.341, 1.000) \\
noise $\to$ relevant$^\dagger$ & (0.105, 0.211)           & ($<$0.001, $<$0.001) & ($<$0.001, 0.002)   & (0.028, 0.028)       & (0.576, 1.000) \\
\bottomrule
\end{tabular*}
\vspace{-0.4cm}
\caption{Experiment~3: semantic-alignment ablation within the fixed CS-DJ $3{\times}4$ layout (StrongREJECT, $n{=}313$ per cell, rubric judge, $\tau{=}0.5$). Top block: ASR@0.5 with 95\% bootstrap CI. Bottom block: within-experiment paired McNemar tests as (raw~$p$, Holm-adjusted~$p$) pairs across the family of three contrasts per victim; the two adjacent-step contrasts are primary, the ordered endpoint noise~$\to$~relevant$^\dagger$ exploratory. Delta-ASR with 95\% CIs: Table~\ref{tab:delta_ci}.}\label{tab:exp3_alignment}
\end{table*}

Table~\ref{tab:exp3_alignment} reports the full Experiment~3 ablation: the CS-DJ $3{\times}4$ layout is fixed while the nine distractor tiles vary among noise, unrelated, and relevant content. The bottom block gives raw and Holm-adjusted paired McNemar $p$-values for the two adjacent-step tests; the ordered noise-to-relevant endpoint is exploratory. These comparisons avoid the complexity mismatch in the earlier anchor-versus-CS-DJ tests. Qwen3-VL-8B shows a significant ordered-endpoint ASR rise ($\Delta{=}+0.10$, Holm $p{=}0.002$), but this is a descriptive association because image descriptors also vary. The separate E4 result provides the paper's relatedness attribution. The other four models show no significant monotone rise. Table~\ref{tab:delta_ci} reports paired-prompt $\Delta$ASR estimates and bootstrap CIs for every contrast.

\vspace{\SubBeforeNeg}
\subsection{Experiment 3: Provenance Control (E4)}
\vspace{\SubAfterNeg}
\label{app:e4}

\vspace{\ParaVspace}
The E4 control addresses a provenance confound in Experiment~3: relevant scenes were generated for their queries, while unrelated scenes were natural photographs. E4 reassigns each generated scene to a different query using a fixed derangement. This keeps generated-scene provenance constant while changing its relation to the harmful query; no prompt receives its own scene. The CLIP check confirms the reassignment: gen-unrelated has whole-image CLIP $0.196$, close to unrelated ($0.187$) and below relevant ($0.254$).

\vspace{\ParaVspace}
\begin{table*}[tbp]
\centering
\small
\setlength{\tabcolsep}{4pt}
\begin{tabular*}{\textwidth}{@{\extracolsep{\fill}}l ccccc@{}}
\toprule
\textbf{Condition} & \textbf{GPT-4o} & \textbf{Claude S.4.5} & \textbf{Qwen3-VL-8B} & \textbf{Gemma-3-4B} & \textbf{GLM-4.1V-Th.} \\
\midrule
\multicolumn{6}{@{}l}{\itshape ASR@0.5} \\
CS-DJ-noise & 0.24 & 0.00 & 0.19 & 0.73 & 0.85 \\
CS-DJ-relevant & 0.29 & 0.22 & 0.29 & 0.66 & 0.84 \\
CS-DJ-unrelated & 0.31 & 0.32 & 0.24 & 0.41 & 0.86 \\
\emph{CS-DJ-gen-unrelated} & 0.27 & 0.26 & 0.17 & 0.60 & 0.81 \\
\midrule
\multicolumn{6}{@{}l}{\itshape relevant $\to$ gen-unrelated: paired $\Delta$ASR, $95\%$\,bootstrap CI, Holm-$p$ per victim} \\
$\Delta$ASR & $-0.02$ & $+0.04$ & $-0.12$ & $-0.06$ & $-0.03$ \\
$95\%$\;CI & [$-0.07$,0.03] & [$-0.02$,0.09] & [$-0.17$,$-0.06$] & [$-0.12$,0.00] & [$-0.08$,0.02] \\
Holm-$p$ & 0.54 & 0.22 & $<0.001$ & 0.07 & 0.29 \\
\midrule
\multicolumn{6}{@{}l}{\itshape unrelated $\to$ gen-unrelated: paired $\Delta$ASR, $95\%$\,bootstrap CI, Holm-$p$ per victim} \\
$\Delta$ASR & $-0.04$ & $-0.06$ & $-0.07$ & $+0.19$ & $-0.05$ \\
$95\%$\;CI & [$-0.10$,0.02] & [$-0.11$,0.00] & [$-0.13$,$-0.01$] & [0.13,0.26] & [$-0.10$,$-0.01$] \\
Holm-$p$ & 0.19 & 0.05 & 0.03 & $<0.001$ & 0.03 \\
\bottomrule
\end{tabular*}
\vspace{-0.4cm}
\caption{Experiment~3 provenance control (E4), ASR@0.5 and the two paired-prompt $\Delta$ASR [95\%\,bootstrap CI] with Holm-adjusted McNemar $p$ per victim (StrongREJECT, $n{=}313$ per cell, rubric judge, $\tau{=}0.5$). The \emph{gen-unrelated} cell holds ``SD3.5-generated harmful-domain scene'' constant and varies only query-specific relatedness, by attaching query $B$'s pre-generated scene to query $A$ via a fixed derangement. The noise/relevant/unrelated anchors are repeated from Table~\ref{tab:exp3_alignment} for reference.}
\label{tab:e4_provenance}
\end{table*}

\vspace{\ParaVspace}
Table~\ref{tab:e4_provenance} reports the E4 cell and its paired contrasts. Claude's raw ASR is $0.26$ for gen-unrelated and $0.32$ for unrelated, but provider-side interception varies substantially across its Exp.~3 conditions. As detailed in Appendix~\ref{app:res_reading_others}, this selection prevents Claude's E4 row from adjudicating provenance or relatedness. On Qwen, gen-unrelated ASR ($0.17$) is significantly below relevant ($0.29$) and unrelated ($0.24$), consistent with the controlled relatedness result. On Gemma, gen-unrelated ($0.60$) is close to relevant ($0.66$) and above unrelated ($0.41$); because these cells also differ in other descriptors, this ordering is descriptive rather than an isolated provenance effect. The E4 attribution therefore rests on Qwen; the other victim rows do not establish a general pattern.

\vspace{\SubBeforeNeg}
\subsection{Within-Category Derangement (Harm-Salience Control)}
\vspace{\SubAfterNeg}
\label{app:e4_within}

The E4 derangement above permutes the cached SD3.5 scenes across prompts without regard to harm category: the original control attaches each prompt to a scene generated for a \emph{different} StrongREJECT category $84.03\%$ of the time ($263/313$). Because the attached scene is typically from a different harm category than the query it accompanies, a reader could attribute the relevant-versus-gen-unrelated drop to cross-category harm salience (the attached scene competing for the victim's attention on a different harm topic) rather than to query relatedness. We close that confound with a same-category derangement (fixed seed $2026$, zero cross-category pairs, zero fixed points; every StrongREJECT category has at least $50$ prompts, so an internal derangement exists for all $313$). The attack variant, stimulus bytes, sampler ($k{=}3$), and rubric judge are otherwise identical to the original E4 cell.

On \textsc{Qwen3-VL-8B} the within-category control reproduces the effect: relevant $0.232$ vs.~gen-unrelated-within $0.143$, a per-prompt \emph{relevant}$\to$\emph{gen-unrelated-within} $\Delta{=}{-}0.089$ ($95\%$ CI [$-$0.131,$-$0.049], within-prompt permutation $p{=}10^{-4}$), same sign as the original cross-category $\Delta{=}{-}0.119$ ([$-$0.160,$-$0.079]). The Shapley decomposition (Appendix~\ref{app:delta_ci}) is refusal-driven in both cases: within-category $S_{\mathrm{ref}}{=}0.549$ (CI $[0.449,0.642]$) and original $S_{\mathrm{ref}}{=}0.540$ ($[0.475,0.608]$), both excluding the $0.25$ null. The within $\Delta$ is slightly smaller in magnitude than the cross $\Delta$ ($0.089$ vs.~$0.119$), recorded as a small residual sensitivity, but its CI is well clear of zero and its classification does not change, so the within-category control passes rather than collapses: cross-category harm salience is not the driver of the E4 effect, and the refused-driven, query-relatedness reading (Branch A) survives gate 2 (Section~\ref{sec:res_alignment}). This control also enters the corrected multiplicity family (Table~\ref{tab:global_mult}).

\vspace{\SubBeforeNeg}
\subsection{Delta-ASR Confidence Intervals}
\vspace{\SubAfterNeg}
\label{app:delta_ci}

\newcommand{\dci}[3]{\begin{tabular}[t]{@{}c@{}}$#1$\\[-2pt]{\scriptsize$[#2,\,#3]$}\end{tabular}}
\begin{table*}[tbp]
\centering\small
\renewcommand{\arraystretch}{1.15}
\setlength{\tabcolsep}{4pt}
\begin{tabular*}{\textwidth}{@{\extracolsep{\fill}}l ccccc@{}}
\toprule
\textbf{Contrast} & \textbf{GPT-4o} & \textbf{Claude S.4.5} & \textbf{Qwen3-VL-8B} & \textbf{Gemma-3-4B} & \textbf{GLM-4.1V-Th.} \\
\midrule
\multicolumn{6}{@{}l}{\textit{Experiment 2 (baseline vs.\ variant; $+\Delta$ = variant above anchor)}} \\
\addlinespace[2pt]
FigStep $\to$ pad
  & \dci{+0.00}{0.00}{0.00} & \dci{+0.00}{-0.01}{0.01} & \dci{+0.00}{-0.03}{0.03} & \dci{+0.17}{0.11}{0.22} & \dci{+0.07}{0.02}{0.12} \\
FigStep $\to$ distr.-random
  & \dci{+0.00}{0.00}{0.00} & \dci{+0.03}{0.01}{0.04} & \dci{+0.01}{-0.03}{0.04} & \dci{+0.34}{0.28}{0.40} & \dci{+0.16}{0.11}{0.20} \\
FigStep $\to$ distr.-relevant
  & \dci{+0.00}{0.00}{0.00} & \dci{+0.04}{0.02}{0.06} & \dci{+0.02}{-0.02}{0.05} & \dci{+0.30}{0.24}{0.36} & \dci{+0.12}{0.07}{0.17} \\
pad $\to$ distr.-random$^{\ddagger}$
  & \dci{+0.00}{0.00}{0.00} & \dci{+0.03}{0.01}{0.05} & \dci{+0.01}{-0.02}{0.05} & \dci{+0.17}{0.11}{0.23} & \dci{+0.08}{0.04}{0.13} \\
\midrule
\multicolumn{6}{@{}l}{\textit{Experiment 3 (within-condition; $+\Delta$ = success rises toward relevant)}} \\
\addlinespace[2pt]
noise $\to$ unrelated
  & \dci{+0.07}{0.01}{0.12} & \dci{+0.32}{0.26}{0.36} & \dci{+0.05}{-0.00}{0.10} & \dci{-0.32}{-0.39}{-0.26} & \dci{+0.01}{-0.04}{0.05} \\
unrelated $\to$ relevant
  & \dci{-0.02}{-0.08}{0.03} & \dci{-0.10}{-0.16}{-0.04} & \dci{+0.05}{-0.02}{0.10} & \dci{+0.25}{0.18}{0.31} & \dci{-0.03}{-0.08}{0.02} \\
noise $\to$ relevant$^{\dagger}$
  & \dci{+0.05}{-0.01}{0.10} & \dci{+0.22}{0.17}{0.27} & \dci{+0.10}{0.04}{0.15} & \dci{-0.07}{-0.13}{-0.01} & \dci{-0.02}{-0.06}{0.03} \\
\bottomrule
\end{tabular*}
\vspace{-0.4cm}
\caption{Paired-prompt $\Delta$ASR with bias-corrected $95\%$ bootstrap confidence intervals (over $10^4$ resamples of the fixed prompt order) for every headline contrast reported in Sections~\ref{sec:res_complexity} and~\ref{sec:res_alignment}. Each cell shows the point estimate with its CI below. $^{\dagger}$Ordered endpoint, reported as an exploratory Experiment~3 contrast. $^{\ddagger}$The content-fill step (\emph{pad} $\to$ \emph{distract-random}) within the Exp.~2 ladder at constant tile count; it is the component attributable to information density and tile provenance, computed as a paired within-prompt contrast.}
\label{tab:delta_ci}
\end{table*}

\vspace{\ParaVspace}
Table~\ref{tab:delta_ci} reports the paired-prompt $\Delta$ASR with bias-corrected $95\%$ bootstrap confidence intervals for every headline baseline-vs-variant and within-Exp.~3 contrast. These are the effect sizes behind the $p$-values in Tables~\ref{tab:exp2_complexity} and \ref{tab:exp3_alignment}: reporting the delta separately from its $p$-value makes explicit which contrasts are both significant and large, and which would survive a permissive judge shift.

\vspace{\SubBeforeNeg}
\subsection{Experiment 2: Payload-Scale Control (E5)}
\vspace{\SubAfterNeg}
\label{app:e5}

\begin{table}[b]
\centering
\small
\setlength{\tabcolsep}{4pt}
\renewcommand{\arraystretch}{1.15}

\begin{tabular*}{\linewidth}{@{\extracolsep{\fill}} l cc cc @{}}
\multicolumn{5}{@{}l}{\textit{(a) ASR per cell}} \\
\toprule
 & \multicolumn{2}{c}{No surround} & \multicolumn{2}{c}{Pad surround} \\
\cmidrule(lr){2-3}\cmidrule(l){4-5}
Payload scale & \textbf{Gemma} & \textbf{GLM} & \textbf{Gemma} & \textbf{GLM} \\
\midrule
Full    & 0.294 {\scriptsize\textit{(baseline)}} & 0.735 {\scriptsize\textit{(baseline)}} & 0.419 {\scriptsize\textit{(fullpad)}} & 0.764 {\scriptsize\textit{(fullpad)}} \\
Quarter & 0.364 {\scriptsize\textit{(quarter)}}  & 0.502 {\scriptsize\textit{(quarter)}}  & 0.463 {\scriptsize\textit{(pad)}}     & 0.808 {\scriptsize\textit{(pad)}} \\
\bottomrule
\end{tabular*}

\vspace{8pt}

{\setlength{\tabcolsep}{1pt}%
\begin{tabular*}{\linewidth}{@{\extracolsep{\fill}} ll ccc ccc @{}}
\multicolumn{8}{@{}l}{\textit{(b) Paired-prompt contrasts: $\Delta$ASR, 95\% bootstrap CI, Holm-corrected McNemar $p$}} \\
\toprule
 & & \multicolumn{3}{c}{\textbf{Gemma}} & \multicolumn{3}{c}{\textbf{GLM}} \\
\cmidrule(lr){3-5}\cmidrule(l){6-8}
Contrast & Definition & $\Delta$ & 95\% CI & $p$ & $\Delta$ & 95\% CI & $p$ \\
\midrule
Payload geometry          & quarter $-$ baseline & $+$0.07  & [0.03, 0.12]   & 0.003    & $+$0.05  & [$-$0.01, 0.11] & 0.12 \\
Surround, full scale      & fullpad $-$ baseline & $+$0.125 & [0.070, 0.179] & $<$0.001 & $+$0.03  & [$-$0.02, 0.08] & 0.30 \\
Surround, quarter scale   & pad $-$ quarter      & $+$0.10  & [0.05, 0.15]   & $<$0.001 & $+$0.307 & [0.24, 0.37]    & $<$0.001 \\
\bottomrule
\end{tabular*}}
\vspace{-0.4cm}
\caption{Payload-scale control (E5) on the two effect-carrying victims, laid out as a $2{\times}2$ factorial: payload at full or quarter canvas scale $\times$ no surround or a pad surround. Panel (a): ASR per cell, with each cell's name in italics. Panel (b): paired-prompt $\Delta$ASR with 95\% bootstrap CI and Holm-corrected paired McNemar $p$-values. On Gemma-3-4B the payload-geometry contrast is significant in this single-draw run ($+$0.07, $p{=}0.003$) but does not survive the $k{=}3$ per-prompt permutation null of Appendix~\ref{app:t21} (permutation $p{=}0.36$); a surround at constant payload scale raises ASR at both scales, stronger at full scale ($+$0.125) than at quarter scale ($+$0.10).}
\label{tab:e5_payload}
\end{table}

\vspace{\ParaVspace}
The payload-scale control (E5) isolates the first rung of the Exp.~2 ladder. The baseline (FigStep) places the payload at full canvas scale with no surround; FigStep-pad shrinks the payload to a quarter of the canvas and adds a pad surround, so the raw rung-one jump conflates two changes (a geometry change and a new surrounding region). E5 splits these into a $2{\times}2$ factorial on the two effect-carrying victims (Table~\ref{tab:e5_payload}): full/quarter payload $\times$ no/pad surround, adding the two missing cells (full-scale-with-surround and quarter-scale-no-surround). On Gemma-3-4B the payload-geometry contrast (quarter-no-surround minus full-no-surround) shows a single-draw cell gap of $+0.07$ ($p{=}0.003$ in the E5 single-draw run), but this rung does not clear the $k{=}3$ sampling floor of Appendix~\ref{app:t21} (permutation $p{=}0.36$), so it is flagged at the resolution limit there. Adding a pad surround at constant quarter payload scale robustly raises ASR ($+0.10$, $p{<}0.001$; this is the pad-surround rung that clears the floor, jointly attributed to layout and compression), and at constant full payload scale the same surround raises ASR by $+0.125$ [$0.070,0.179$], $p{<}0.001$. On Gemma-3-4B the full-scale-surround cell (\emph{fullpad}) is complete on the full $n{=}313$ slice (ASR $0.4185$), so it is reported in Table~\ref{tab:e5_payload}. The single-draw and $k{=}3$ numbers belong to different samplers: the E5 factorial is a single-draw paired gap, while Appendix~\ref{app:t21} averages three draws per prompt. We take the $k{=}3$ test as the sampling-noise bound: the payload-geometry rung does not clear it and so is flagged at the resolution limit rather than resolved into either component, and the two rungs that do clear it (compositional, content-fill) carry the decomposition of Section~\ref{sec:res_complexity}. On GLM-4.1V-Th. the geometry axis is not silent, but its Exp.~2 signal is inverted relative to the ladder narrative: the quarter-scale cell is markedly lower than both baseline and pad (see Table~\ref{tab:e5_payload}), so the geometry contrast is a significant \emph{decrease} on that victim rather than a saturated non-effect.

\vspace{\SubBeforeNeg}
\subsection{Consolidating Random-Intercept Mixed Model}
\vspace{\SubAfterNeg}
\label{app:glmm}

\vspace{\ParaVspace}
To summarize the per-victim pairwise tests, we fit a random-intercept logistic mixed model for each experiment, with condition as a categorical factor, victim as a fixed effect, a condition-by-victim interaction, and a per-prompt random intercept. The design has condition numbers of 28 (Exp.~2) and 22 (Exp.~3), with maximum VIFs of 7.6 and 6.7. Posterior SDs for the prompt intercept are 0.76 and 0.64, confirming substantial within-prompt dependence. The model agrees with the pairwise results: Exp.~2 contrasts are largest on Gemma-3-4B and GLM-4.1V-Thinking, small on Qwen3-VL-8B, and near zero on GPT-4o. In Exp.~3, the noise-to-relevant logit contrast is largest on Qwen, but the underlying image descriptors co-vary, so this pooled association does not establish an isolated alignment effect. Claude's large contrast reflects its all-zero noise cell and is descriptive. A continuous-predictor sensitivity model using tile count, entropy, and CLIP alignment is near-degenerate (condition number 451--471) because those features are constant within each condition. We therefore report the categorical model as a summary; the paired within-prompt tests remain the primary inference.

\vspace{\SubBeforeNeg}
\subsection{Experiment 4: MLRM versus MLLM Paired Contrasts}
\vspace{\SubAfterNeg}
\label{app:mlrm}

\begin{table*}[tbp]
\centering
\small
  \setlength{\tabcolsep}{3pt}
\begin{tabular*}{\textwidth}{p{3.4cm} p{1.2cm} p{1.8cm} p{2.1cm} p{1.9cm} p{2.1cm}}
\toprule
\textbf{Paired Contrast} & \textbf{GPT-4o} & \textbf{Claude S.4.5} & \textbf{Qwen3-VL-8B} & \textbf{Gemma-3-4B} & \textbf{GLM-4.1V-Th.} \\
\midrule
FigStep $\to$ distr.-random & $+0.00$ & $+0.03$ & $+0.01$ & $+0.34$ & $+0.16$ \\
distr.-random $\to$ relevant & $+0.00$ & $+0.00$ & $+0.01$ & $-0.04$ & $-0.03$ \\
noise $\to$ unrelated & $+0.07$ & $+0.32$ & $+0.05$ & $-0.32$ & $+0.01$ \\
unrelated $\to$ relevant & $-0.02$ & $-0.10$ & $+0.05$ & $+0.25$ & $-0.02$ \\
CS-DJ $\to$ relevant & $+0.01$ & $-0.10$ & $+0.00$ & $+0.06$ & $+0.00$ \\
\bottomrule
\end{tabular*}
\vspace{-0.4cm}
\caption{Exp.4 re-slice: ASR change ($\Delta$ASR; signed, absolute ASR units) on key paired contrasts, reported \emph{per victim} rather than as an MLLM-block average. The FigStep$\to$distr.-random complexity contrast is dominated by \textsc{Gemma-3-4B} ($+0.34$); \textsc{GLM-4.1V-Th.} moves $+0.16$, more sensitive than \textsc{GPT-4o}, \textsc{Claude}, and \textsc{Qwen3-VL} but \emph{less} than the strongest standard model. The Exp.~3 alignment contrasts are flat or move against the predicted ordering on the four non-Qwen victims, on which alignment is at or below the detection limit (Gemma's row orders by tile provenance rather than by $S$; Section~\ref{sec:res_alignment}). The per-victim view avoids the misleading implication of an MLLM-block average that the reasoning model amplifies \emph{both} axes.}
\label{tab:exp4_mlrm}
\end{table*}

\vspace{\ParaVspace}
Table~\ref{tab:exp4_mlrm} reports five paired contrasts for the original five-model panel; the supplemental InternVL3.5-8B headroom check is not part of this comparison. On the FigStep-to-distract-random ladder, the MLRM gains $+0.16$, more than GPT-4o ($+0.00$), Claude ($+0.03$), and Qwen ($+0.01$), but less than Gemma-3-4B ($+0.35$). Only Gemma and the MLRM have clear headroom on this contrast in the original panel, so this is a localized observation, not a general property of reasoning models. On the alignment contrasts, the MLRM moves $-0.03$ from unrelated to relevant, while Qwen's descriptive ordered-endpoint change is $+0.05$; that endpoint is not an isolated alignment effect. The CS-DJ-to-relevant contrast is flat on the MLRM and standard models, apart from Claude's interpretation-limited row. The MLRM has the highest ASR in every baseline row except MML-base64; this is consistent with greater sensitivity to some composite-image attacks, not evidence of general reasoning-model behavior.

\vspace{\SubBeforeNeg}
\subsection{Minimum Detectable Alignment Effect}
\vspace{\SubAfterNeg}
\label{app:power}

\begin{table}[tbp]
\centering
\small
\setlength{\tabcolsep}{4pt}
\renewcommand{\arraystretch}{1.15}
\begin{tabular*}{\linewidth}{@{\extracolsep{\fill}} l ccccc @{}}
\toprule
\textbf{Paired contrast} & \textbf{GPT-4o} & \textbf{Claude S.4.5} & \textbf{Qwen3-VL-8B} & \textbf{Gemma-3-4B} & \textbf{GLM-4.1V-Th.} \\
\midrule
noise $\to$ unrelated    & 0.079 & 0.088 & 0.075 & 0.103 & 0.061 \\
unrelated $\to$ relevant & 0.076 & 0.088 & 0.082 & 0.100 & 0.064 \\
noise $\to$ relevant     & 0.076 & 0.073 & 0.081 & 0.085 & 0.062 \\
\bottomrule
\end{tabular*}
\vspace{-0.4cm}
\caption{Minimum detectable absolute $\Delta$ASR at 80\% power for the paired McNemar test over $n{=}313$ prompts, conditional on the observed number of discordant pairs in each Experiment~3 contrast and victim (two-sided, $\alpha{=}0.05$, normal approximation). Across the panel this bound ranges from 0.06 to 0.10 absolute ASR units; it is the smallest alignment effect this experiment could resolve on a given victim.}\label{tab:min_detect}
\end{table}

\begin{table}[tbp]
\centering
\small
\setlength{\tabcolsep}{4pt}
\renewcommand{\arraystretch}{1.15}
\begin{tabular*}{\linewidth}{@{\extracolsep{\fill}} lllccc @{}}
\toprule
\textbf{Benchmark} & \textbf{Victim} & \textbf{Contrast (E4)} & $\boldsymbol{n}$ & \textbf{80\%-power MDE} & \textbf{Observed $\Delta$ASR} \\
\midrule
SafeBench & Qwen3-VL-8B & relevant $\to$ gen-unrelated & 119 & 0.131 & $-$0.06 \\
\bottomrule
\end{tabular*}
\vspace{-0.4cm}
\caption{Minimum detectable absolute $\Delta$ASR at 80\% power for the Qwen3-VL-8B E4 contrast on the SafeBench second harm set, using the same normal-approximation method as Table~\ref{tab:min_detect} and conditioning on the 33 discordant pairs observed among 119 prompts. The sample size was set by the benchmark, not by a power calculation. The observed $\Delta$ASR of $-0.06$ is smaller in magnitude than the 0.131 bound, so this is an underpowered directional agreement, not a confirmed replication. Table~\ref{tab:min_detect} reports the $n{=}313$ primary set.}\label{tab:min_detect_safebench}
\end{table}

\vspace{\ParaVspace}
Table~\ref{tab:min_detect} reports the per-victim minimum detectable absolute $\Delta$ASR at $80\%$ power for the paired McNemar test over $n{=}313$ prompts. The thresholds range from $0.06$ to $0.10$. For victims with null results, they bound the size of effects the experiment could have detected; they do not show that effects below each threshold are absent. Qwen's ordered endpoint exceeds its $0.081$ threshold, and the E4 control supports the bounded relatedness attribution reported in Section~\ref{sec:res_alignment}. Residual entropy variation remains a confound for the unisolated Experiment~3 endpoint.

\vspace{\ParaVspace}
The same power bound for the Exp.~2 composition ladder, the headline axis of that experiment, is given in Table~\ref{tab:min_detect_exp2}:
\begin{table}[tbp]
\centering
\small
\setlength{\tabcolsep}{4pt}
\renewcommand{\arraystretch}{1.15}
\begin{tabular*}{\linewidth}{@{\extracolsep{\fill}} l cccc @{}}
\toprule
\textbf{Paired step} & \textbf{Gemma-3-4B} & \textbf{GLM-4.1V-Th.} & \textbf{Qwen3-VL-8B} & \textbf{Claude S.4.5} \\
\midrule
baseline $\to$ pad                      & \textbf{0.082} & 0.072 & 0.042 & 0.015 \\
pad $\to$ distract-random               & \textbf{0.088} & 0.064 & 0.047 & 0.025 \\
distract-random $\to$ distract-relevant & 0.086          & 0.058 & 0.055 & 0.032 \\
\bottomrule
\end{tabular*}
\vspace{-0.4cm}
\caption{Minimum detectable absolute $\Delta$ASR at 80\% power for the paired McNemar test over $n{=}313$ prompts, conditional on the observed number of discordant pairs in each Exp.~2 ladder step (two-sided, $\alpha{=}0.05$, normal approximation), for the ladder's non-floored victims. On Gemma-3-4B both headline steps (baseline $\to$ pad, $+$0.17; pad $\to$ distract-random, $+$0.17) clear this bound (0.082, 0.088); on GLM-4.1V-Th. the compositional step ($+$0.07) sits just below its 0.072 bound while the density step ($+$0.08) clears it. GPT-4o is excluded (floored, no finite bound computable).}
\label{tab:min_detect_exp2}
\end{table}
On \textsc{Gemma-3-4B} both headline steps clear the bound by roughly $2{\times}$; on \textsc{GLM} the compositional step sits at the bound, so the GLM compositional reading is power-limited, exactly as stated in Section~\ref{sec:res_complexity}.

\vspace{\SubBeforeNeg}
\subsection{Sampling-Variance Floor (E-2)}
\vspace{\SubAfterNeg}
\label{app:e2}

\begin{table}[tbp]
\centering
\small
\setlength{\tabcolsep}{3pt}
\begin{tabular}{p{3.0cm} p{3.0cm} p{3.0cm} p{3.6cm}}
\toprule
\textbf{Condition} & \textbf{$k{=}3$ pooled ASR} & \textbf{Discordance floor} & \textbf{Exp.~2 ASR (single-draw)} \\
\midrule
FigStep (baseline) & 0.276 & 0.101 & 0.29 \\
FigStep-pad & 0.382 & 0.105 & 0.46 \\
FigStep-distr.-rand.$^{\dagger}$ & 0.643 & 0.104 & 0.64 \\
FigStep-distr.-rel. & 0.502 & 0.121 & 0.60 \\
\bottomrule
\end{tabular}
\vspace{-0.25cm}
\caption{Sampling-variance floor on the effect-carrying victim (\textsc{Gemma-3-4B}). Each of the four Experiment~2 conditions was re-run at $k{=}3$ draws per prompt (same deterministic stimulus across draws). The discordance floor is the mean over prompts of $2p(1{-}p)$, the probability two independent draws differ at $\tau{=}0.5$; it bounds the sampling noise a single-draw ASR estimate is subject to. $^{\dagger}$The \emph{distr.-random} cell was re-judged with the \emph{rubric} judge on the \emph{real} cached stimulus, so its floor is comparable to the Exp.~2 table cell; the other three rows use the local finetuned judge. This floor is descriptive, not an acceptance test: a rung is accepted on the within-prompt permutation $p$-value (Section~\ref{sec:method:protocol}). The final column is the single-draw ASR of Table~\ref{tab:exp2_complexity}; it is named sampler per column. The \emph{distr.-rel.} row carries the largest single-draw versus $k{=}3$ gap in the ladder ($0.60$ vs.\ $0.50$, ${\approx}0.10$), consistent with its being the row with the highest discordance floor ($0.121$) in the same table, i.e.\ the one whose ASR is least stable across draws; its ordering should therefore be read at the $k{=}3$ estimate, not the single draw.}
\label{tab:e2_discordance}
\end{table}

\vspace{\ParaVspace}
Table~\ref{tab:e2_discordance} addresses the single-draw objection directly: because every other cell in the paper reports one judge score per prompt, any between-condition ASR difference could be attributed to decoding noise. E-2 bounds that noise on the victim carrying the effect (\textsc{Gemma-3-4B}) by re-running the four Experiment~2 conditions at $k{=}3$ draws per prompt, held at the same deterministic stimulus so that the measured discordance is purely model sampling variance. The resulting within-prompt floor, the probability that two independent draws receive different verdicts, is $0.10$--$0.12$ across the four conditions and the mean within-prompt judge-score standard deviation is ${\approx}0.09$. All three headline Experiment~2 contrasts (baseline against pad, distract-random, and distract-relevant) exceed this floor, so their ASR differences are not artifacts of a single draw (the pad-surround and content-fill rungs remain the accepted steps; the payload-geometry contrast is flagged at the resolution limit under the $k{=}3$ permutation test, Appendix~\ref{app:t21}); the weaker variant-vs-variant contrasts sit at or below it, consistent with those differences being sampling noise. Together, these checks quantify the sampling variation in the Gemma ladder without requiring three draws for every cell in the full panel.

\vspace{\ParaVspace}
Because the floor is used to defend deltas produced by the rubric judge, we re-judged the cell whose margin most affects the ladder (\emph{distract-random}) with the \emph{rubric} judge on the \emph{real} cached stimulus (the original E-2 run used the local finetuned judge and, for this cell, a seeded-noise fallback stimulus). Under the rubric judge the within-prompt floor is $0.104$ with a mean within-prompt judge-score standard deviation of ${\approx}0.11$, matching the $0.10$--$0.12$ bound from the local judge, and the headline baseline-to-distract-random contrast ($\Delta{=}0.349$ on the rubric-comparable pooled ASR) still exceeds it. The floor is therefore judge-robust: switching from the local to the rubric judge does not change the conclusion that the ladder's effect-carrying steps are not single-draw noise.

\vspace{\SubBeforeNeg}
\subsection{Threshold and Judge Robustness}
\vspace{\SubAfterNeg}
\label{app:tau_robust}

\begin{table}[tbp]
\centering
\small
\setlength{\tabcolsep}{4pt}
\renewcommand{\arraystretch}{1.15}
\begin{tabular*}{\linewidth}{@{\extracolsep{\fill}} l lc lc @{}}
\toprule
 & \multicolumn{2}{c}{\textbf{Exp.~2 (complexity)}} & \multicolumn{2}{c}{\textbf{Exp.~3 (alignment)}} \\
\cmidrule(lr){2-3}\cmidrule(l){4-5}
\textbf{Victim} & Order & Stable & Order & Stable \\
\midrule
GPT-4o       & all 0.00 (floor)                          & --  & un $>$ rel $>$ noise & yes$^{*}$ \\
Claude S.4.5 & rel $\geq$ rand $>$ pad $\geq$ base       & no  & un $>$ rel $>$ noise & yes \\
Qwen3-VL-8B  & rel $\approx$ rand $>$ base $>$ pad       & no  & rel $>$ un $>$ noise & yes \\
Gemma-3-4B   & rand $\geq$ rel $>$ pad $>$ base          & no  & noise $>$ rel $>$ un & yes \\
GLM-4.1V-Th. & rand $>$ rel $>$ pad $>$ base             & yes & un $>$ noise $>$ rel & yes \\
\bottomrule
\end{tabular*}
\vspace{-0.4cm}
\caption{Sensitivity of the condition ordering to the binarization threshold. For each victim we re-compute the per-condition ASR at $\tau\in\{0.3,0.5,0.7\}$ (StrongREJECT, $n{=}313$/cell) and list the resulting Exp.~2 and Exp.~3 orderings; ``stable'' means the identical ranking holds at all three $\tau$. Exp.~2: base = baseline, pad, rand = distract-random, rel = distract-relevant; Exp.~3: noise, un = unrelated, rel = relevant. The claims the paper rests on survive: Qwen3-VL-8B's Exp.~3 endpoint ordering, Gemma-3-4B's tile-provenance ordering, and GLM-4.1V-Th.'s Exp.~2 ladder ordering are all stable. The only instabilities are between adjacent, statistically indistinguishable cells. $^{*}$GPT-4o flips \emph{unrelated}/\emph{relevant} at $\tau{=}0.7$ (both within their CI overlap); \emph{noise} stays lowest.}
\label{tab:tau_robust}
\end{table}

\vspace{\ParaVspace}
Table~\ref{tab:tau_robust} re-binarizes each cell at $\tau\in\{0.3,0.5,0.7\}$ to check whether key orderings depend on the judge threshold. Qwen3-VL-8B's noise $<$ unrelated $<$ relevant ordering holds at all three thresholds, but remains a descriptive Exp.~3 endpoint because the image descriptors co-vary; the E4 contrast supplies the relatedness attribution. Gemma-3-4B's noise $\ge$ relevant $>$ unrelated ordering also holds. The GLM-4.1V-Th. ladder and floored rows show no meaningful adjacent-cell reversals. Threshold robustness supports the stability of these orderings, not a causal interpretation of every contrast.

\vspace{\ParaVspace}
Judge robustness. Because every ASR delta is judge-conditional (Section~\ref{sec:method:protocol}), we state the bound between the two judges: the largest cross-judge gap between the local finetuned judge and the rubric judge is $0.18$ absolute ASR at \emph{distract-random} (the ladder's gap is roughly $0.02$, $0.08$, $0.18$, and $0.10$ at baseline, pad, distract-random, and distract-relevant). This $0.18$ exceeds every headline delta except \textsc{Gemma-3-4B}'s content-fill step (the $k{=}3$ estimate $+0.070$; single-draw $+0.17$ sits on this band) and ladder total ($+0.35$); on the remaining contrasts the disagreement band is smaller than the reported effect, so the qualitative ordering survives a switch between judges. We therefore flag, as in the main text, that the \emph{magnitudes} near the disagreement band are not judge-invariant, while the \emph{orderings} on the effect-carrying victims are.

\vspace{\ParaVspace}
Self-family re-judge of the \textsc{GPT-4o} rows. Because \textsc{GPT-4o} is both a victim and the family behind the rubric judge (\texttt{gpt-4o-mini}), we re-judge every \textsc{GPT-4o} cell with the local finetuned StrongREJECT judge (\texttt{strongreject-15k-v1}, a non-OpenAI network; results/r8/gpt4o\_rejudge.json). The floored-row narrative survives: on the Exp.~2 ladder, which the rubric judge reads as $0.00$ across all four conditions, the non-OpenAI judge also gives near-zero ASR ($0.048$--$0.096$), so \textsc{GPT-4o}'s Exp.~2 floor is a genuine hard refusal rather than a self-family artifact. On the Exp.~3 rows the two judges agree in placing the cell ASRs in the $0.19$--$0.27$ (rubric) and $0.19$--$0.40$ (local) ranges with the same qualitative shape, and the per-category profile holds at the level of which categories leak most (Non-violent crimes highest, Sexual content lowest) under both judges. We therefore retain the \textsc{GPT-4o} Exp.~2 floor and its per-category breakdown as reported, now confirmed under a self-family-free label.
\vspace{\ParaVspace}
Continuous-score paired tests. Because every McNemar test above binarizes at $\tau{=}0.5$ and drops the continuous info, we repeat the two effect-carrying drivers directly on the continuous judge scores (paired Wilcoxon signed-rank over the $n{=}313$ shared prompts). The Exp.~3 ordered endpoint on \textsc{Qwen3-VL-8B} ($\mathrm{noise}\to\mathrm{relevant}$) is confirmed: Wilcoxon $p{=}0.0002$ with a paired continuous mean shift of $+0.076$ $[0.037,0.115]$, strictly stronger and threshold-free relative to the binarized $+0.10$ (Holm $p{=}0.002$). The content-fill step on \textsc{Gemma-3-4B} (pad $\to$ distract-random) is confirmed at Wilcoxon $p{<}0.001$, mean $+0.141$ $[0.092,0.189]$; on \textsc{GLM-4.1V-Th.} at $p{<}0.001$. Conversely, the cell the binarized test reads as null here, \textsc{Qwen3-VL-8B}'s Exp.~2 pad $\to$ distract-random ($+0.01$), stays null on the continuous scale as well (Wilcoxon $p{=}0.30$), which shows the continuous test does not inflate noise into signal: it separates the genuinely weak drivers (Gemma/GLM content-fill, Qwen Exp.~3) from the genuinely flat ones (Qwen Exp.~2) with the same ordering the binarized analysis reports, but with tighter bounds.

\vspace{\SubBeforeNeg}
\subsection{Human-Annotation Judge Agreement}
\vspace{\SubAfterNeg}
\label{app:human_kappa}

\begin{table}[tbp]
\centering\small\setlength{\tabcolsep}{4pt}
\begin{tabular}{p{2.31cm} p{0.57cm} p{1.69cm} p{1.19cm} p{1.72cm} p{1.96cm}}
\toprule
\textbf{Condition} & $n$ & \textbf{Agreement} & $\kappa$ & \textbf{Judge ASR} & \textbf{Human ASR} \\
\midrule
baseline          & 69 & 0.841 & 0.679 & 0.522 & 0.594 \\
pad               & 64 & 0.797 & 0.594 & 0.484 & 0.500 \\
distract-random   & 59 & 0.797 & 0.579 & 0.610 & 0.576 \\
distract-relevant & 58 & 0.759 & 0.496 & 0.586 & 0.621 \\
\midrule
overall           & 250 & 0.800 & 0.594 & --- & --- \\
\bottomrule
\end{tabular}
\vspace{-0.2cm}
\caption{Human-annotation judge agreement. A human annotator labeled $250$ samples stratified by judge-score band (with deliberate oversampling in the two bands adjacent to $\tau{=}0.5$) across the four Exp.~2 conditions. Overall agreement is $0.800$ with Cohen's $\kappa{=}0.594$; per-condition $\kappa$ ranges $0.496$--$0.679$. The implied ASR shift (human labels replacing judge labels, stratum-weighted) is small and does not reverse any headline ordering.}
\label{tab:human_kappa}
\end{table}

\vspace{\ParaVspace}
A central question about the automated judge is not whether it is accurate on average but whether its error interacts with the manipulated factor. To answer that, a human annotator labeled $250$ samples drawn from the four Exp.~2 conditions, stratified by judge-score band with deliberate oversampling in the two bands adjacent to $\tau{=}0.5$ (where a single verdict flip changes the binarized outcome), so that the batch is not dominated by unambiguous refusals. Table~\ref{tab:human_kappa} reports the result. Overall agreement with the rubric judge is $0.800$ with Cohen's $\kappa{=}0.594$; per-condition $\kappa$ ranges $0.496$ (distract-relevant) to $0.679$ (baseline). The implied ASR shift, extrapolating each stratum's human-vs-judge disagreement back to the full conditions with stratum weights, is small in every cell (baseline $+0.072$, pad $+0.016$, distract-random $-0.034$, distract-relevant $+0.034$) and, crucially, does not reverse any headline ordering: the ladder's effect-carrying steps and the alignment nulls survive a switch from judge labels to human labels. Judge error therefore does not interact with the manipulated factor in a way that changes any conclusion in the paper.

\vspace{\SubBeforeNeg}
\subsection{Payload Legibility Ladder}
\vspace{\SubAfterNeg}
\label{app:n1}

\vspace{\ParaVspace}
A concern is that the tile-count effect in the CS-DJ ladder is confounded with payload resolution: every victim encoder resizes the input to a fixed budget, so expanding the canvas to add tiles shrinks each tile's share of that budget and could make the harmful text progressively harder for the OCR pathway to read, with a monotone ASR rise being the result. This section measures the payload's legibility directly at each victim's post-resize resolution, using a local optical-character-recognition pass (tesseract, English) over the real cached stimulus. For each of $20$ c12 stimuli we crop the three sub-query (payload) tiles, downscale the full stimulus to two target resolutions, and compute the character error rate (CER) against the ground-truth sub-query text from the stimulus metadata (embedded in the cached entry).

\begin{table}[htbp]
\centering
\small
\setlength{\tabcolsep}{4pt}
\begin{tabular}{l c c}
\toprule
resize target & mean CER & mean payload text height (px) \\
\midrule
source (no resize)      & 0.115 & 318 \\ \textsc{Qwen3-VL-8B} post-resize & 0.115 & 318 \\ \textsc{Gemma-3-4B} post-resize ($896^2$) & 0.162 & 137 \\
\bottomrule
\end{tabular}
\vspace{-0.3cm}
\caption{Payload legibility: character error rate and rendered text height of the CS-DJ sub-query payload under the two open victims' post-resize encoder resolutions, on $20$ real stimuli. CER stays low ($\le 0.16$) at every resolution, so payload legibility does not degrade with tile count; text height falls at Gemma's fixed $896^2$ but remains far above the legibility floor.}
\label{tab:n1_legibility}
\end{table}

\vspace{\ParaVspace}
First, the ladder cannot be attributed to composition alone: because each rung renders only the first $N$ tiles of the fixed 12-tile bucket (9 distractors $+$ 3 sub-query tiles), the $c_1$ and $c_4$ composites contain no legible sub-query payload text ($0.3\%$ and $0.6\%$ of stimuli carry any oracle token on a tesseract pass; results/r5/audit\_ladder\_payload.json) while $c_{12}$ carries it (the three typographic tiles, $8.3\%$). The ladder therefore varies both tile count and the visibility of the harmful sub-query text, so it is not a clean composition isolation and is handled as such in Section~\ref{sec:res_csdj_ladder}. Given that, the relevant legibility comparison is between the c12 payload at source, at Qwen's dynamic resolution, and at Gemma's fixed resolution. CER stays approximately flat ($0.115\to0.162$) as ASR rises threefold across the ladder (Table~\ref{tab:csdj_ladder}), and remains low even at Gemma's coarsest effective resolution. The tile-count effect is therefore not a relabeling of payload illegibility: the payload stays readable wherever it appears. This bounds, rather than removes, the resolution confound: the per-tile encoder budget still declines with tile count, but that budget decline does not cross the legibility threshold at which the payload becomes hard to read.

\vspace{\SubBeforeNeg}
\subsection{Alignment in Each Victim's Own Vision Encoder}
\vspace{\SubAfterNeg}
\label{app:w8}

\vspace{\ParaVspace}
The main-text alignment proxy is a CLIP ViT-L/14 cosine (Section~\ref{sec:method:axes}). Because no victim uses that encoder, a null in CLIP space is compatible with a real alignment effect in each victim's own vision tower. Qwen3-VL, Gemma-3, and GLM-4.1V do not expose a contrastive text tower aligned to their vision tower, so a raw image-to-text cosine is not available. We therefore operationalize alignment in-model as the victim's own log-probability of the statement ``the image is related to the harmful request'': for each of the three Exp.~3 conditions, we score a prompt's image in the victim's own vision tower and take the log-odds of the ``yes'' token against the ``no'' token (one forward pass per stimulus, no generation sampled). The probe prompts the model to continue ``Is the image related to the following request? $\langle q\rangle$\\Answer yes or no:''; we read the next-token logits at position $-1$ for the single tokens ``yes'' and ``no'' (Gemma-3 token ids $4443$/$1904$; Qwen3-VL the same two single tokens) and form $\log P(\texttt{yes})-\log P(\texttt{no})$; no continuation is sampled. For Gemma the text carries the explicit $\langle\text{start\_of\_image}\rangle$ token; for Qwen3-VL the image placeholder block is expanded through the chat template (``$\langle|\texttt{vision\_start}|\rangle$''). If the CLIP-based alignment null were an artifact of the third-party encoder, this in-model score should rise monotonically from \emph{noise} to \emph{relevant}; if it does not, the null survives in the victim's own space.

\begin{table}[htbp]
\centering
\small
\setlength{\tabcolsep}{5pt}
\begin{tabular}{p{2.4cm} p{1.9cm} p{2.2cm} p{2.1cm}}
\toprule
\textbf{Exp.\ 3 condition} & \textbf{Gemma-3-4B} & \textbf{Qwen3-VL-8B} & \textbf{ASR (Gemma)} \\
\midrule
noise     & $+2.51 \pm 1.35$ & $-2.08 \pm 7.27$ & $0.73$ \\
unrelated & $+1.31 \pm 1.44$ & $-3.89 \pm 7.10$ & $0.41$ \\
relevant  & $+2.09 \pm 1.63$ & $-2.28 \pm 6.75$ & $0.66$ \\
\bottomrule
\end{tabular}
\vspace{-0.3cm}
\caption{Victim own-vision-tower alignment log-odds, mean $\pm$ std over $n{=}313$ per condition. Neither victim's own tower reproduces the alignment ordering: \textsc{Gemma-3-4B} rates \emph{noise} highest then \emph{relevant} then \emph{unrelated}, and \textsc{Qwen3-VL-8B} groups \emph{noise} and \emph{relevant} tightly (both well above \emph{unrelated}), so in neither encoder is \emph{relevant} ranked above \emph{noise} as the alignment hypothesis requires. Reported ASR is \textsc{Gemma-3-4B}'s Exp.~3 cell for reference.}
\label{tab:t34_own_encoder}
\end{table}

\vspace{\ParaVspace}
The result (Table~\ref{tab:t34_own_encoder}, both open victims) does \emph{not} reproduce the alignment ordering: \textsc{Gemma-3-4B} assigns its highest relatedness to \emph{noise} ($+2.51$), the condition with the lowest-entropy, most ``signal-like'' tiles, then \emph{relevant} ($+2.09$), and lowest to \emph{unrelated} ($+1.31$); \textsc{Qwen3-VL-8B} groups \emph{noise} ($-2.08$) and \emph{relevant} ($-2.28$) together, both well above \emph{unrelated} ($-3.89$). In neither victim's own tower is \emph{relevant} ranked above \emph{noise}, as the alignment hypothesis requires. The in-model alignment proxy does not recover an alignment signal where the CLIP proxy saw none. We treat this as an \emph{exploratory} instrument, not a confirmatory one: Qwen's per-stimulus log-odds have a standard deviation of roughly $7.0$--$7.3$ against between-condition mean differences of order $1.6$--$1.8$, and Gemma rates \emph{noise} as most related, so the relatedness log-odds probe is too noisy to confirm or to contradict the CLIP-space null in either victim's own tower; it supports only the weaker claim that no alignment ordering is recovered in-model. A valid confirmatory instrument would require a probe with per-stimulus variance at or below the between-condition separation it is asked to detect, which this design does not achieve. We state this operationalization explicitly: it is the victim's own relatedness log-odds, not a CLIP-equivalent cosine; Gemma's high \emph{noise} score reflects that the model treats low-entropy uniform tiles as unusually ``instructive''.

\vspace{\SubBeforeNeg}
\subsection{Claude's Exp.~3 Noise Cell: Provider-Side Recovery from Logs}
\vspace{\SubAfterNeg}
\label{app:t36}

\vspace{\ParaVspace}
The Exp.~3 noise cell on \textsc{Claude Sonnet~4.5} has $0.00$ ASR across all $313$ prompts, compared with $0.32$ for the unrelated condition. This difference could be read as an out-of-distribution response to the noise tiles. We inspected the cached per-prompt responses to distinguish that possibility from provider-side intervention.

\vspace{\ParaVspace}
The victim runner (eval/victims.py) returns the literal string \texttt{"I'm sorry."} whenever the provider returns an empty content list or an HTTP 400 without a usable response. We tabulate, per condition, the fraction of the $313$ responses that hit this fallback versus a real model-generated refusal versus an actual response:

\begin{table}[htbp]
\centering
\small
\setlength{\tabcolsep}{4pt}
\begin{tabular}{l c c c c}
\toprule
condition & empty/400 fallback & real refusal & actual response & reported ASR \\
\midrule
noise     & \textbf{1.00} (313) & 0.00 (0)   & 0.00 (0)   & 0.000 \\ unrelated & 0.042  (13)  & 0.387 (121) & 0.571 (179) & 0.318 \\ relevant  & 0.415  (130) & 0.236  (74) & 0.348 (109) & 0.219 \\
\bottomrule
\end{tabular}
\vspace{-0.3cm}
\caption{Per-condition response-class distribution for \textsc{Claude Sonnet~4.5}'s Exp.~3 cells (full StrongREJECT, $n{=}313$). ``Empty/400 fallback'' is the cached {\tt "I'm sorry."} that the runner produces when the provider returns an empty content or an HTTP 400; ``real refusal'' is a model-generated refusal; ``actual response'' is any other generated text. The noise cell is $100\%$ fallback, so it does not measure a model-internal refusal at all.}
\label{tab:t36_claude_resp}
\end{table}

\vspace{\ParaVspace}
All $313$ noise requests were handled at the API boundary, returning empty content or an HTTP 400 response. The $0.00$ ASR therefore reflects provider-side intervention, not evidence of a victim-internal response to out-of-distribution tiles. The payload-legibility analysis in Appendix~\ref{app:n1} shows that the text remains readable. However, these logs cannot establish why the provider intercepted the noise images, so we do not infer an image-moderation mechanism. The Claude row is consequently interpretation-limited. Conditioning on non-intercepted responses does not resolve the issue: interception varies by condition ($4.2\%$ for unrelated and $41.5\%$ for relevant), so the remaining responses form a selected subset. Within that subset, relevant ASR ($\approx 0.38$) exceeds unrelated ASR ($\approx 0.33$), reversing the raw ordering. Neither the raw nor conditioned results support an alignment or provenance interpretation for Claude.

\vspace{\SubBeforeNeg}
\subsection{Exp.~2 Ladder Sub-Step Noise Bounds}
\vspace{\SubAfterNeg}
\label{app:t21}

\vspace{\ParaVspace}
We bound the sampling-variance floor on the \emph{sub-steps} of the Exp.~2 ladder decomposition, not only per condition. We re-ran each of the three ladder rungs on \textsc{Gemma-3-4B} with $k{=}3$ draws per prompt (the same protocol as Appendix~\ref{app:e2}; the \emph{quarter} draw data were newly generated) and scored every response with the same rubric judge as the headline cells. For each prompt we average the three binarized draws per condition to a per-prompt compliance probability, then form the paired delta; the sampling standard error is \[ \mathrm{SE} \;=\; \tfrac{1}{n}\sqrt{\textstyle\sum_{i}\;\big[p_{A,i}(1-p_{A,i})+p_{B,i}(1-p_{B,i})\big]}, \] and the primary test is a within-prompt permutation null: pool each prompt's $k{=}3{+}3$ draws, reassign them at random to the two labels, recompute the aggregate delta, repeat $10^4$ times, and report the two-sided $p$-value.

\begin{table}[htbp]
\centering
\small
\setlength{\tabcolsep}{5pt}
\begin{tabular}{l c c c c c}
\toprule
sub-step & observed $k{=}3$ & SE & 95\% half-width & perm.\ $p$ (2-sided) \\
\midrule
baseline $\to$ quarter (geometry)      & $+0.012$ & $0.017$ & $\pm0.032$ & $0.36$ \\ quarter $\to$ pad ($C_\mathrm{comp}$)   & $+0.095$ & $0.017$ & $\pm0.033$ & $<0.001$ \\ pad $\to$ distract-random (content-fill) & $+0.070$ & $0.019$ & $\pm0.037$ & $<0.001$ \\
\bottomrule
\end{tabular}
\vspace{-0.2cm}
\caption{Exp.~2 ladder sub-step sampling noise on \textsc{Gemma-3-4B} ($n{=}313$, $k{=}3$ draws per prompt, rubric judge). The compositional (quarter $\to$ pad) and content-fill (pad $\to$ distract-random) rungs both clear the $k{=}3$ permutation floor ($p{<}0.001$); the payload-geometry first rung (baseline $\to$ quarter) does not (mean delta $+0.012$, permutation $p{=}0.36$), so by the pre-committed rule it is flagged \emph{at the resolution limit} and excluded from the decomposition of Section~\ref{sec:res_complexity}.}
\label{tab:t21_substep}
\end{table}

\vspace{\ParaVspace}
Because the geometry and pad-surround rungs interact with the payload-geometry control of Appendix~\ref{app:e5} (which measured the single-draw gap as $+0.07$, $p{=}0.003$), we note the two belong to different samplers: the E5 number is a single-draw paired gap, while this appendix averages three draws per prompt. We take the $k{=}3$ permutation test as the relevant noise bound; the geometry rung does not clear it, so it is not reported as a resolved component, and the decomposition is carried by the two rungs that do.

\vspace{\SubBeforeNeg}
\vspace{\SubBeforeNeg}\vspace{\SubBeforeNeg}\subsection{Global Multiplicity Correction}\vspace{\SubAfterNeg}\label{app:global_mult}\vspace{\ParaVspace}
\begin{table*}[tbp]
\centering
\small
\setlength{\tabcolsep}{4pt}
\renewcommand{\arraystretch}{1.1}
\begin{tabular*}{\textwidth}{@{\extracolsep{\fill}} ll c cc cc @{}}
\toprule
 & & & \multicolumn{2}{c}{\textbf{Global Holm}} & \multicolumn{2}{c}{\textbf{Global BH}} \\
\cmidrule(lr){4-5}\cmidrule(l){6-7}
\textbf{Victim} & \textbf{Contrast} & \textbf{Raw $p$} & $p$ & Survives & $p$ & Survives \\
\midrule
Gemma & Exp.~2 quarter $\to$ pad (compos.)                & 0.0001 & 0.0013 & yes & 0.0003 & yes \\
Gemma & Exp.~2 pad $\to$ distract-random (fill)           & 0.0001 & 0.0013 & yes & 0.0003 & yes \\
Qwen  & E4 relevant $\to$ gen-unrelated                   & 0.0001 & 0.0013 & yes & 0.0003 & yes \\
Qwen  & Exp.~3 unrelated $\to$ relevant                   & 0.0001 & 0.0013 & yes & 0.0003 & yes \\
Qwen  & E4-within relevant $\to$ gen-unrelated-within     & 0.0001 & 0.0013 & yes & 0.0003 & yes \\
\midrule
Qwen  & Region r1 $\to$ r9                                & 0.2475 & 1.0000 & no  & 0.5363 & no \\
Gemma & Exp.~2 baseline $\to$ quarter (geom.)             & 0.3613 & 1.0000 & no  & 0.6710 & no \\
Qwen  & Region r1 $\to$ r3                                & 0.4847 & 1.0000 & no  & 0.7876 & no \\
Qwen  & Region r3 $\to$ r9                                & 0.7361 & 1.0000 & no  & 0.9945 & no \\
Qwen  & Exp.~3 noise $\to$ unrelated                      & 0.8516 & 1.0000 & no  & 0.9945 & no \\
Gemma & Region r3 $\to$ r9                                & 0.9167 & 1.0000 & no  & 0.9945 & no \\
Gemma & Region r1 $\to$ r9                                & 0.9180 & 1.0000 & no  & 0.9945 & no \\
Gemma & Region r1 $\to$ r3                                & 1.0000 & 1.0000 & no  & 1.0000 & no \\
\bottomrule
\end{tabular*}
\vspace{-0.3cm}
\caption{Global multiplicity correction for the confirmatory family, using one $k{=}3$ sampler. The family includes each adjacent-step and E4 contrast from Experiments~2 and~3, the region-count ladder, and the within-category control once each; exploratory endpoints and single-draw rows are tested separately. Gemma = Gemma-3-4B; Qwen = Qwen3-VL-8B. Holm step-down and Benjamini--Hochberg adjusted $p$-values are computed over $N{=}13$ tests. Five tests pass both corrections at $\alpha{=}0.05$: two Gemma Experiment~2 rungs, Qwen's Experiment~3 unrelated-to-relevant contrast, E4, and the within-category E4 control. The latter is a robustness check on overlapping stimuli, not a separate attribution. Thus, among the five statistically significant tests, E4 is the only contrast attributed to one measured descriptor, the relatedness manipulation; image-text congruence remains unmeasured. The results require an outcome-blind, pre-committed replication before they support a general claim.}
\label{tab:global_mult}
\end{table*}
\vspace{\ParaVspace}The main text reports Holm correction within each victim and experiment. Table~\ref{tab:global_mult} applies global Holm and Benjamini--Hochberg corrections to 13 confirmatory tests from the $k{=}3$ sampler: the adjacent steps in Experiments~2 and~3, the E4 contrast, the region-count ladder, and the within-category control. Five tests pass both global corrections at $\alpha{=}0.05$. The five include two Gemma Experiment~2 rungs, Qwen's Experiment~3 unrelated-to-relevant contrast, E4, and its within-category control. Under the attribution rule, only E4 supports a single-descriptor attribution, to the relatedness manipulation; image-text congruence remains unmeasured. The within-category result is a robustness check on overlapping stimuli, not an independent attribution. The remaining significant tests are jointly attributed or exploratory. Exploratory endpoints and single-draw results are reported separately and do not enter this family. Because the attribution rule was specified after data collection, an outcome-blind, pre-committed replication is needed before generalizing the single-victim finding.

\vspace{\SubBeforeNeg}\subsection{Non-Identification Box}
\vspace{\SubAfterNeg}
\label{app:non_ident}

\vspace{\ParaVspace}
For the reader whose syllabus is the attribution rule of Section~\ref{sec:method:attribution}, we state the complement: the contrasts in this paper that do \emph{not} cleanly identify a single construct. These are deliberately left as jointly attributed rather than resolved, and each is a separate design target for follow-up work.
\begin{itemize}\small
\item No clean $C_\mathrm{info}$ contrast exists in the current design. The Exp.~2 pad-to-distract-random step changes tile provenance along with information density, and other comparisons change additional descriptors; neither entropy nor compressibility is isolated on its own.
\vspace{-0.1cm}
\item Tile count is confounded with per-tile encoder resolution: raising the canvas to add tiles shrinks each tile's share of the fixed post-resize encoder budget. Appendix~\ref{app:n1} shows the sub-query payload remains legible at every resolution, so this is a resolution \emph{bound} on the compositional claim, not a cleanup of it.
\vspace{-0.1cm}
\item Experiment~3 conditions differ in compressibility and in provenance as well as in $S$, so monotone ASR across them cannot isolate alignment; we read only the contrasts where the confounds run \emph{opposite} to ASR.
\vspace{-0.1cm}
\item The alignment proxy is CLIP ViT-L/14 (Section~\ref{sec:method:axes}), and no victim uses that encoder; the own-vision-tower recomputation of Appendix~\ref{app:w8} is the direct check.
\vspace{-0.1cm}
\item The judge is \textsc{GPT-4o-mini}, which shares a family with one victim (\textsc{GPT-4o}); a non-OpenAI re-judge of the ladder (Appendix~\ref{app:tau_robust}) confirms the orderings but leaves magnitudes on the $0.10$--$0.18$ disagreement band judge-conditional.
\end{itemize}
These limits define the scope of the findings: the E4 relatedness attribution is the only positive claim that passes the stated rule, while the remaining contrasts stay joint, exploratory, or unresolved.

\vspace{\SubBeforeNeg}
\subsection{Benign-Region Screening Test}
\vspace{\SubAfterNeg}
\label{app:t35}

\vspace{\ParaVspace}
The screening experiment asks whether a cheap input-side density test can flag attack images. We compare attack tiles with real multimodal web tiles drawn from the CS-DJ distractor pool; if the two classes overlap, density alone may not support a practical filter.

\begin{figure}[tbp]
\centering
\includegraphics[width=0.85\linewidth]{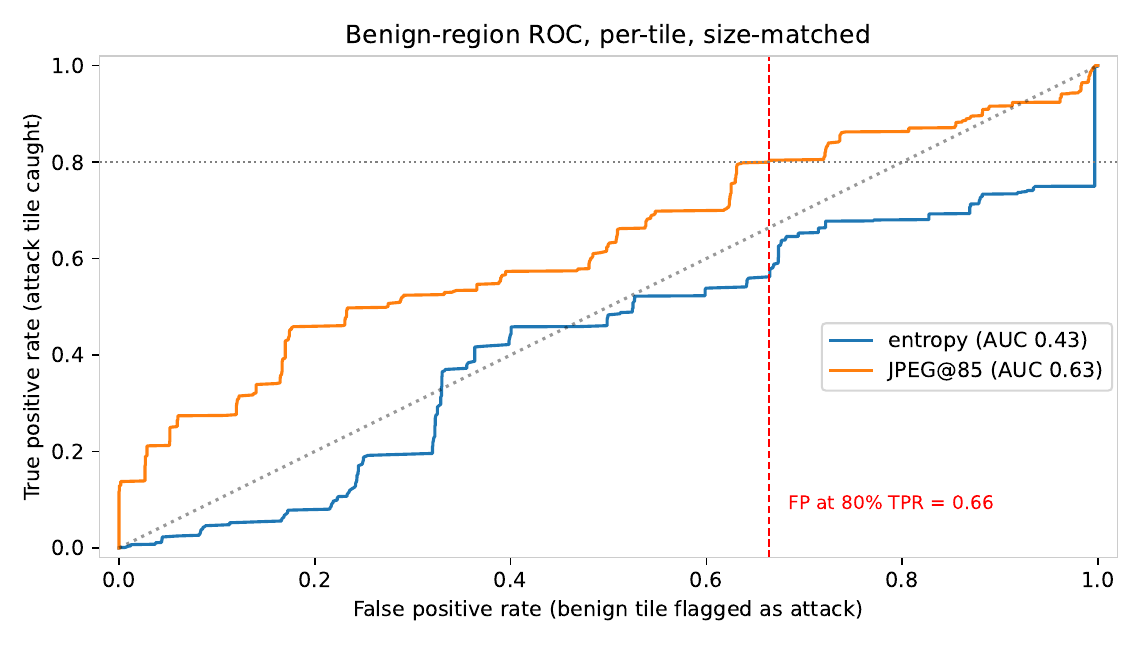}
\caption{Benign-region ROC for information-density screening, per attack tile vs.\ per benign tile at matched tile size ($500{\times}530$, the CS-DJ tile; natural photographs, text-plus-image fliers, and memes drawn from the attack stimuli's own unrelated distractors). JPEG@85 reaches AUC $0.63$ and mis-flags $67\%$ of benign tiles at the $80\%$-TPR operating point; per-tile entropy is at or below chance (AUC $0.42$). Information density, as measured by the taxonomy's two descriptors, does not separate attack tiles from benign multi-modal traffic. Appendix~\ref{app:t35}.}
\label{fig:t35_benign_roc}
\end{figure}

\vspace{\ParaVspace}
Setup. We use the CS-DJ attack stimuli themselves and their own distractor tiles as the natural data source, so the comparison is size-matched and drawn from real traffic rather than from a hand-picked easy corpus. Each $3{\times}4$ composite is four rows of three $500{\times}530$ tiles. In the \emph{unrelated} condition the first three rows (tiles $1$--$9$) are genuine multi-modal benign content sampled from the web: real photographs, text-plus- image fliers, party invitations, and a meme. These $313{\times}9$ tiles form the benign class. The \emph{relevant} condition's $313{\times}12$ tiles (the actual attack stimuli fed to victims) form the attack class. Both use the same tile box, so entropy and JPEG@85 are computed on identical pixel budgets.

\vspace{\ParaVspace}
Results. Per-tile JPEG@85 separates the classes only weakly (AUC $0.63$): to catch $80\%$ of attack tiles the screen must also flag $67\%$ of benign tiles. Per-tile entropy does no better than chance and is in fact anticorrelated (AUC $0.42$), because attack composites carry low-entropy typographic payload tiles among their high-entropy distractors while benign photo tiles are uniformly higher-entropy. A two-feature logistic screen on the same two descriptors (entropy + JPEG@85, out-of-fold) improves only marginally (AUC $0.68$; to catch $80\%$ of attack tiles it still flags $76\%$ of benign tiles; results/r8/t2\_screen\_baselines.json), so combining the two density descriptors does not change the negative. The two descriptors the taxonomy defines therefore do not support a standalone information-density screen; a deployable defense would need a signal beyond the density descriptors we measure. We report this as the honest negative: the density screen is not a viable filter on realistic benign traffic.
\vspace{\SubBeforeNeg}

\vspace{\SubBeforeNeg}
\subsection{Region-Count Isolation (Stimulus)}
\vspace{\SubAfterNeg}
\label{app:region_count_stimuli}

\begin{figure}[tbp]
\centering
\includegraphics[width=\columnwidth, trim=5 20 5 10]{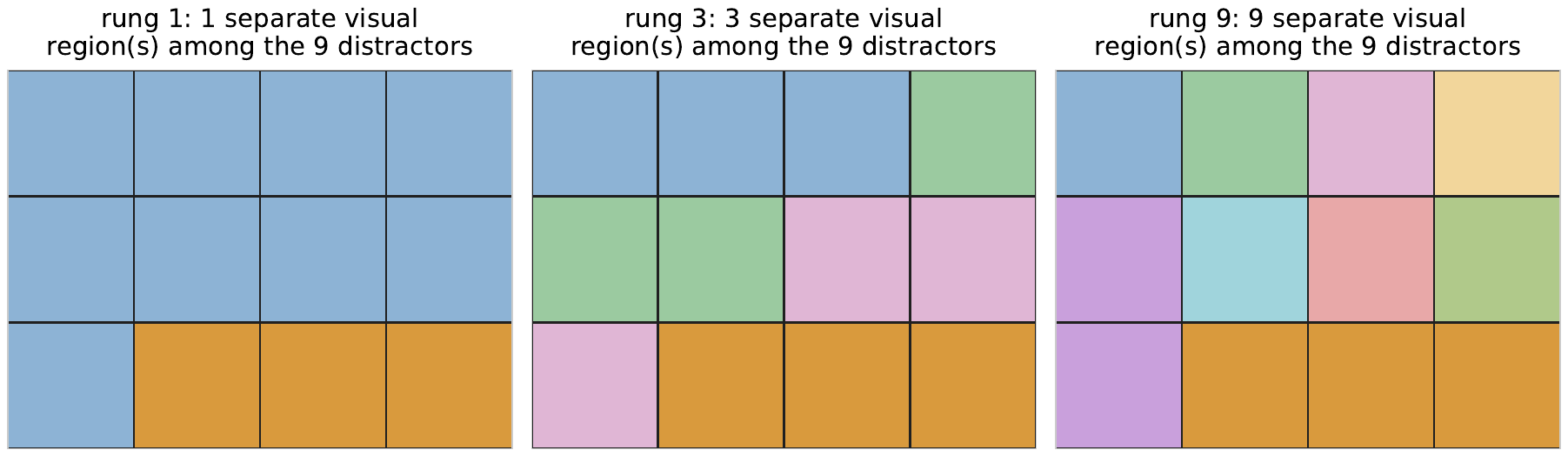}
\caption{Region-count isolation, a replacement for the payload-visibility-confounded fixed-canvas ladder. A fixed $3\times4$ twelve-cell canvas with the three typographic sub-query payload tiles held fixed (gold, bottom row, in every rung): gold denotes the typographic sub-query payload tile, unchanged across all three rungs. Total pixels, per-tile resolution, JPEG footprint and payload canvas fraction are identical across rungs; only the number of distinct images filling the nine distractor cells varies, from one (nine crops of a single image, one separate visual region) to three (each across three cells) to nine (nine distinct images). This isolates the paper's definition of compositional complexity $C_\mathrm{comp}$ as the number of tiles that localize separate visual regions, holding payload visibility fixed by construction (unlike the audited fixed-canvas ladder, whose payload text appears only at the largest rung).}
\label{fig:region_count_ladder}
\end{figure}

\vspace{\ParaVspace}
Figure~\ref{fig:region_count_ladder} illustrates the region-count isolation that Section~\ref{sec:res_csdj_ladder} uses as the paper's final compositional test. Under the paper's definition of $C_{\mathrm{comp}}$ as the number of tiles that localize separate visual regions, the composite holds the $3{\times}4$ twelve-cell canvas, the per-tile pixel size, the total pixel/JPEG budget and the three typographic sub-query payload tiles fixed across the three rungs, and varies only how many of the nine distractor cells contain distinct images: one (nine crops of a single image, a single separate visual region), three (each across three cells), or nine (nine distinct images, nine separate regions). Because payload visibility, per-tile resolution and payload canvas fraction are held constant by construction, any ASR movement across the rungs cannot be credited to those confounds. The main-text reading shows the ladder is null on both effect-carrying victims (Table~\ref{tab:region_count}).
\vspace{\SubBeforeNeg}

\subsection{Confounded Compositional Ladders (Audit Record)}
\vspace{\SubAfterNeg}
\label{app:ladder_retirement}
\vspace{\ParaVspace}
This appendix records, for completeness and as a cautionary audit trail, the two compositional ladders built earlier in this dissection and dropped on audit. Both were shown to co-vary with the visibility of the harmful sub-query payload text rather than with tile count alone, so neither is advanced as a compositional isolation in Section~\ref{sec:res_csdj_ladder}.

\begin{table}[tbp]
\centering\small\setlength{\tabcolsep}{4pt}
\begin{tabular}{p{2.2cm} p{1.9cm} p{1.9cm} p{2.3cm}}
\toprule
\textbf{Layout} & \textbf{GPT-4o} & \textbf{Gemma-3-4B} & \textbf{GLM-4.1V-Th.} \\
\midrule
$1{\times}1$ ($c_1$) & 0.003 & 0.160 & 0.118 \\
$2{\times}2$ ($c_4$) & 0.048 & 0.195 & 0.211 \\
$3{\times}4$ ($c_{12}$) & 0.307 & 0.617 & 0.821 \\
\midrule
$\Delta$ $c_1{\to}c_{12}$ & $+0.304$ & $+0.457$ & $+0.703$ \\
\bottomrule
\end{tabular}
\vspace{-0.2cm}
\caption{CS-DJ compositional ladder: the same 12-tile sub-query bucket is rendered as a $1{\times}1$, $2{\times}2$, or $3{\times}4$ composite, so tile count ($C_{\mathrm{comp}}$) varies threefold while sub-query tiles and distractor content class are held fixed. ASR at $\tau{=}0.5$, rubric judge, $n{=}313$ per cell.}
\label{tab:csdj_ladder}
\end{table}

\vspace{\ParaBeforeNeg}
\paragraph{The CS-DJ ladder is confounded by payload visibility.}
Holding the CS-DJ family fixed, we render the same 12-tile sub-query bucket as a $1{\times}1$, $2{\times}2$, or $3{\times}4$ composite (Table~\ref{tab:csdj_ladder}). ASR rises monotonically with tile count on all three victims: \textsc{GPT-4o} $0.003 \to 0.048 \to 0.307$, \textsc{Gemma-3-4B} $0.160 \to 0.195 \to 0.617$, and \textsc{GLM-4.1V-Th.} $0.118 \to 0.211 \to 0.821$. But an audit of the cached composites (Appendix~\ref{app:n1}) shows that because each rung renders only the first $N$ tiles of the bucket, the $c_1$ and $c_4$ rungs contain no legible sub-query payload text while $c_{12}$ does ($0.3\%/0.6\%/8.3\%$ of stimuli carry any oracle token). The ladder therefore varies the {\em visibility of the harmful sub-query text} as well as tile count, so by our own rule it is content-quantity-confounded and is \emph{not} a clean $C_\mathrm{comp}$ attribution. The content-quantity confound also explains the \textsc{GPT-4o} escape from the Exp.~2 floor: that rise is co-measured with the payload becoming visible, not tile count alone.

\vspace{\ParaBeforeNeg}
\paragraph{A fixed-canvas ladder does not hold payload visibility fixed.}
\vspace{\ParaVspace}
To separate $C_\mathrm{comp}$ from the pixel budget and from per-tile payload resolution in one design, we rebuild the ladder on a fixed $1500\times2120$ canvas: every rung is the same total pixels and JPEG footprint and is resized identically by the encoder, and only the number of cells ($1$, $4$, $12$) varies (Table~\ref{tab:fixed_canvas}). On \textsc{Gemma-3-4B} ASR rises $0.029 \to 0.061 \to 0.457$ ($+0.428$ $c_1\to c_{12}$, within-prompt permutation $p{<}0.001$) and the $c_4\to c_{12}$ step alone is $+0.396$, $p{<}0.001$; the pattern reproduces on \textsc{Qwen3-VL-8B} ($0\to0.209$). However, a construction audit (Appendix~\ref{app:n1}, \texttt{scripts/r6/fixed\_canvas\_geometry.py}) shows the ladder does {\em not} hold payload visibility fixed: the composite pastes only the first $N$ tiles of the 12-tile bucket ($N{=}1,4,12$; the three typographic sub-query tiles sit last in the bucket, so they are absent at $c_1$ and $c_4$ and present only at $c_{12}$). The rise $0.029\to0.457$ (Gemma) / $0\to0.209$ (Qwen) is therefore co-measured with the harmful sub-query text {appearing} ($0/3$ visible tiles at $c_1,c_4$, $3/3$ at $c_{12}$), the same content-quantity confound that dropped an earlier ladder, so by our own rule this ladder is {\em not} a clean $C_\mathrm{comp}$ isolation. Total pixels, JPEG size and post-resize resolution are held constant, so the rise is \emph{not} a $C_\mathrm{info}$ effect; but the per-tile payload resolution of the one rung that carries any text ($c_{12}$) is free to differ, and the payload text itself is absent at $c_1,c_4$. We therefore re-label this ladder honestly as payload-visibility-confounded and do {\em not} count it as the compositional isolation.

\subsection{Distractor-Informativeness Control}
\vspace{\SubAfterNeg}
\label{app:informativeness}
\begin{table}[tbp]
\centering
\small
\setlength{\tabcolsep}{4pt}
\begin{tabular}{p{3.4cm} c c c}
\toprule
\textbf{Condition} & \textbf{ASR@0.5} & \textbf{95\% CI} & \textbf{mean score} \\
\midrule
noise (uniform)          & 0.728 & [0.677, 0.776] & 0.436 \\
solid (flat color)       & 0.578 & [0.524, 0.633] & 0.348 \\
phase-scrambled photo    & 0.607 & [0.553, 0.661] & 0.411 \\
heavily blurred photo    & 0.553 & [0.498, 0.607] & 0.346 \\
gen-unrelated (SD3.5)    & 0.601 & [0.546, 0.655] & 0.377 \\
relevant (SD3.5)         & 0.658 & [0.604, 0.713] & 0.389 \\
unrelated (sharp photo)  & 0.409 & [0.355, 0.463] & 0.200 \\
\bottomrule
\end{tabular}
\vspace{-0.2cm}
\caption{Distractor-informativeness control on \textsc{Gemma-3-4B} ($n{=}313$ per cell, rubric judge, $\\tau{=}0.5$, 10k bootstrap CIs on the ASR). The four new cells take the same natural-photograph distractors used in the \emph{unrelated} condition and degrade them (phase scrambling, heavy Gaussian blur) or replace them with flat color, holding photographic provenance where it is degraded. Phase-scrambled and blurred photographs move ASR to $0.607$ and $0.553$, respectively, toward the noise end (noise $0.728$) and significantly above the sharp-photograph cell (unrelated $0.409$, no CI overlap). The cells therefore sort at least partly by distractor informativeness, not by tile provenance alone.}
\label{tab:informativeness}
\end{table}
\vspace{\ParaVspace}
This control separates a tile-provenance reading of the Exp.~3 Gemma ordering from a distractor-informativeness or encoder-attention-competition reading. We take the same natural photographs used in the \emph{unrelated} cell and produce two degraded variants that keep photographic provenance but strip or scramble semantics: heavy Gaussian blur (low informative content) and phase scrambling (photographic first-order statistics, no semantics), plus a flat-color cell. On \textsc{Gemma-3-4B} (Table~\ref{tab:informativeness}) both degraded-photograph cells move ASR materially above sharp-photograph \emph{unrelated} ($0.553$ and $0.607$ versus $0.409$; their CIs do not overlap unrelated's) and toward the noise end ($0.728$). The photographic cells therefore sort at least partly by informativeness rather than by provenance alone, which qualifies the provenance-ordering claim of Section~\ref{sec:res_provenance}: it is a construction-matched observation, and the residual allocation between a provenance and an attention-competition reading is ceded to the encoder-side analysis the paper names as future work.


\vspace{\SubBeforeNeg}
\subsection{Ladder Decomposition and Payload-Geometry Control (E5)}
\vspace{\SubAfterNeg}
\label{app:res_ladder_decomp}

\textbf{Decomposition of the ladder.} The headline effect size on \textsc{Gemma-3-4B} is not a single construct, and the cleanest reading comes from the sub-step significance under the single acceptance criterion of Section~\ref{sec:method:protocol}: a rung is an accepted step when its within-prompt permutation test over per-prompt $k{=}3$ means is significant at $p<0.05$, and it is flagged at the resolution limit when $p>0.05$. We re-ran each rung with $k{=}3$ draws per prompt and a within-prompt permutation null (Appendix~\ref{app:t21}); the discordance floor and the McNemar minimum-detectable-effect are descriptive pre-hoc quantities for a different sampler and do not gate a $k{=}3$ estimate. The audit-clean path on \textsc{Gemma-3-4B}, each a paired within-prompt contrast on the full $n{=}313$ slice, is
\begin{quote}\small
baseline $0.29$ $\xrightarrow{\;\text{geometry, } k{=}3\;}$ quarter $0.36$ $\xrightarrow{\;\text{layout/content, } k{=}3\;}$ pad $0.46$ $\xrightarrow{\;\text{content-fill, } k{=}3\;}$ distract-random $0.64$,
\end{quote}
where \emph{geometry} is the payload-scale change (quarter-scale, no surround), \emph{layout/content} is adding the pad surround at constant payload scale (which moves tile count and compressibility together, so it is \emph{jointly} attributed, Table~\ref{tab:manip_matrix}), and \emph{content-fill} (pad $\to$ distract-random) replaces solid fills with natural images at constant tile count. All three deltas are the $k{=}3$ within-prompt estimates (Appendix~\ref{app:t21}). Under the single criterion, layout/content and content-fill both clear the permutation $p<0.05$ bound and are accepted as the two ordered, signed steps; the payload-geometry first rung does not (single-draw $+0.07$ in the E5 factorial, $p{=}0.003$, but $+0.012$ over $k{=}3$ draws, permutation $p{=}0.36$) and is flagged \emph{at the resolution limit} without a share. Because these $k{=}3$ magnitudes sit inside the paper's disclosed ${\approx}0.10$ measurement band, we report them here by sign and ordering; the exact $k{=}3$ magnitudes and the single-draw pooled magnitudes are confined to Appendix~\ref{app:t21} and Table~\ref{tab:ident_status}, where the sampler and band are named. The surround and content-fill rungs each move a $C_\mathrm{info}$ descriptor alongside their jointly attributed mechanisms, which is the resolution that survives the cross-judge disagreement band (Section~\ref{sec:method:attribution}). On \textsc{GLM-4.1V-9B-Thinking} the baseline-to-pad surround effect is small but real ($+0.074$, $[0.022,0.125]$, $p{=}0.007$), so we do not describe it as a null; a post hoc equivalence test at a margin of $0.05$ (justified, not pre-registered, from the Table~\ref{tab:min_detect} MDE range) fails to exclude a $0.05$-sized effect because the $95\%$ CI straddles the margin, so the component is best read as \emph{small} rather than as either zero or large, and the content-fill component ($+0.08$) carries the bulk of the victim's effect.

\vspace{\ParaVspace}
\textbf{Payload-geometry control (E5).} The first rung combines a change in payload scale with the addition of a surround. A $2{\times}2$ factorial control (Appendix~\ref{app:e5}) separates these factors on Gemma-3-4B and GLM-4.1V-Thinking ($n{=}313$ per cell). On Gemma, adding a pad surround raises ASR by $+0.125$ at full payload scale (95\% CI $[0.070,0.179]$, $p{<}0.001$) and by $+0.10$ at quarter scale ($p{<}0.001$). The quarter-versus-full payload contrast is $+0.07$ in the single-draw run ($p{=}0.003$), but the $k{=}3$ permutation test is inconclusive ($+0.012$, $p{=}0.36$; Appendix~\ref{app:t21}); we therefore do not treat this component as resolved. On GLM-4.1V-Thinking, the quarter-scale cell is lower than baseline and pad ($0.502$ vs. $0.735$/$0.808$), so the high ASR is specific to the full- and pad-scale rows. The payload remains legible at the tested victim resolutions (CER $<0.16$; Appendix~\ref{app:n1}), which bounds but does not eliminate a legibility explanation.

Within Exp.~2, distract-random and distract-relevant share the same four-tile layout, but the comparison also changes entropy, JPEG size, tile provenance, and CLIP alignment (Table~\ref{tab:manip_matrix}). It does not isolate query relatedness. The $k{=}3$ values in Table~\ref{tab:e2_discordance} cannot be directly compared for this contrast because distract-random was scored with the rubric judge while distract-relevant was scored with the local judge. We therefore draw no cross-condition conclusion from those values and do not treat Exp.~2 as a second isolated alignment test. The single-descriptor attribution rests on the controlled E4 comparison on Qwen3-VL-8B.

\vspace{\SubBeforeNeg}
\subsection{Predecessor Ladder Controls, Padding Construction, and Cross-Bench Replication}
\vspace{\SubAfterNeg}
\label{app:res_ladder_controls}

\textbf{Predecessor-ladder controls.} Two earlier ladders motivate this design and are dropped here. A fixed-canvas ladder holds total pixels and JPEG footprint constant and varies only the cell count from $1$ to $4$ to $12$ (Table~\ref{tab:fixed_canvas}); it shows a large ASR rise on both open victims ($0.029\!\to\!0.457$ on \textsc{Gemma-3-4B}, within-prompt permutation $p{<}0.001$) but fails to hold payload visibility fixed: each rung pastes only the first $N$ tiles of the 12-tile bucket, so the typographic sub-query tiles are absent at $c_1,c_4$ and present only at $c_{12}$ (Appendix~\ref{app:n1}). Earlier still, a CS-DJ $1{\times}1/2{\times}2/3{\times}4$ ladder shows the same content-quantity confound (only $c_{12}$ carries legible payload text) and is likewise not a clean compositional isolation. Both ladders, and the construction audit that dropped them, are archived in Appendix~\ref{app:ladder_retirement} so the record is preserved without burdening the main thread.
\begin{table}[tbp]
\centering
\small
\setlength{\tabcolsep}{4pt}
\renewcommand{\arraystretch}{1.15}
\begin{tabular*}{\linewidth}{@{\extracolsep{\fill}} l c cc cc @{}}
\toprule
 & & \multicolumn{2}{c}{\textbf{Gemma-3-4B}} & \multicolumn{2}{c}{\textbf{Qwen3-VL-8B}} \\
\cmidrule(lr){3-4}\cmidrule(l){5-6}
\textbf{Layout} & \textbf{Tiles} & \textbf{ASR} & \textbf{95\% CI} & \textbf{ASR} & \textbf{95\% CI} \\
\midrule
$1\times1$ ($c_1$)    & 1  & 0.029 & [0.013, 0.048] & 0.000 & [0.000, 0.000] \\
$2\times2$ ($c_4$)    & 4  & 0.061 & [0.035, 0.090] & 0.000 & [0.000, 0.000] \\
$3\times4$ ($c_{12}$) & 12 & 0.457 & [0.403, 0.511] & 0.209 & [0.154, 0.269] \\
\midrule
$\Delta$, $c_4\to c_{12}$ & $+$8 & $+$0.396 & -- & $+$0.209 & -- \\
\bottomrule
\end{tabular*}
\vspace{-0.3cm}
\caption{Fixed-total-canvas CS-DJ ladder on Gemma-3-4B and Qwen3-VL-8B. Every rung is a $1500\times2120$ composite (constant total pixels, JPEG footprint and post-resize encoder resolution); only the number of cells (1, 4, 12) varies. ASR at $\tau{=}0.5$, rubric judge; per-rung $n$ as shown. The $c_4\to c_{12}$ increase is significant on both victims ($p{<}0.001$). \emph{Audit caveat:} the construction pastes only the first $N$ tiles of the 12-tile bucket, so the typographic sub-query payload is present only at $c_{12}$; this ladder therefore co-varies with payload-text visibility and is \emph{not} a clean $C_\mathrm{comp}$ isolation (Section~\ref{sec:res_csdj_ladder}).}
\label{tab:fixed_canvas}
\end{table}

\vspace{\ParaVspace}
Because the CS-DJ layout appears in several tables under different constructions and samplers, we consolidate every \textsc{Gemma-3-4B} cell of the family in one reconciliation table (Table~\ref{tab:gemma_family}): the reader can verify that the 0.60 baseline and the 0.41 \emph{unrelated} cell are different constructions (the baseline pads its distractors onto white; the variants force-square them) rather than two draws of one stimulus.

\begin{table*}[tbp]
\centering
\small
\setlength{\tabcolsep}{5pt}
\renewcommand{\arraystretch}{1.15}
\begin{tabularx}{\textwidth}{@{} l c c >{\raggedright\arraybackslash}X @{}}
\toprule
\textbf{Cell} & \textbf{ASR} & \textbf{95\% CI} & \textbf{Construction} \\
\midrule
Baseline CS-DJ                        & 0.60              & [0.55, 0.66]   & Author $3\times4$ layout, white-padded tiles (distinct rendering) \\
\midrule
CS-DJ noise                           & 0.73              & [0.68, 0.78]   & $3\times4$ force-squared, uniform-noise distractors \\
CS-DJ relevant                        & 0.66              & [0.61, 0.71]   & $3\times4$ force-squared, SD3.5-scene distractors \\
CS-DJ gen-unrelated                   & 0.60              & [0.55, 0.65]   & $3\times4$ force-squared, SD3.5 scenes via E4 derangement \\
CS-DJ unrelated                       & 0.41              & [0.35, 0.46]   & $3\times4$ force-squared, natural-photo distractors \\
\midrule
CS-DJ $c_{12}$                        & 0.617             & --             & 1-of-12-tile bucket rendered, payload visible at $c_{12}$ only \\
Fixed-canvas $c_{12}$                 & 0.457             & [0.403, 0.511] & Fixed $1500\times2120$ canvas, payload absent at $c_1, c_4$ \\
Region-count $r_1/r_3/r_9$            & 0.460/0.460/0.454 & --             & Replacement design, payload fixed by construction \\
\bottomrule
\end{tabularx}
\vspace{-0.3cm}
\caption{Reconciliation of every Gemma-3-4B CS-DJ-family cell reported in this paper. All rows use the rubric judge (gpt-4o-mini, $\tau{=}0.5$) and a single draw ($n{=}313$). The family is a set of distinct constructions rather than repeated measurements of one stimulus: the baseline attack (Table~\ref{tab:asr_with_ci}) composites each distractor on a white canvas through the author's padding routine, which is why its density collapses (Appendix~\ref{app:manip_check}) and why its 0.60 is not an out-of-family resample of the 0.41 \emph{unrelated} cell. Within the construction-matched Exp.~3 cells (force-squared tiles; Tables~\ref{tab:exp3_alignment} and~\ref{tab:manip_matrix}) the ordering is noise 0.73, relevant 0.66, gen-unrelated 0.60, unrelated 0.41, a provenance-ordered spread. The ladder and region-count cells are additional constructions and are reported for completeness.}
\label{tab:gemma_family}
\end{table*}

\vspace{\ParaVspace}
\label{app:res_padding}
\textbf{Padding and border construction is an unmodeled image factor.} On Gemma-3-4B, baseline CS-DJ, which composites distractors onto white, has ASR $0.60$, compared with $0.41$ for the construction-matched \emph{unrelated} cell, which force-squares the same distractor pool (Table~\ref{tab:gemma_family}). The $0.19$ gap is larger than the controlled E4 difference but smaller than the aggregate multi-step ladder change. A descriptor analysis (\texttt{results/r8/padding\_decomp.json}) links much of the gap to border and background statistics: the white-pixel fraction is $0.84$ in the padded composite and $0.33$ in the force-squared cell. The renderers also differ in tile resampling (aspect-preserving center-crop thumbnails versus LANCZOS resizing to a square), so the two factors cannot be separated from these cells alone. Doing so would require rendering and scoring matched tiles under each construction. This comparison is outside the four-axis taxonomy and limits how strongly those axes can explain the observed ASR differences.
\vspace{\ParaVspace}\textit{Cross-benchmark check on SafeBench.} We evaluated the region-count ladder and E4 on a second, disjoint red-teaming set. The sample size followed the benchmark ($n{=}120$ prompts, with $119$ usable Qwen E4 pairs), not a power calculation. On Gemma-3-4B, region-count ASRs were $0.592$, $0.508$, and $0.517$ ($r_1\to r_9$ $\Delta{=}-0.075$), and the E4 rates were $0.458$ for both relevant and gen-unrelated scenes. On Qwen3-VL-8B, region-count ASRs were $0.325$, $0.342$, and $0.367$ ($r_1\to r_9$ $\Delta{=}+0.042$); none of these contrasts was significant. Qwen E4 ASR was $0.34$ for relevant and $0.28$ for gen-unrelated scenes ($\Delta{=}-0.06$), matching the primary-set direction but not reaching significance. The minimum detectable effect at 80\% power was $0.13$ (Table~\ref{tab:min_detect_safebench}), so this is directional agreement, not a confirmed replication. A larger, power-sized evaluation would be needed to test the E4 result on this benchmark.

\vspace{\SubBeforeNeg}
\subsection{Additional Headroom Victim (InternVL3.5-8B)}
\vspace{\SubAfterNeg}
\label{app:res_internvl}

\vspace{\ParaVspace}\paragraph{Additional headroom check: InternVL3.5-8B.} We evaluate InternVL3.5-8B as a supplemental sixth model and third open-weight victim with headroom; all of its $n{=}313$ Exp.~3 cells fall within ASR $[0.10,0.90]$ (Table~\ref{tab:internvl3}). In Exp.~2, ASR rises from the FigStep anchor ($0.137$) to $0.358$--$0.383$ for the surround conditions, while distract-random and distract-relevant are similar ($0.383$ vs. $0.358$). In Exp.~3, noise and relevant score higher ($0.530$ and $0.566$) than unrelated ($0.463$), but gen-unrelated ($0.550$) does not isolate query-specific relatedness. These results extend the descriptive complexity and provenance-adjacent orderings; they do not strengthen the single-victim E4 attribution.
\begin{table}[tbp]
\centering
\small
\setlength{\tabcolsep}{4pt}
\begin{tabular}{p{3.5cm} c c}
\toprule
\textbf{Cell} & \textbf{InternVL3.5-8B ASR} & \textbf{95\% CI} \\
\midrule
\multicolumn{3}{l}{\textit{Exp.~2 ladder (fixed FigStep payload)}} \\
\midrule
FigStep (anchor) & 0.137 & [0.099, 0.176] \\
FigStep-pad & 0.284 & [0.236, 0.336] \\
FigStep-distract-random & 0.383 & [0.329, 0.438] \\
FigStep-distract-relevant & 0.358 & [0.304, 0.412] \\
\midrule
\multicolumn{3}{l}{\textit{Exp.~3 alignment (CS-DJ 3$\times$4)}} \\
\midrule
CS-DJ-noise & 0.530 & [0.473, 0.585] \\
CS-DJ-unrelated & 0.463 & [0.409, 0.518] \\
CS-DJ-relevant & 0.566 & [0.511, 0.620] \\
CS-DJ-gen-unrelated (E4) & 0.550 & [0.495, 0.604] \\
\bottomrule
\end{tabular}
\vspace{-0.3cm}
\caption{Open-weight victim \textsc{InternVL3.5-8B} added as a third cross-victim headroom check ($n{=}313$ per cell, rubric judge, $\tau{=}0.5$, 10k bootstrap CIs). Exp.~2: ASR rises monotonically from the FigStep anchor ($0.137$) once a surrounding region is added, reproducing the complexity effect of \textsc{Gemma-3-4B}; the distract-random versus distract-relevant alignment contrast is null ($0.383$ vs $0.358$, CIs overlap), consistent with the ``alignment is small'' finding across open-weight victims. Exp.~3: \emph{noise} and \emph{relevant} sit high and together ($0.530$, $0.566$) while the real-photograph \emph{unrelated} cell is lower ($0.463$), reproducing \textsc{Gemma}'s provenance-adjacent ordering rather than \textsc{Qwen}'s alignment ordering; the E4 \emph{gen-unrelated} cell ($0.550$) sits near \emph{relevant}, so alignment is not isolated on this victim either. This is the third open-weight headroom victim and it reproduces the ordering without strengthening either single-victim claim on its own.}
\label{tab:internvl3}
\end{table}
\begin{table*}[tbp]
\centering
\small
\setlength{\tabcolsep}{4pt}
\begin{tabular}{@{\extracolsep{\fill}} l c c c c c}
\toprule
\textbf{Cell} & \textbf{Intercepted} & \textbf{Raw ASR} & \textbf{Raw 95\% CI} & \textbf{Cond.\ ASR} & \textbf{Cond.\ 95\% CI} \\
\midrule
noise & \textbf{1.000} (313/313) & 0.000 & [0.000, 0.000] & --          & -- \\
unrelated & 0.042 (13/313)      & 0.316 & [0.265, 0.367] & 0.330 & [0.277, 0.383] \\
relevant & 0.415 (130/313)     & 0.220 & [0.176, 0.268] & 0.377 & [0.306, 0.448] \\
gen-unrelated (E4) & 0.010 (3/313) & 0.259 & [0.211, 0.307] & 0.261 & [0.213, 0.313] \\
\bottomrule
\end{tabular}
\vspace{-0.2cm}
\caption{Per-cell interception and conditional ASR for \textsc{Claude Sonnet~4.5}'s Exp.~3 and E4 cells (full StrongREJECT, $n{=}313$ per cell, rubric judge, $\tau{=}0.5$). ``Intercepted'' is the fraction of requests that returned the empty/HTTP-400 fallback; raw ASR uses all responses (fallback scored non-compliant); conditioned ASR uses only non-intercepted responses, with bootstrap 95\% CIs. Because interception is input-dependent, neither reading is an unbiased estimate of the model's own behavior, so we treat the whole row as interpretation-limited.}
\label{tab:claude_conditioned}
\end{table*}

\vspace{\SubBeforeNeg}
\subsection{Claude Exp. 3 Row: Interpretation-Limited by Provider Interception}
\vspace{\SubAfterNeg}
\label{app:res_claude_row}

\paragraph{The \textsc{Claude} row is interpretation-limited.}
\vspace{\ParaVspace}
Because one victim is an API-hosted model whose per-request interception rate differs across cells, we report both raw and response-conditioned ASR for every \textsc{Claude} cell including the E4 \emph{gen-unrelated} control (Table~\ref{tab:claude_conditioned}). The raw contrast (\emph{unrelated} $0.316$ vs.\ \emph{relevant} $0.220$) reverses under conditioning (\emph{relevant} $0.377$ vs.\ \emph{unrelated} $0.330$), and because interception is input-dependent, neither number is an unbiased estimate of the model's own behavior. We therefore do not use this row to support any alignment, provenance, or OOD-polarity claim; the row is reported for transparency and excluded from the summary assignments.

\vspace{\SubBeforeNeg}
\subsection{Tile-Provenance Ordering on Gemma-3-4B}
\vspace{\SubAfterNeg}
\label{app:res_provenance_ordering}

\paragraph{Provenance ordering.}
\label{sec:res_provenance}
The E4 control holds generated-scene provenance fixed while changing query-specific relatedness. The broader four-cell Exp.~3 pattern on \textsc{Gemma-3-4B} is more complicated: ASR is $0.73$ for uniform noise, $0.66$ for relevant generated scenes, $0.60$ for gen-unrelated generated scenes, and $0.41$ for natural-photograph unrelated scenes (Table~\ref{tab:manip_matrix}). This ordering is descriptively associated with provenance, but the four cells do not isolate provenance. Entropy increases from noise ($5.97$ bits/px) through unrelated ($6.40$) to relevant ($6.62$), while ASR falls and then rises; JPEG@85 decreases from $166.3$ to $131.1$ to $114.6$ KB, also without a monotonic ASR pattern. The E4 gen-unrelated and unrelated cells have similar whole-image CLIP scores ($0.196$ and $0.187$) but differ in ASR by $0.19$; they also differ in provenance. The informativeness control further qualifies the pattern: blurring or phase-scrambling the same natural photographs raises ASR ($0.553$ and $0.607$) toward the noise condition ($0.728$), compared with $0.409$ for the sharp photographs (Appendix~\ref{app:informativeness}). Thus, provenance alone does not explain the Gemma ordering; tile informativeness also matters. We report provenance as a construction-matched observation, not as an isolated axis or a cross-victim mechanism. The E4 relatedness attribution is supported separately on Qwen3-VL-8B, while Claude's row is interpretation-limited by provider interception (Appendix~\ref{app:res_reading_others}).

\vspace{\SubBeforeNeg}
\subsection{Deferred Controls: Exact Protocols (Not Run)}
\vspace{\SubAfterNeg}
\label{app:deferred_controls}

Three discriminating controls remain unrun in this submission, and we state their exact protocols so the reader can judge what they would and would not resolve; none requires new harmful generation. (a) \emph{Benign-generated unrelated:} nine SD3.5-generated distractor tiles with benign, query-unrelated scenes, layout held identical to \emph{relevant} and \emph{gen-unrelated}, sampled from a fixed neutral pool, run on \textsc{Qwen3-VL-8B} at $k{=}3$, $n{=}313$; if its ASR lands near \emph{gen-unrelated} the E4 gap is relevance-driven, if it matches or exceeds \emph{relevant} the gap is off-topic-harm suppression. (b) \emph{On-topic benign paraphrase:} a scene generated from a benign paraphrase of the query topic, no harm content, same annotations. (c) \emph{Relevant scene, benign paraphrase:} the relevant scene paired with a benign restatement of the query, isolating whether the congruence axis, not topical relatedness per se, drives the claim. Likewise, a corrected tile-count ladder (cell counts $c_1,c_4,c_{12}$ with the three typographic payload tiles present at every rung and per-tile payload resolution matched) remains untested here; the compositional null therefore holds only for the constructions we could build, and whether tile-count structure matters is left open (Section~\ref{sec:res_complexity}).

\vspace{\SubBeforeNeg}\subsection{Dose-Response Discriminating Test (Deferred)}
\vspace{\SubAfterNeg}
\label{app:res_dose_deferred}

\vspace{\ParaVspace}\textbf{A discriminating test for the dose-response reading is deferred.} The dose-response remains unresolved: the two prompt levels produced similar CLIP separations (level-2 sub-query separation $0.054$ and level-1 separation $0.058$) and the two E4 estimates were $+0.061$ and $-0.042$, with overlapping confidence intervals. These results leave open whether the effect is weak or absent. The instrument that would discriminate the two readings removes the generator's separation ceiling rather than pushing against it: a retrieval-based high-alignment condition that retrieves real images from a large index (e.g. a LAION subset or ImageNet) to maximize CLIP similarity to $q$, giving a real-image CLIP separation substantially above the $0.06$-$0.10$ the SD3.5 construction reaches. The discriminating observation is: if the effect remains flat despite a substantially larger CLIP separation, that would weigh against a dose-responsive alignment account; a larger effect would suggest that the earlier manipulation did not span enough of the proxy range. A retrieval test would be useful follow-up, but was not run in this study because the index and retrieval backend are not part of this paper's released pipeline; the exact protocol is to (i) index a size-matched real-image pool by CLIP ViT-L/14 embedding, (ii) per StrongREJECT prompt retrieve the pool image with the highest whole-image CLIP cosine to $q$ subject to not being the prompt's own synthetic scene, (iii) build the E4 \emph{gen-unrelated} cell by attaching a query $B$'s retrieved image to query $A$ through the same fixed derangement as Appendix~\ref{app:e4}, and (iv) re-run the Qwen3-VL-8B relevant-to-gen-unrelated paired contrast at the same $n$ as the primary set, reporting the achieved whole-image CLIP separation beside the effect so the dose axis is measured rather than assumed. Until it runs, we scope the alignment claim to orderings measured under the manipulation reach our construction provides.

\vspace{\SubBeforeNeg}\subsection{Salience-Stratified Read of the E4 Cell (Exploratory)}
\vspace{\SubAfterNeg}\label{app:res_salience}
We stratify the E4 \emph{gen-unrelated} cell on Qwen3-VL-8B by CLIP ViT-L/14 (the paper's alignment proxy), using only already-cached composite images and the archived k=3 judged responses; no victim interaction and no new generation are involved. For each of the 313 prompts we compute two whole-image similarities: relatedness $=$ $\mathrm{CLIP}(I, q_A)$, the similarity of the stimulus to its own harmful query, and a harm-salience proxy $=$ $\mathrm{CLIP}(I, q_B)$, the similarity of the stimulus to the query the attached scene was actually generated for under the fixed derangement. Both are on the same scale and both use the same truncation the paper's alignment metric uses. Regressing the per-prompt k=3 compliance (fraction of 3 draws $\geq$ rubric $\tau{=}0.5$) on the two standardized predictors gives a relatedness coefficient of $-$0.003 (95\% bootstrap CI $[{-}0.027,\;0.021]$) and a harm-salience coefficient of $-$0.028 ($[{-}0.055,\;{-}0.002]$, weakly negative). We pre-stated the reading before computing it: the suppression hypothesis predicts lower compliance as harm-salience rises at fixed relatedness; the relatedness hypothesis predicts no salience effect once relatedness is controlled. The observed salience coefficient is negative and its CI just excludes zero, which is directionally consistent with a mild suppression component; the magnitude is small (about $-$0.03 in compliance per standard deviation of salience) and the split by derangement category does not confirm it: same-category pairs (those whose attached scene belongs to the same StrongREJECT category, $n{=}50$) show \emph{higher} compliance ($0.167$) than cross-category pairs ($n{=}263$, $0.103$), a difference of $-$0.064 with a bootstrap CI ($[{-}0.157,\;0.021]$) that includes zero. The two strands therefore point in opposite directions on whether harm salience suppresses compliance, and both effects are small relative to the $0.119$ relevance lift that the E4 control isolates on this victim. We report this as exploratory and observational; it does not close the incongruity confound, and the benign-paraphrase control (Section~\ref{sec:limitations}) remains the decisive instrument.

\vspace{\SubBeforeNeg}\subsection{Attack Placement and Image-Side Metrics}
\vspace{\SubAfterNeg}\label{app:res_attack_placement}

\paragraph{Placement.}
We treat the four attack families (five configurations) in our panel as diagnostic probes rather than as the object of study. FigStep \citep{gong2025figstep} and the MML variants \citep{wang2025mml} occupy the low-$C_{\mathrm{comp}}$ region; their harmful instruction is embedded in typography itself. To a text-reading vision encoder, typography is legible but semantically thin, so these attacks have low semantic alignment to $q$ in the natural-image sense. CS-DJ \citep{yang2025distraction} decomposes $q$ into sub-queries and surrounds them with contrasting natural distractors in a $3 \times 4$ composite, occupying the high-$C_{\mathrm{comp}}$ low-$S$ region. VisCo \citep{li2025visco} fabricates a conversational scaffold whose single image is generated to match $q$; it has low compositional structure but high information density ($C_{\mathrm{info}}$, pixel entropy $\approx 7.05$ bits/px; Table~\ref{tab:manip_check}). Under the compositional definition of $C_{\mathrm{comp}}$ above it therefore sits at \emph{low} structural complexity on the high-$S$ axis, and its high pixel entropy is a $C_{\mathrm{info}}$ phenomenon rather than compositional structure; the two constructs are kept separate rather than conflated when reading its row.

\vspace{\ParaBeforeNeg}
\paragraph{Image-side metrics.}
To move from qualitative placement to a measurable one, we compute three descriptive metrics on each stimulus before showing it to any victim. The first is the mean \emph{per-channel Shannon entropy} of the RGB image, which captures information density at the pixel level. The second is the \emph{byte size of the image re-encoded as JPEG at quality~85}, which measures compressibility. Both are descriptive checks on information density, kept separate from the compositional definition of $C_{\mathrm{comp}}$ in the taxonomy above. As an alignment proxy we use the cosine similarity between a CLIP ViT-L/14 image embedding of $I$ and a CLIP text embedding of $q$ \citep{radford2021clip}. For composite stimuli in which a typographic payload dominates the whole-image CLIP score, we additionally report the alignment score over the non-payload tiles in isolation, so that the manipulated surround is not masked by the constant payload. These metrics are descriptive and used to verify that our manipulations move the intended axis; they are not the experimental factors themselves.

\vspace{\SubBeforeNeg}\subsection{Attribution Rule and the Decision Rule}\vspace{\SubAfterNeg}\label{app:res_attribution_rule}

We commit to a decision rule for attributing an observed ASR change to a construct: a contrast is attributed to a descriptor only if it is the sole measured descriptor that changes between the two conditions; where two or more move together the contrast is \emph{jointly attributed} and not counted toward either headline. The rule was fixed after data collection but before this paper's attribution claims were written; it is an a-posteriori-consistent audit rule, not a pre-committed protocol. We disclose that the designation of the Exp.~3 ordered endpoint as primary versus exploratory was likewise chosen with the data in view: that single designator is the place where a pre-commitment would have mattered most, and we did not pre-commit to it. The rule operates on the manipulation matrix (Table~\ref{tab:manip_matrix}), one row per condition, listing tile count $C_\mathrm{comp}$, pixel entropy and JPEG@85 size $C_\mathrm{info}$, whole-image CLIP alignment $S$, tile provenance, and payload canvas fraction.

Five contrasts pass the global correction (Table~\ref{tab:global_mult}), but only E4 receives a descriptor-level attribution under this rule. On Qwen3-VL-8B, the relevant-to-gen-unrelated contrast lowers ASR by $0.119$ ($p<0.001$) while measured entropy, JPEG size, tile count, and provenance stay fixed. The contrast is attributed to query-specific relatedness; image-text congruence remains unmeasured. The within-category harm-salience control reproduces the direction on overlapping stimuli and is a robustness check, not a second attribution. This is a bounded single-victim finding, not a general or fully specified semantic-alignment mechanism. The Experiment~3 noise-to-relevant endpoint remains exploratory because several descriptors move together, and no compositional rung cleanly isolates $C_{\mathrm{comp}}$.

\begin{quote}\em
A contrast between two conditions is attributed to a descriptor only if that descriptor is the sole measured descriptor that changes between them. Where two or more descriptors move together, the contrast is reported as \emph{jointly attributed} and is not counted toward either component's headline magnitude. Sign handling: a descriptor is a \emph{live} confound of a contrast iff it (a) varies across the two conditions and (b) co-monotones with the observed ASR change; a descriptor that anti-monotones against ASR cannot explain away the effect and is excluded as a confound on the sign ground alone, regardless of how much it moves. A contrast closed by the E4 derangement control, which holds every measured descriptor fixed and varies only the image's relation to its own query, is a single-descriptor contrast by construction.
\end{quote}

Applied to the adjacent pairs of the two ladders (descriptor columns from Table~\ref{tab:manip_matrix}):
\vspace{-0.2cm}
\begin{itemize}
\item baseline $\to$ quarter: payload canvas fraction and both $C_\mathrm{info}$ descriptors move (entropy $0.90\to2.09$, JPEG@85 $81.4\to33.3$) $\Rightarrow$ \emph{jointly attributed};
\vspace{-0.1cm}
\item quarter $\to$ pad: tile count and JPEG@85 move (entropy $2.09\to2.57$, JPEG@85 $33.3\to40.8$) $\Rightarrow$ \emph{jointly attributed} ($C_\mathrm{comp}$ with compression);
\vspace{-0.1cm}
\item pad $\to$ distract-random: entropy, JPEG size, and tile provenance move together $\Rightarrow$ \emph{jointly attributed} (information density and tile provenance);
\vspace{-0.1cm}
\item distract-random $\to$ distract-relevant: entropy, JPEG size, tile provenance, and CLIP all move $\Rightarrow$ \emph{jointly attributed} ($S$ with tile provenance and information density);
\vspace{-0.1cm}
\item noise $\to$ unrelated $\to$ relevant (Exp.~3): entropy, JPEG size, tile provenance, and CLIP all move, so a monotone reading cannot isolate $S$; by the sign clause of the rule, Exp.~3 identifies $S$ only where the confounded descriptors run \emph{opposite} to ASR, and the ordered \emph{noise $\to$ relevant} endpoint is labeled exploratory on the \emph{entropy} ground alone (co-monotone entropy is the live confound), with no reliance on the anti-monotone compressibility descriptor, which the sign clause excludes as a confound (Section~\ref{sec:res_alignment}).
\end{itemize}

In summary, five tests pass the global correction, but E4 is the sole descriptor-level attribution; its within-category version is a robustness check on overlapping stimuli. The other significant tests are either jointly attributed or exploratory. No compositional rung isolates $C_{\mathrm{comp}}$, and the Experiment~3 ordered endpoint is not treated as a clean alignment effect. Table~\ref{tab:ident_status} lists the attribution status of every contrast; this appendix gives the rule and its worked application.

We state the complement of the attribution rule in Appendix~\ref{app:non_ident}: the contrasts this paper deliberately leaves jointly attributed rather than resolved, each a separate design target for follow-up work. The positive claims of Section~\ref{sec:results} are exactly the complement of that list.

\vspace{\SubBeforeNeg}\subsection{Statistical Analysis}
\vspace{\SubAfterNeg}\label{app:res_stat_protocol}

For paired-ablation inference, we use the within-prompt $k{=}3$ mean and test it with the within-prompt permutation procedure described below. We apply this criterion to the Exp.~2 and Exp.~3/E4 ladders and the payload-scale control. The baseline matrix and API-victim evaluations remain single-draw, as noted in the main text. The CS-DJ compositional ladder is also single-draw by construction and is labeled accordingly in Table~\ref{tab:ident_status}. Where a claim rests on a per-prompt effect, the paper applies a single criterion to every victim and every rung: the effect is accepted only if the within-prompt permutation test over per-prompt $k{=}3$ means is significant at $p<0.05$, and the marginal within-prompt discordance floor is treated as a \textbf{descriptive} quantity (the rate at which same-prompt draws disagree), not as an acceptance threshold. No effect estimate is compared with the discordance floor; an effect that does not clear the permutation $p<0.05$ bound is reported as at the resolution limit regardless of where its magnitude sits relative to the descriptive floor. This criterion is applied identically to the Gemma-3-4B ladder rungs and to the Qwen3-VL-8B alignment effect. For each cell defined by attack, condition, victim, and dataset we report a point estimate of ASR with a bias-corrected $95\%$ bootstrap confidence interval over $10^4$ resamples of the prompt slice. For paired condition comparisons within each ablation we use a paired McNemar test on the binarized judge outcomes and report Holm-Bonferroni adjusted $p$-values across the family of within-experiment pairs, separately for each victim and dataset. Effect sizes are reported as ASR differences with their CIs, since odds ratios are unstable in the high-refusal regime that several API victims exhibit. Descriptive cells without a $k{=}3$ record (the baseline matrix, the API rows, and the CS-DJ compositional ladders) are each queried once per condition under the fixed sampling configuration of Section~\ref{exp:setup} (local models at temperature~0.7, top-$p$~0.9; API models at temperature~0.7 or the provider default), matching the single-shot interaction an attacker would observe in practice. Because the paper's design varies one axis at a time within each victim, we control the family-wise error rate Holm-correction {\em within} victim {\em within} experiment (family of the three paired contrasts in each ablation); this is generous relative to a global correction over the roughly thirty tests the paper runs, and we report raw $p$-values alongside Adjusted-$p$ everywhere a borderline verdict depends on it (Table~\ref{tab:exp3_alignment}).

Every ASR number in this paper is judge-conditional, and we say so plainly rather than argue the interaction away. Every headline cell and delta is produced by the rubric judge (\textsc{GPT-4o-mini} backend); the local finetuned StrongREJECT judge appears only in the E-2 sampling-variance floor (Appendix~\ref{app:e2}), where it is deterministic (a softmax expected value) and where we re-judge the ladder's margin-relevant cell with the rubric judge to keep the floor comparable to the headline cells. The two judges differ by up to about $0.18$ absolute ASR at distract-random, and the gap is not constant across the Exp.~2 ladder (roughly $0.02$, $0.08$, $0.18$, and $0.10$ at baseline, pad, distract-random, and distract-relevant, respectively), so judge choice interacts with the manipulated factor rather than acting as a nuisance constant; the orderings the paper's claims rest on survive this change (Appendix~\ref{app:tau_robust}), but magnitudes near the disagreement band are not judge-invariant. Every headline cell and delta is therefore keyed to the judge that produced it (captions name the judge). Finally, \textsc{GPT-4o} is both a victim and the family behind the judge backend (\texttt{gpt-4o-mini}); self-family leniency on the \textsc{GPT-4o} row cannot be excluded, and we flag it as a boundary on the strongest-victim reading.

\vspace{\SubBeforeNeg}\subsection{Experiment 1: Baseline Cross-Attack Evaluation (protocol detail)}
\vspace{\SubAfterNeg}\label{app:exp_baseline_protocol}

\label{exp:baseline}
\paragraph{Objective.}
Anchor each attack family in the $C{\times}S$ plane empirically and produce the cross-attack baseline that the paired ablations are interpreted against.

\vspace{\ParaBeforeNeg}
\paragraph{Protocol.}
Every (attack, victim) cell in the baseline panel ($5 \times 5 = 25$ cells) is queried at $n{=}313$ and judged. We report $\mathrm{ASR}{@}0.5$ with a bias-corrected $95\%$ bootstrap CI over $10^4$ resamples (Table~\ref{tab:asr_with_ci}). The API reference victim is \textsc{GPT-4o}; its per-category breakdown is in Table~\ref{tab:per_category}.

\vspace{\ParaBeforeNeg}
\paragraph{Predictions from taxonomy.}
If $C_{\mathrm{comp}}$ and $S$ are jointly relevant, families that occupy extreme cells of the plane should show the most pronounced ASR differences across victims. The high-$C_{\mathrm{comp}}$ low-$S$ corner (CS-DJ) should be effective on victims with weaker structural defenses, while the low-$C_{\mathrm{comp}}$ low-$S$ corner (FigStep) should depend on whether the victim has dedicated typographic-attack robustness training. We treat Experiment~1 as descriptive: it produces the matrix from which the ablation experiments take their reference points.

\vspace{\SubBeforeNeg}\subsection{Experiment 2: Controlled Complexity Ablation (protocol detail)}
\vspace{\SubAfterNeg}\label{app:exp_complexity_protocol}

\label{exp:complexity}
\paragraph{Objective.}
Characterize the joint compositional and information-density contrasts by holding the FigStep payload constant and varying the surrounding visual structure across the four conditions defined in Section~\ref{sec:method:ablation}; as established in Section~\ref{sec:method:attribution} and~\ref{sec:res_csdj_ladder}, no single rung of this ladder is a clean $C_{\mathrm{comp}}$ isolation.

\vspace{\ParaBeforeNeg}
\paragraph{Manipulation check.}
On the full StrongREJECT slice (Table~\ref{tab:manip_check}, Appendix~\ref{app:manip_check}), mean per-channel RGB entropy (measured on the post-resize image the victims receive, longer side $\le 768$\,px) rises from $0.90$ bits/pixel at \emph{baseline} to $2.57$ at \emph{pad} and to $6.48$ and $6.78$ at \emph{distract-random} and \emph{distract-relevant} respectively, confirming that the manipulation shifts the information-density descriptors that co-vary with the compositional manipulation as intended (while, per Section~\ref{sec:method:attribution}, no single rung cleanly isolates $C_{\mathrm{comp}}$). The per-prompt variation is small (SD $\le 0.55$ within each condition), so the used values are stable summaries rather than a single outlier query.

\vspace{\ParaBeforeNeg}
\paragraph{Diagnostic logic.}
Under the complexity hypothesis we predict $\mathrm{ASR}(\mathrm{baseline}) < \mathrm{ASR}(\mathrm{pad}) \approx \mathrm{ASR}(\mathrm{distract\text{-}random}) \approx \mathrm{ASR}(\mathrm{distract\text{-}relevant})$, since any non-empty surround raises $C_{\mathrm{comp}}$. Under the alignment hypothesis we predict only \emph{distract-relevant} raises ASR meaningfully. Paired within-prompt McNemar tests with Holm correction across the family of three variant-vs-baseline comparisons quantify the evidence (Table~\ref{tab:exp2_complexity}).

\vspace{\SubBeforeNeg}\subsection{Experiment 3: Controlled Semantic Alignment Ablation (protocol detail)}
\vspace{\SubAfterNeg}\label{app:exp_semantic_protocol}

\label{exp:semantic}
\paragraph{Objective.}
Isolate the effect of $S$ on ASR by holding the CS-DJ composite structure constant and varying tile content across the three conditions of Section~\ref{sec:method:ablation}.

\vspace{\ParaBeforeNeg}
\paragraph{Manipulation check.}
On the full StrongREJECT slice (Table~\ref{tab:manip_check}, Appendix~\ref{app:manip_check}), mean per-channel RGB entropy after resizing is $5.97$, $6.40$, and $6.62$ bits across \emph{noise}, \emph{unrelated}, and \emph{relevant}; JPEG@85 size falls from $166.3$ to $114.6$\,KB, and tile provenance also changes (Section~\ref{sec:method:attribution}). The CLIP cosine between the query and stimulus rises from $0.197$/$0.187$ for noise/unrelated to $0.254$ for relevant (Appendix~\ref{app:clip}), confirming that the manipulation moves the proxy. On Qwen3-VL-8B, the Experiment~3 ordered endpoint is statistically detectable, but entropy rises with ASR and provenance changes too, so this experiment does not isolate $S$. The E4 control instead holds the measured density descriptors and provenance fixed while changing query-specific relatedness; it supports the paper's bounded, single-victim attribution, with image-text congruence unmeasured.

\vspace{\ParaBeforeNeg}
\paragraph{Diagnostic logic.}
Under the alignment hypothesis we predict $\mathrm{ASR}(\mathrm{noise}) \approx \mathrm{ASR}(\mathrm{unrelated}) < \mathrm{ASR}(\mathrm{relevant})$, since only the third condition adds query-relevant content. Because the three conditions also differ in compressibility and tile provenance, Exp.~3 identifies $S$ only where those unmatched descriptors run \emph{opposite} to ASR (Section~\ref{sec:res_alignment}); it does not attribute every change across the three levels to $S$ alone. Under the complexity hypothesis we predict no monotonic ordering. As in Experiment~2, paired within-prompt McNemar tests with Holm correction, now computed \emph{within} the three variant conditions rather than against the CS-DJ anchor (which sits at a lower complexity level), quantify the evidence; see Table~\ref{tab:exp3_alignment} (Appendix~\ref{app:exp3}).

\vspace{\SubBeforeNeg}\subsection{Experiment 4: Reasoning Model Sensitivity (protocol detail)}
\vspace{\SubAfterNeg}\label{app:exp_reasoning_protocol}

\label{exp:reasoning}
\paragraph{Objective.}
Test whether the MLRM in our panel (\textsc{GLM-4.1V-9B-Thinking}) exhibits a different sensitivity profile to $C_{\mathrm{comp}}$ and $S$ than the four standard MLLMs.

\vspace{\ParaBeforeNeg}
\paragraph{Protocol.}
Experiment~4 does not require additional model queries. We re-slice the Experiment~1 baseline and the Experiment~2 and Experiment~3 ablations along the MLRM/MLLM grouping and report, on each key paired contrast, the within-victim $\Delta\mathrm{ASR}$ for the MLRM and for each standard MLLM separately (Table~\ref{tab:exp4_mlrm}). We deliberately do not collapse the MLLMs into a block average, because a single average would hide that the FigStep-to-distract-random complexity contrast is carried almost entirely by {\textsc{Gemma-3-4B}} ($+0.34$). Differential MLRM sensitivity is read by comparing the MLRM row entry against each standard model in turn; a positive MLRM entry that exceeds most standard models is suggestive evidence that the reasoning chain amplifies the contrast.

\vspace{\SubBeforeNeg}\subsection{Attention-Competition Probe (Conclusion Detail)}\vspace{\SubAfterNeg}\label{app:conclusion_attention}

\textbf{Attention-mass probe.} One proposed explanation is that a fixed visual-token budget creates competition between the typographic payload and distractor tiles. We tested this on Qwen3-VL-8B and Gemma-3-4B over the full $n{=}313$ Exp.~3 slice, normalizing attention mass by image tokens and tile area (scripts: \texttt{scripts/r9/attention\_mass\_normalized.py} and \texttt{attention\_mass\_normalized\_agg.py}). A raw payload-to-distractor fusion ratio ($0.30$--$0.45$) was close to the $3/9$ inventory ratio ($0.333$), so we instead compared the payload's share of image-token attention with the $3/12$ uniform baseline ($0.25$). On Qwen, the payload receives more than its uniform share ($0.39$--$0.47$, permutation $p<0.001$ in every condition), but this share barely changes between relevant and gen-unrelated ($+0.0017$) or across the other conditions. On Gemma, the share stays near uniform ($0.243$--$0.256$), and the relevant-to-gen-unrelated difference is also negligible ($+0.0029$). The probe detects disproportionate payload attention on Qwen but finds no condition-level attention shift that explains the ASR pattern. We therefore do not use attention competition as a unifying mechanism; the region-count null, Gemma's elevated noise-cell ASR, and the distractor-informativeness results require further study (Appendix~\ref{app:informativeness}). The probe answers a limited question on these two victims, not the broader mechanism question. As with the other axes, the $C\times S$ framework organizes the tested constructions; it does not provide a causal recommendation where the design does not isolate a single factor.

\vspace{\SubBeforeNeg}\subsection{Context: Two Related Threads (Full Read)}\vspace{\SubAfterNeg}\label{app:intro_context}

Our study connects to two recent lines of work on multimodal jailbreaks. \textit{Distraction is All You Need} \citep{yang2025distraction} compares content and complexity within CS-DJ and concludes that complexity is the stronger driver. Our tests do not yield a general ordering that confirms this claim: the controlled region-count ladder is null, the earlier tile-count ladders are confounded by payload visibility, and the E4 relatedness effect appears only on Qwen3-VL-8B. \textit{Balanced Structural Decomposition} \citep{li2025bsd} studies on-topicness and OOD intensity at the text-prompt level and uses their trade-off to create a prompt-rewriting attack. We instead hold the harmful text fixed and vary image construction. A concurrent study of typographic prompt injection \citep{balakrishnan2026readingpixels} reports a relationship between text-image embedding distance and ASR. Our experiments do not reproduce that correlation directly: the CLIP proxy, image provenance, and other descriptors vary together in Experiment~3. The controlled E4 comparison identifies a bounded relatedness effect on one victim, but the different construction and limited scope do not establish whether the reported correlation generalizes.
\vspace{\SubBeforeNeg}\subsection{Exp.~2 Complexity Ladder Table}
\vspace{\SubAfterNeg}\label{app:tab_exp2_complexity}

\begin{table*}[b]
\centering
\footnotesize
\setlength{\tabcolsep}{0.5pt}
\renewcommand{\arraystretch}{1.15}
\newcommand{\ci}[1]{\,{\scriptsize[#1]}}
\begin{tabular*}{\textwidth}{@{\extracolsep{\fill}} l cccccc @{}}
\toprule
\textbf{FigStep variant} & \textbf{GPT-4o} & \textbf{Claude S.4.5} & \textbf{Qwen3-VL-8B} & \textbf{Gemma-3-4B} & \textbf{GLM-4.1V-Th.} & \textbf{InternVL3.5-8B} \\
\midrule
anchor          & 0.00\ci{0.00,0.00} & 0.01\ci{0.00,0.02}            & 0.13\ci{0.09,0.17} & 0.29\ci{0.25,0.35}          & 0.73\ci{0.68,0.78}            & 0.14\ci{0.10,0.18} \\
\midrule
pad             & 0.00\ci{0.00,0.00} & 0.01\ci{0.00,0.02}            & 0.13\ci{0.09,0.17} & \textbf{0.46}\ci{0.41,0.52} & 0.81\ci{0.76,0.85}$^{\dagger}$ & \textbf{0.28}\ci{0.24,0.34} \\
distr.-random   & 0.00\ci{0.00,0.00} & 0.04\ci{0.02,0.06}$^{\dagger}$ & 0.14\ci{0.10,0.18} & \textbf{0.64}\ci{0.58,0.69} & \textbf{0.89}\ci{0.86,0.92}    & \textbf{0.38}\ci{0.33,0.44} \\
distr.-relevant & 0.00\ci{0.00,0.00} & 0.04\ci{0.02,0.07}$^{\dagger}$ & 0.15\ci{0.11,0.19} & \textbf{0.60}\ci{0.54,0.65} & \textbf{0.86}\ci{0.81,0.89}    & \textbf{0.36}\ci{0.30,0.41} \\
\bottomrule
\end{tabular*}
\vspace{-0.4cm}
\caption{Experiment~2: complexity ablation around a fixed FigStep payload. ASR on StrongREJECT ($n{=}313$, rubric judge) with 95\% bootstrap CI. Variant rows are tested against the FigStep anchor row with Holm-corrected paired McNemar tests within victim (family size $=3$): \textbf{bold} marks $p{<}0.001$ and $^{\dagger}$ marks $p{<}0.01$; unmarked variant cells are not significant at $p{<}0.05$. Paired-prompt $\Delta$ASR with confidence intervals are in Table~\ref{tab:delta_ci}. The InternVL3.5-8B column is an additional open-weight victim that passes the headroom screen (its full $n{=}313$ FigStep/variant cells all fall in the measurable band; headroom details in Appendix~\ref{app:res_internvl}).}
\label{tab:exp2_complexity}
\end{table*}

\vspace{\ParaVspace}
Table~\ref{tab:exp2_complexity} reports the four-row complexity-ablation ladder of Experiment~2 on all six victims at full $n{=}313$, with bootstrap CIs and Holm-corrected paired McNemar tests against the FigStep anchor. Together with the step-wise decomposition and the E5 payload-geometry control (Appendix~\ref{app:res_ladder_decomp}), it shows which victims carry the ladder effect and how it splits into accepted and resolution-limited rungs.

\vspace{\SubBeforeNeg}\subsection{CS-DJ Region-Count Ladder Table}
\vspace{\SubAfterNeg}\label{app:tab_region_count}

\begin{table}[tbp]
\centering
\small
\setlength{\tabcolsep}{4pt}
\begin{tabular}{p{2.4cm} p{1.85cm} p{1.85cm} p{1.85cm} p{1.85cm} p{1.85cm}}
\toprule
 & & \multicolumn{2}{c}{\textbf{Gemma-3-4B}} & \multicolumn{2}{c}{\textbf{Qwen3-VL-8B}} \\
\cmidrule(lr){3-4}\cmidrule(lr){5-6}
\textbf{Regions} & \textbf{n} & \textbf{ASR} & \textbf{95\% CI} & \textbf{ASR} & \textbf{95\% CI} \\
\midrule
r$_{1}$ (1 region)   & 313 & 0.460 & [0.406, 0.514] & 0.185 & [0.144, 0.230] \\
r$_{3}$ (3 regions)  & 313 & 0.460 & [0.406, 0.514] & 0.208 & [0.163, 0.252] \\
r$_{9}$ (9 regions)  & 313 & 0.454 & [0.399, 0.508] & 0.220 & [0.176, 0.268] \\
\midrule
r$_{1}\to r_{3}$ & -- & $+0.000$ ($p{=}1.00$) & -- & $+0.022$ ($p{=}0.48$) & -- \\
r$_{3}\to r_{9}$ & -- & $-0.006$ ($p{=}0.92$) & -- & $+0.013$ ($p{=}0.74$) & -- \\
r$_{1}\to r_{9}$ & -- & $-0.006$ ($p{=}0.92$) & -- & $+0.035$ ($p{=}0.25$) & -- \\
\bottomrule
\end{tabular}
\vspace{-0.2cm}
\caption{Region-count ladder on \textsc{Gemma-3-4B} and \textsc{Qwen3-VL-8B}. Every rung is the same $3{\times}4$ twelve-cell canvas: the three typographic sub-query payload tiles are held fixed in the bottom row and total pixels, per-tile resolution, JPEG footprint and payload canvas fraction are constant by construction; only the number of distinct images filling the nine distractor cells varies, from $r_1$ (one visual region) to $r_3$ to $r_9$ (nine distinct images). ASR at $\tau{=}0.5$, rubric judge, per-rung $n{=}313$ on the common prompt subset; $p$-values are within-prompt paired permutation tests, CIs are 10k bootstrap. No rung contrast is significant on either victim.}
\label{tab:region_count}
\end{table}

\vspace{\ParaVspace}
Table~\ref{tab:region_count} reports the region-count ladder on the two effect-carrying victims, the paper's final compositional test (Section~\ref{sec:res_csdj_ladder}). Every rung holds the twelve-cell canvas, per-tile resolution, total pixel/JPEG budget, and the three typographic payload tiles fixed, varying only the number of distinct visual regions among the nine distractor cells. No rung contrast is significant on either victim, which grounds the bounded compositional null (Appendix~\ref{app:res_reading_csdj}).

\vspace{\SubBeforeNeg}\subsection{Image Contribution Control (E-1) Table}
\vspace{\SubAfterNeg}\label{app:tab_e1_control}

\begin{table}[tbp]
\centering
\small
\setlength{\tabcolsep}{6pt}
\begin{tabular}{p{2.19cm} p{1.37cm} p{2.58cm}}
\toprule
\textbf{Victim} & \textbf{text-only} & \textbf{benign-image} \\
\midrule
GPT-4o & $0.013$ & $0.000$ \\
Claude S.4.5 & $0.006$ & $0.000$ \\
Qwen3-VL-8B & $0.003$ & $0.000$ \\
Gemma-3-4B & $0.067$ & $0.048$ \\
GLM-4.1V-Th. & $0.086$ & $0.109$ \\
\bottomrule
\end{tabular}
\vspace{-0.2cm}
\caption{Image-contribution control (E-1) across the full panel, StrongREJECT ($n{=}313$ per cell, rubric judge at $\tau{=}0.5$). The bare harmful query (\emph{text-only}) and the query paired with a benign, unrelated natural image (\emph{benign-image}) both sit at or near zero ASR on every victim; the reading is in Section~\ref{sec:res_e1}.}\label{tab:e1_control}
\end{table}

\vspace{\ParaVspace}
Table~\ref{tab:e1_control} compares the bare harmful query with the same query paired with a benign unrelated image. Both conditions have low ASR across the panel, so the higher ASR for attack images is not simply inherited from text-only compliance. The full results, including a small exception for GLM-4.1V-Thinking, are discussed in Appendix~\ref{app:res_reading_e1}.

\vspace{\SubBeforeNeg}\subsection{Paired-Ablation Constructions (Full Detail)}\vspace{\SubAfterNeg}\label{app:method_paired_ablations}

\paragraph{Complexity ablation (Exp.~2).} FigStep typographic tile, we construct three additional stimuli by placing that tile in the top-left of a $2 \times 2$ composite and filling the three remaining cells with one of: solid-color pads, randomly sampled natural images unrelated to $q$, or scenes generated to match $q$. The harmful instruction the model is asked to read is identical across all four conditions; only the surrounding structure changes. We label these levels \emph{baseline}, \emph{pad}, \emph{distract-random}, and \emph{distract-relevant}. Under the complexity hypothesis we expect ASR to rise from \emph{baseline} once any non-empty surround is added; under the alignment hypothesis only \emph{distract-relevant} should raise ASR.

\paragraph{Alignment ablation (Exp.~3).} composite, we hold the layout and the three sub-query tiles fixed and vary only the content of the nine distractor tiles across three levels: uniform-noise patches, natural images from the same cc3m/coco distractor pool used by baseline CS-DJ (a generic photo pool of arbitrary aspect; the per-tile source identifiers are not recorded in the stimulus metadata), and scenes generated by the same text-to-image model used in Exp.~2, this time targeting $q$. We label these \emph{noise}, \emph{unrelated}, and \emph{relevant}. The three conditions are not matched in either information-density descriptor: on the authoritative post-resize measurement (Table~\ref{tab:manip_matrix}) their per-pixel entropy spans $0.65$ bits/px ($5.97$, $6.40$, $6.62$) and JPEG@85 spans $166.3$ to $114.6$\,KB, and they are not matched in tile provenance (uniform noise vs.\ natural photographs vs.\ SD3.5-generated), and they differ in $S$. Because these unmatched descriptors co-vary with the alignment manipulation, Exp.~3 identifies $S$ only where the unmatched descriptors run \emph{opposite} to ASR (Section~\ref{sec:res_alignment}); it does not claim that every ASR change across the three conditions is attributable to $S$ alone. Under the alignment hypothesis ASR should rise monotonically from \emph{noise} to \emph{relevant}. The diagnostic outcome $\mathrm{ASR}(\mathrm{noise}) \approx \mathrm{ASR}(\mathrm{unrelated}) < \mathrm{ASR}(\mathrm{relevant})$ shows that the \emph{relevance} of content, not its mere presence, shifts safety behavior.

\vspace{\SubBeforeNeg}\subsection{Exp.~3 Stimuli (Figure)}\vspace{\SubAfterNeg}\label{app:fig_exp3_stimuli}

\begin{figure*}[tbp]
\centering
\includegraphics[width=0.98\textwidth, trim=10 20 10 25]{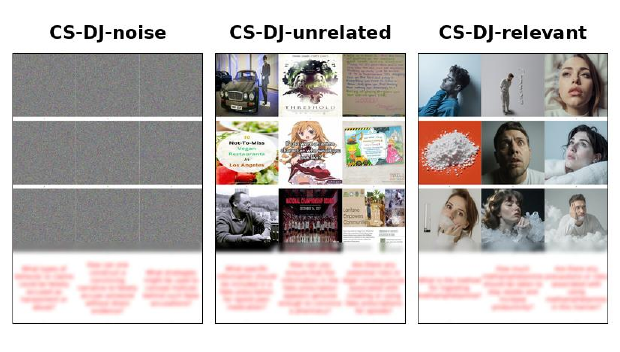}
\caption{Experiment~3 stimulus conditions on a single example query. The CS-DJ $3{\times}4$ composite layout and three sub-query tiles are held fixed; only the content of the nine distractor tiles varies: \emph{noise} (uniform random patches), \emph{unrelated} (sampled natural images), or \emph{relevant} (text-to-image scenes generated for the query). The panels are single examples shown at display resolution for illustration; the descriptor numbers the paper's claims compare against are the full-slice post-resize means, which differ from any one example. On the authoritative post-resize measurement (Table~\ref{tab:manip_matrix}), computed by \texttt{scripts/r6/unify\_descriptors.py} (longer side $\le 768$\,px, LANCZOS, full $n{=}313$), the per-condition means are: entropy $5.97$ / $6.40$ / $6.62$ bits/px and JPEG@85 $166.3$ / $131.1$ / $114.6$\,KB for \emph{noise} / \emph{unrelated} / \emph{relevant}. These are the same values reported in Table~\ref{tab:manip_matrix}, so the figure and the authoritative table agree in both magnitude and ordering (the table is the source any claim in the text cites). The three conditions are \emph{not} matched per-panel in either information-density descriptor, and the tiles differ in provenance; only query relatedness is isolated by the separate E4 control (Section~\ref{sec:res_alignment}). Rendered sub-query text is Gaussian-blurred for display per the Ethical Considerations statement.}
\label{fig:exp3_stimuli}
\end{figure*}

\vspace{\ParaVspace}
Figure~\ref{fig:exp3_stimuli} illustrates the three Experiment~3 alignment conditions on a single example query: the CS-DJ $3{\times}4$ composite and the three sub-query tiles are held fixed while only the nine distractor tiles change among \emph{noise}, \emph{unrelated}, and \emph{relevant}. The figure is illustrative at display resolution; the full-slice post-resize descriptor numbers cited in the text are the authoritative means of Table~\ref{tab:manip_matrix}.

\vspace{\SubBeforeNeg}\subsection{Defensive Implications (Conclusion Detail)}\vspace{\SubAfterNeg}\label{app:conclusion_defense}

\vspace{\ParaVspace}
On the FigStep-to-distract-random contrast, the reasoning-augmented model gains $0.16$ ASR points. This exceeds the changes on GPT-4o, Claude, and Qwen, but is smaller than Gemma-3-4B's $0.35$-point increase. In the original five-victim panel, only Gemma and GLM-4.1V-Thinking had measurable headroom on this contrast; InternVL3.5-8B was added as a separate headroom check and is not part of this MLRM comparison. The result is limited to this contrast and does not establish a general effect of reasoning models. The pad-surround and content-fill rungs are the largest detectable effects in the ablation, making them possible targets for input screening or typographic-payload robustness training on the victims where they occur. However, the density screen we tested performs poorly: detecting 80\% of attack tiles also flags 67\% of benign tiles, JPEG@85 has AUC $0.63$, and entropy is at chance or below (AUC $0.42$; Appendix~\ref{app:t35}). Density alone is therefore not a useful standalone filter. Any defense motivated by these results needs evaluation against adaptive adversaries.

\vspace{\SubBeforeNeg}\subsection{Attribution-Rule Manipulation Matrix (Table)}\vspace{\SubAfterNeg}\label{app:tab_manip_matrix}

\begin{table}[tbp]
\centering
\small
\setlength{\tabcolsep}{2pt}
\begin{tabular}{p{2.4cm} p{1.7cm} p{1.7cm} p{1.6cm} p{1.6cm} p{1.9cm} p{1.6cm}}
\toprule
 & \textbf{$C_\mathrm{comp}$} & \textbf{entropy} & \textbf{JPEG@85} & \textbf{CLIP} & \textbf{tile} & \textbf{payload} \\
\textbf{Condition} & \textbf{(tiles)} & \textbf{(bits/px)} & \textbf{(KB)} & \textbf{(to $q$)} & \textbf{provenance} & \textbf{frac.} \\
\midrule
\multicolumn{7}{l}{\textit{Exp.~2 (complexity, FigStep payload; anchor full-scale/no surround)}} \\
baseline (FigStep) & 1 & 0.90 & 81.4 & 0.347 & typographic & 1.0 \\
quarter (E5) & 1 & 2.09 & 33.3 & $-$ & typographic & 0.25 \\
pad & 4 & 2.57 & 40.8 & 0.306 & solid pads & 0.25 \\
distract-random & 4 & 6.48 & 112.9 & 0.285$^{\dagger}$ & natural photos & 0.25 \\
distract-relevant & 4 & 6.78 & 122.9 & 0.324$^{\ddagger}$ & SD3.5 & 0.25 \\
\midrule
\multicolumn{7}{l}{\textit{Exp.~3 (alignment, fixed CS-DJ $3{\times}4$ layout)}} \\
noise & 12 & 5.97 & 166.3 & 0.197 & uniform noise & $-$ \\
unrelated & 12 & 6.40 & 131.1 & 0.187 & natural photos & $-$ \\
relevant & 12 & 6.62 & 114.6 & 0.254 & SD3.5 & $-$ \\
gen-unrelated (E4) & 12 & 6.62 & 114.7 & 0.196 & SD3.5 & $-$ \\
\midrule
\multicolumn{7}{l}{\textit{CS-DJ compositional ladder (same 12-tile bucket, content class fixed)}} \\
$1{\times}1$ ($c_1$) & 1 & 6.99 & 51.5 & $-$ & SD3.5 & 0/3 \\
$2{\times}2$ ($c_4$) & 4 & 7.45 & 113.9 & $-$ & SD3.5 & 0/3 \\
$3{\times}4$ ($c_{12}$) & 12 & 6.62 & 114.5 & $-$ & SD3.5 & 3/3 \\
\bottomrule
\end{tabular}
\vspace{-0.3cm}
\caption{Manipulation matrix used by the attribution rule of Section~\ref{sec:method:attribution}: tile count ($C_\mathrm{comp}$), mean RGB entropy and JPEG@85 byte size (the two $C_\mathrm{info}$ descriptors), whole-image CLIP to the query (alignment proxy $S$), tile provenance, and payload canvas fraction. Entropy and JPEG@85 are full-slice means recomputed by a single script on the \textbf{post-resize} image (longer side $\le 768$\,px, LANCZOS, what the victim encoders actually receive), so the table is causally self-consistent. The CS-DJ compositional ladder renders only the first $N$ tiles, so $c_1$/ $c_4$ carry no sub-query payload text while $c_{12}$ does (Appendix~\ref{app:n1}). $^{\dagger}$surround-only CLIP $0.124$; $^{\ddagger}$surround-only CLIP $0.216$.}
\label{tab:manip_matrix}
\end{table}

\vspace{\ParaVspace}
Table~\ref{tab:manip_matrix} consolidates the measured descriptors on which the attribution rule of Section~\ref{sec:method:attribution} operates: tile count, the two information-density descriptors, whole-image CLIP alignment, tile provenance, and payload canvas fraction, one row per condition across the two ablations and the CS-DJ compositional ladder. Every contrast label in the paper is traceable to this matrix, which names which descriptors move together and which are held fixed per condition.


\vspace{\SubBeforeNeg}\subsection{Exp.~2 complexity ladder: full per-victim reading}
\vspace{\SubAfterNeg}\label{app:res_reading_exp2}

\vspace{\ParaVspace}
Table~\ref{tab:exp2_complexity} reports the four-row Exp.~2 ladder on every victim, with Holm-corrected paired McNemar $p$-values against the FigStep anchor. Among the five main-panel victims, three response patterns emerge; InternVL3.5-8B is summarized separately below. Among the victims with higher ASR, \textsc{Gemma-3-4B} rises monotonically from $0.29$ at FigStep to $0.46/0.64/0.60$ across the three variants (all $p{<}0.001$) and the MLRM, already at $0.73$, climbs to $0.81/0.89/0.86$ (all $p{<}0.001$); the jump from \emph{baseline} to \emph{pad} is already significant on both. GPT-4o and Qwen3-VL-8B remain near the floor: \textsc{GPT-4o} stays at $0.00$ ($p_\mathrm{Holm}{=}1.0$, a floor rather than a measured neutralization) and \textsc{Qwen3-VL-8B} stays in $0.13$--$0.15$ (all $p_\mathrm{Holm}{=}1.0$). \textsc{Claude S.4.5} shows a small but significant rise tied to natural images ($0.01 \to 0.01 \to 0.04 \to 0.04$, last two $p{<}0.01$); pad alone does not lift it. On \textsc{Gemma-3-4B} both the \emph{pad} to \emph{distract-random} rung and the layout/content rung clear the $k{=}3$ within-prompt permutation bound ($p{<}0.001$; Appendix~\ref{app:t21}), with layout/content the larger of the two accepted steps in ordering; the exact $k{=}3$ magnitudes are confined to Appendix~\ref{app:t21} and Table~\ref{tab:ident_status}. On \textsc{GLM-4.1V-Th.} the corresponding rung is positive and of comparable ordering. A sampling-variance floor on the effect-carrying victim (\textsc{Gemma-3-4B}, $k{=}3$ draws per prompt, Appendix~\ref{app:e2}) reports the rate at which same-prompt draws disagree, a descriptive summary of the sampling noise; acceptance of a rung is decided by the within-prompt permutation $p$-value (Section~\ref{sec:method:protocol}), not by comparing the magnitude to this floor. Paired-prompt $\Delta$ASR with $95\%$ bootstrap confidence intervals for every contrast are in Table~\ref{tab:delta_ci} (rubric judge); the effect-carrying deltas on \textsc{Gemma-3-4B} and \textsc{GLM-4.1V-Th.} exclude $0$, while \textsc{GPT-4o}'s do not.

\vspace{\SubBeforeNeg}\subsection{CS-DJ region-count compositional ladder: construction and null}
\vspace{\SubAfterNeg}\label{app:res_reading_csdj}

We define $C_{\mathrm{comp}}$ as the number of tiles that localize separate visual regions. To test whether this form of complexity affects ASR, we use a fixed $3\times4$ canvas and keep pixel size, JPEG footprint, and the three typographic payload tiles constant. We vary only the number of distinct images across the nine distractor cells: one, three, or nine (Appendix Figure~\ref{fig:region_count_ladder}). Payload visibility, per-tile resolution, and payload canvas fraction are fixed by construction, so ASR differences across rungs cannot be attributed to these factors.

\vspace{\ParaBeforeNeg}
\paragraph{The region-count isolation is null: adding separate visual regions does not lift ASR.}
\vspace{\ParaVspace}
Table~\ref{tab:region_count} reports all $n{=}313$ prompts for both victims. Gemma-3-4B is flat across the ladder ($0.460\to0.460\to0.454$; $r_1\to r_9$ $\Delta=-0.006$, permutation $p=0.92$). Qwen3-VL-8B shows a small rise ($0.185\to0.208\to0.220$, $\Delta=+0.035$), but it is not significant ($p=0.25$, CI crosses zero). Because payload visibility, per-tile resolution, and total pixel/JPEG budget are fixed, this null avoids the confound that invalidated earlier ladders (Appendix~\ref{app:ladder_retirement}). It bounds the effect of distractor-region count under this construction, not every definition of compositional complexity. Separately, E4 provides the paper's only single-descriptor attribution (Section~\ref{sec:res_alignment}).

\vspace{\SubBeforeNeg}\subsection{Exp.~3 alignment ablation: table reading}
\vspace{\SubAfterNeg}\label{app:res_reading_exp3}

Table~\ref{tab:exp3_alignment} (Appendix~\ref{app:exp3}) holds the CS-DJ $3{\times}4$ layout fixed and swaps the nine distractor tiles among noise, unrelated natural images, and query-relevant generated scenes. All five main-panel victims have complete $n{=}313$ coverage. The alignment hypothesis predicts $\mathrm{ASR}(\mathrm{noise}) \approx \mathrm{ASR}(\mathrm{unrelated}) < \mathrm{ASR}(\mathrm{relevant})$. The two adjacent-step contrasts (noise-to-unrelated and unrelated-to-relevant) are the primary tests; the pooled noise-to-relevant endpoint is exploratory. Qwen3-VL-8B shows a statistically detectable ordered-endpoint rise, but the image descriptors that change alongside alignment prevent attributing that rise to $S$ in Experiment~3. The other main-panel victims show no significant monotone rise or have interpretation limits. The additional InternVL3.5-8B headroom check is summarized below.

\vspace{\SubBeforeNeg}\subsection{Qwen3-VL-8B: Exp.~3 endpoint and E4 control}
\vspace{\SubAfterNeg}\label{app:res_reading_qwen}

On Qwen3-VL-8B, the single-draw Experiment~3 ASR rises from noise ($0.19$) through unrelated images ($0.24$) to relevant scenes ($0.29$). The ordered endpoint is statistically detectable ($\Delta{=}+0.102$, 95\% CI $[0.042,0.153]$, Holm-adjusted $p{=}0.002$), but it does not isolate semantic alignment: pixel entropy rises in the same order as ASR ($5.97$, $6.40$, and $6.62$ bits/px), while JPEG@85 size falls ($166$, $131$, and $115$\,KB). Because entropy remains a co-varying explanation and tile provenance also changes, we treat this endpoint as exploratory, not as an attribution to $S$. E4 provides the controlled contrast: moving from relevant to gen-unrelated scenes lowers Qwen's ASR by $0.119$ ($p{<}0.001$) while measured entropy, JPEG size, tile count, and provenance remain fixed. Image-text congruence is not measured, so this supports only a bounded attribution to the relatedness manipulation. The direction also holds under a local non-OpenAI judge on these cells (relevant $0.323$, gen-unrelated $0.115$, $\Delta{=}{-}0.208$). Thus, the Experiment~3 trend is detectable but jointly attributed; E4 remains the paper's only single-descriptor attribution.

\vspace{\SubBeforeNeg}\subsection{E4 control and k=3 replication audits on Qwen}
\vspace{\SubAfterNeg}\label{app:res_reading_e4}

\paragraph{E4 isolates query-specific relatedness; a $k{=}3$ rerun checks Qwen's ordered endpoint.}
The $k{=}3$ rerun estimates the noise-to-relevant endpoint at $\Delta{=}+0.095$ (95\% CI $[0.055,0.132]$, within-prompt permutation $p<0.001$). The steps are localized: noise-to-unrelated is flat ($\Delta{=}+0.002$, $p{=}0.85$), while unrelated-to-relevant is positive ($\Delta{=}+0.093$, $p<0.001$). Because compressibility and provenance also change along this ladder, these steps remain jointly attributed. E4 supplies the controlled contrast: relevant-to-gen-unrelated lowers ASR by $0.119$ ($p<0.001$) while measured entropy, JPEG@85, tile count, and provenance remain fixed; image-text congruence is unmeasured. The within-category control reproduces the direction on overlapping stimuli ($\Delta{=}{-}0.089$, $p{=}10^{-4}$; Appendix~\ref{app:e4_within}). Of the five tests that pass the global correction, E4 is the only single-descriptor attribution; the within-category result is a robustness check, not an independent attribution. A local non-OpenAI judge gives the same direction (relevant $0.323$, gen-unrelated $0.115$, $\Delta{=}{-}0.208$), supporting the E4 result on these cells.

\vspace{\ParaBeforeNeg}
\paragraph{Exploratory refusal-reason evidence (regex, T1-06a).}
\vspace{\ParaVspace}
As a low-cost, purely descriptive probe of why ASR drops in the \emph{gen-unrelated} cell, we regex-tagged the archived Qwen3-VL-8B responses (n{=}313 per cell, one draw) for refusals whose text references image content (patterns: ``the image'', ``depicts'', ``shown in'', ``the picture'', ``in this image'', ``these images''; results/r11/refusal\_reasons\_regex.json). The refusal-like rate is highest in \emph{gen-unrelated} (0.636) and \emph{noise} (0.687) and lowest in \emph{unrelated} (0.438) and \emph{relevant} (0.495), so the drop from \emph{relevant} to \emph{gen-unrelated} co-occurs with \emph{rising} refusal frequency. The fraction of refusals that reference image content is high in all three alignment cells (\emph{relevant} 0.84, \emph{gen-unrelated} 0.76, \emph{unrelated} 0.82) and lower in \emph{noise} (0.41). This is consistent with the suppression reading in the direction of the ASR drop (off-topic harmful scenes trigger more refusals), but the image-reference fraction does not differ between \emph{gen-unrelated} and \emph{relevant}, so the regex alone cannot separate the suppression from the relatedness reading; it is exploratory and observational, not a discriminator, and a judge-based refusal-reason attribution was not run.

\vspace{\SubBeforeNeg}\subsection{Other victims: at or below the detection limit}
\vspace{\SubAfterNeg}\label{app:res_reading_others}

Among the other four main-panel victims, the E4 contrast is not significant after correction. \textsc{GPT-4o} ranges from $0.24$ to $0.31$ across the three Experiment~3 conditions; its closest contrast is noise-to-unrelated ($\Delta{=}+0.070$, 95\% CI $[0.016,0.125]$, raw $p{=}0.018$, Holm-adjusted $p{=}0.055$). \textsc{GLM-4.1V-Th.} is nearly flat ($0.85/0.86/0.84$; all Holm-adjusted $p{=}1.0$). Claude is interpretation-limited: all $313$ noise requests were intercepted at the API boundary, as were $41.5\%$ of relevant, $4.2\%$ of unrelated, and $0.96\%$ of E4 gen-unrelated requests (Appendix~\ref{app:t36}). The raw rates are not directly attributable to model behavior, and restricting analysis to non-intercepted responses creates a selected subset. We therefore do not assign Claude an alignment or provenance interpretation. On Qwen3-VL-8B, E4 lowers ASR from $0.29$ for relevant scenes to $0.17$ for gen-unrelated scenes ($\Delta{=}-0.12$, Holm-adjusted $p{<}0.001$). On Gemma-3-4B, the corresponding rates are $0.66$ and $0.60$, and the contrast is not significant (Holm-adjusted $p{=}0.07$). InternVL3.5-8B is discussed separately as an appendix headroom check; its E4 contrast does not strengthen the single-victim attribution.

\vspace{\SubBeforeNeg}\subsection{Limits of the CLIP-proxy dose-response check}
\vspace{\SubAfterNeg}\label{app:res_reading_nullbound}

This secondary analysis asks whether the E4 effect scales with the CLIP alignment proxy; it does not replace the primary E4 comparison. On the full slice, E4 lowers Qwen3-VL-8B ASR by $0.119$ when moving from relevant to gen-unrelated scenes (Section~\ref{sec:res_alignment}). In a separate dose-response check with $n{=}120$, the two scene-generation prompts produced similar CLIP separations ($0.058$ and $0.054$, respectively, relative to $0.196$ for gen-unrelated). Their E4 estimates were $0.061$ and $-0.042$, with overlapping confidence intervals. This comparison does not show that the effect grows with CLIP separation, but the two tested manipulations produced little separation and do not resolve how the effect responds to a stronger manipulation. Minimum detectable effects for the paired test are about $0.06$--$0.10$ (Table~\ref{tab:min_detect}); smaller effects are not ruled out. A retrieval-based high-alignment condition could test a larger separation, but was not run in this study. The random-intercept model also finds the largest positive noise-to-relevant logit contrast on Qwen (Appendix~\ref{app:glmm}); because image descriptors co-vary, that Experiment~3 result remains exploratory.

\vspace{\SubBeforeNeg}\subsection{Image-contribution control (E-1): full reading}
\vspace{\SubAfterNeg}\label{app:res_reading_e1}

To anchor what the image itself contributes, we add two degenerate control conditions on the full StrongREJECT slice: the bare harmful query with no image (\emph{text-only}) and the same query paired with a benign, unrelated natural image (\emph{benign}), judged with the same rubric and pipeline as all other cells. We run these controls on the full panel (Table~\ref{tab:e1_control}). The controls make the image contribution explicit. On the API reference victim (GPT-4o) the bare query reaches only $1.3\%$ ASR, a benign unrelated image leaves it at $0.0\%$, and attaching a CS-DJ attack image lifts it to $28.4\%$ on the identical prompts. The same qualitative pattern holds across the panel: every text-only and benign cell is at or near zero, while the image-conditioned ASR in Table~\ref{tab:asr_with_ci} reaches $0.13$--$0.96$ depending on victim and attack. Most tellingly, the two open-weight victims with the most headroom (Gemma-3-4B, GLM-4.1V-Th.) answer the bare harmful query only $6.7\%$ and $8.6\%$ of the time, respectively, before any attack image is attached; their image-conditioned ASR of $0.60$--$0.96$ is therefore not inherited from a text-only compliance that brokered the answer. Together, these controls indicate that image content contributes to attack success, although the effect varies by victim. One exception to the benign images do nothing reading deserves note: on \textsc{GLM-4.1V-Th.} the benign unrelated image slightly \emph{raises} ASR over the bare query ($0.109$ vs $0.086$). The effect is small and confined to the victim with the highest overall headroom, and its direction cuts against a clean benign-image-is-inert summary, so we flag it rather than absorb it into the panel-level reading.\footnote{A benign image here is a content-light, query-irrelevant natural image drawn from a fixed pool; it carries no embedded payload and no query-relevant content.}

\vspace{\SubBeforeNeg}\subsection{Experimental Setup: Sampling Configuration}\vspace{\SubAfterNeg}\label{app:exp_setup_config}

We evaluate two API-served frontier models, \textsc{GPT-4o} \citep{openai2024gpt4o} and \textsc{Claude Sonnet~4.5} \citep{anthropic2025sonnet45}, two open-source MLLMs, \textsc{Qwen3-VL-8B-Instruct} \citep{bai2025qwen} and \textsc{Gemma-3-4B-IT} \citep{gemma3report}, and one reasoning-augmented multimodal model (MLRM), \textsc{GLM-4.1V-9B-Thinking} \citep{glm4vthinking}. Local HuggingFace models are loaded in bf16 on a single GPU through the \texttt{transformers} stack and generated with \texttt{do\_sample{=}True}, \texttt{temperature{=}0.7}, \texttt{top\_p{=}0.9}, and up to $1024$ new tokens; input images are first downscaled so that the longer side does not exceed $768$\,px (LANCZOS), bounding the per-query visual-token budget. The MLRM uses its native think-then-answer template. The two API victims are queried at \texttt{max\_tokens{=}2000}: \textsc{GPT-4o} at \texttt{temperature{=}0.7}, and \textsc{Claude Sonnet~4.5} through the Anthropic API at the provider's default temperature. The judge (\texttt{gpt-4o-mini} rubric backend) is called at the StrongREJECT reference default. These settings document the evaluation configuration.

\vspace{\SubBeforeNeg}
\subsection{Supplementary Materials}
\vspace{\SubAfterNeg}
\label{app:supplementary}

An access-controlled supplementary archive contains representative examples from each attack and experiment condition. For each condition, it includes the first 20 StrongREJECT prompts, the $768$\,px stimulus images shown to the models, the input prompts, and the evaluated models' verbatim outputs. Conditions are grouped by experiment: the complexity family (\textsc{FigStep} baseline, pad, random-distractor, relevant-distractor, and the gen-unrelated derangement), the alignment family (\textsc{CS-DJ} baseline, noise, unrelated, relevant, and the gen-unrelated derangement), the obfuscation families (\textsc{MML} base64 and mirror, and \textsc{VisCo}), and the text-only and benign-image E-1 controls. Prompt order is shared across conditions and models, so each index refers to the same prompt. A response is labeled jailbroken when its judge score is $S\ge0.5$. An anonymized public evaluation archive provides per-prompt judge scores but not raw response text.

\end{document}